\documentclass[preprint]{aastex}
\usepackage{hyperref}
\usepackage{xcolor}
\usepackage{graphicx}
\usepackage{booktabs}
\pdfoutput=1
\usepackage{multicol}
\let\oldbibliography\thebibliography
\renewcommand{\thebibliography}[1]{%
  \oldbibliography{#1}%
  \setlength{\itemsep}{0pt}%
}
 {\typeout{Couldn't patch the command}}

\begin{document}

\title{CLOVER: Convnet Line-fitting Of Velocities in Emission-line Regions}

\author{Jared Keown\altaffilmark{1}, James Di Francesco\altaffilmark{1,2}, Hossen Teimoorinia\altaffilmark{2,1}, Erik Rosolowsky\altaffilmark{3}, Michael Chun-Yuan Chen\altaffilmark{1}}

\email{jkeown@uvic.ca}

\altaffiltext{1}{Department of Physics and Astronomy, University of Victoria, Victoria, BC, V8P 5C2, Canada} 

\altaffiltext{2}{NRC Herzberg Astronomy and Astrophysics, 5071 West Saanich Road, Victoria, BC, V9E 2E7, Canada}

\altaffiltext{3}{Department of Physics, University of Alberta, Edmonton, AB, Canada}


\begin{abstract}  
When multiple star-forming gas structures overlap along the line-of-sight and emit optically thin emission at significantly different radial velocities, the emission can become non-Gaussian and often exhibits two distinct peaks.  Traditional line-fitting techniques can fail to account adequately for these double-peaked profiles, providing inaccurate cloud kinematics measurements.  We present a new method called Convnet Line-fitting Of Velocities in Emission-line Regions (CLOVER) for distinguishing between one-component, two-component, and noise-only emission lines using 1D convolutional neural networks trained with synthetic spectral cubes.  CLOVER utilizes spatial information in spectral cubes by predicting on $3\times3$ pixel sub-cubes, using both the central pixel's spectrum and the average spectrum over the $3\times3$ grid as input.  On an unseen set of 10,000 synthetic spectral cubes in each predicted class, CLOVER has classification accuracies of $\sim99\%$ for the one-component class and $\sim97\%$ for the two-component class.  For the noise-only class, which is analogous to a signal-to-noise cutoff of four for traditional line-fitting methods, CLOVER has classification accuracy of $100\%$.  CLOVER also has exceptional performance on real observations, correctly distinguishing between the three classes across a variety of star-forming regions.  In addition, CLOVER quickly and accurately extracts kinematics directly from spectra identified as two-component class members.  Moreover, we show that CLOVER is easily scalable to emission lines with hyperfine splitting, making it an attractive tool in the new era of large-scale NH$_3$ and N$_2$H$^+$ mapping surveys.


\end{abstract}


\section{Introduction}

\begin{figure}[h]
\epsscale{0.67}
\plotone{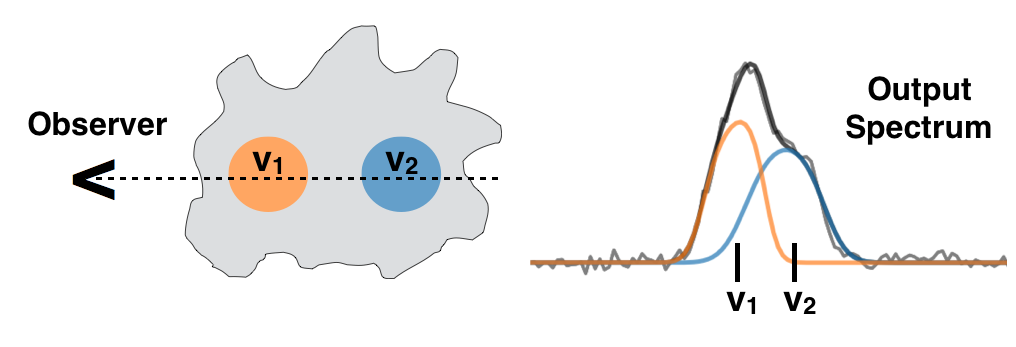}
\caption{Schematic diagram of a molecular cloud observation that would result in a spectrum with two velocity components.  The observer views two cores along the line of sight (dashed line) at slightly offset centroid velocities (v$_1$ and v$_2$).  The combination of the two Gaussian emission line profiles for each core (orange and blue spectra) results in a broadened observed spectrum (black spectrum) with a ``shoulder'' at one side.}
\label{cartoon}
\end{figure}

Kinematics observations of star-forming molecular clouds reveal the turbulent motions of the clouds' gas and provide an understanding of how gas is funneled onto sites of star formation \citep[e.g.,][]{Pineda_2010, Kirk_2013, Friesen_2013}.  Such kinematics measurements are obtained by modeling the emission lines from molecular transitions in the gas.  Typically, emission lines without self-absorption (i.e., optically thin lines) are modeled as Gaussian distributions with a single centroid velocity and velocity dispersion.  A major limitation of this ``single-Gaussian'' line fitting approach is its inability to account for spectra that display multiple velocity components along the line of sight.  For instance, if two slabs of emitting gas with slightly offset centroid velocities lie along our line of sight to a particular cloud, a broadened second peak or ``shoulder'' is produced in the observed spectrum.  Figure \ref{cartoon} shows a schematic of this situation.  Traditional single-Gaussian line fitting pipelines, which assume the observed emission contains a single velocity component, would fit this broadened spectrum with a line width that is much larger than those of the individual line components that produced the observed spectrum.  In addition, the centroid measurement would be skewed to a value in-between those of the individual line components.  These inaccuracies have significant impacts on many analyses of star-forming regions.  For example, virial stability analyses \citep[e.g.,][]{Kauffmann_2013, Pattle_2015, Pattle_2017, Seo_2015, Kirk_2017} and velocity gradient calculations \citep[e.g.,][]{Schneider_2010, Henshaw_2013, Kirk_2013, Peretto_2014} are highly dependent on velocity dispersion and centroid, respectively.

High spatial and spectral resolutions can provide observers with a lower chance of viewing multiple velocity component spectra as there can be less chance of ``smearing'' together slabs of gas that are close to one another in the spatial and spectral dimensions.  When observing clouds at farther distances, however, there can be a higher chance of observing multiple velocity component spectra due to the worsened spatial resolving power.  Thus, modern spectroscopic surveys of molecular clouds at large distances must incorporate a line-fitting strategy that considers multiple velocity components along the line-of-sight to obtain the most accurate kinematics measurements from their data.  


Although several multi-component line fitting methods have been developed for molecular emission line observations, they either require user input and direction during the line fitting procedure \citep[e.g., SCOUSE: Semi-automated multi-COmponent Universal Spectral-line fitting Engine,][]{Henshaw_2016}, or they require several iterations of fitting both single and multiple-component models to test which model produces the ``best'' fit \citep{Riener_2019, Clarke_2018, Sokolov_2017, Lindner_2015, Chen_prep}.  These semi-automated and brute-force methods are plagued by several issues:  1) They are highly dependent on the model's initial parameter guesses and degrees of freedom used for their $\chi^2$-minimization fits.  For example, for $\chi^2$-minimization to converge onto the optimal solution, it must be fed initial conditions for the model parameters (e.g., velocity dispersion and centroid) that are near the ``true'' values of the emission.  This requirement often leads to large amounts of pre-processing the data to obtain estimates for the centroid and dispersion of each velocity component that can be used as initial guesses for the line fitting procedure.  Alternatively, one can blindly repeat the line fitting procedure using a large grid of initial parameter guesses to search for the optimal fit.  2) Due to this pre-processing or grid search requirement, traditional methods tend to be computationally expensive, often requiring hours to fit typical spectral cubes.  3) Furthermore, they tend to neglect spectra in neighboring pixels that could confirm the presence or lack of multiple velocity components. 

This paper provides a solution for efficiently identifying multiple velocity component spectra using artificial neural networks (ANNs).  ANNs are a type of machine learning model that attempts to map input features to output classes or values using hierarchical feature representations that are learned during the training process.  These hierarchical features are learned by stacked layers of artificial neurons that use a weighted function to map the inputs they receive into outputs that are fed into subsequent layers.  The complexity of features learned by each layer increases with depth into the network.  For example, a neural network trained for facial recognition might first detect facial edges and contours, which can then be used to detect facial features such as noses, ears, and eyes, until the final layer is able to build facial templates that can be used to predict which face is being viewed in a given image. 

In terms of astronomy, ANNs are becoming increasingly prevalent due to the advantages they can provide by learning non-linear patterns that traditional methods struggle to reproduce and making quick predictions once trained.  For example, ANNs have been successfully applied to a variety of problems across many different fields of research, such as: detecting planets in the Kepler archive that were missed by the standard Kepler identification pipeline \citep{Shallue_2018}, discriminating galaxies with an active galactic nucleus from star-forming galaxies in Sloan Digital Sky Survey (SDSS) observations \citep{Teimoorinia_2018}, deriving stellar temperature, metallicity, and gravity from SDSS APOGEE stellar spectra \citep{Fabbro_2018}, detecting 72 previously missed fast radio burst (FRB) pulses from the first-discovered repeating FRB \citep{Zhang_2018}, and identifying wind-driven shells in magneto-hydrodynamic molecular cloud simulations \citep{Van_2019}.  ANNs have also been used for multiple-component emission line identification of optical spectra.  For instance, \citep{Hampton_2017} have trained an ANN to classify optical spectra of galaxies using parameters output by a traditional Gaussian line-fitting approach called LZIFU \citep{Ho_2016}.

Many of the ANNs used for astronomy applications rely on a particular type of ANN called a convolutional neural network (CNN or convnet), which preserves the spatial structure of its input features using convolutional kernels that are learned during training.  The convolution involves taking the dot product between the network's input (which can be an image, spectrum, light curve, etc.) and a sliding kernel that is moved across the input in predefined steps.  The output is a convolved feature map that is used in subsequent layers of the network to make a prediction on the input's class (in the case of classification).  The convolved feature map not only preserves the spatial structure in the input image, but also reduces the number of input features into the next layer of the network, which leads to faster training times.  

This paper will utilize the advantages of training a 1D CNN to classify input spectra as either single or multiple velocity components and predict the kinematics of each velocity component.  The method requires no initial parameter guesses, incorporates spectra from nearby pixels to make predictions, and analyzes entire spectral cubes in seconds.  Such improvements are welcome with the advent of large multi-receiver arrays where thousands of spectra can be collected in a reasonable time.  Named Convnet Line-fitting of Velocities in Emission-line Regions (CLOVER), the method is also publicly available as a Python package called \texttt{astroclover}\footnote{\url{https://github.com/jakeown/astroclover/}}.

The paper is organized as follows: Section 2 describes the data used for training CLOVER and testing its performance; Section 3 outlines the CNN architecture of CLOVER; Section 4 compares CLOVER's classification performance to that of a traditional single-Gaussian line fitting method on both synthetic and real data; Section 5 discusses predicting kinematics from two-component spectra with CLOVER; Section 6 describes further applications of CLOVER classifications to emission lines with hyperfine splitting; Section 7 presents CLOVER kinematics predictions for NH$_3$ (1,1) synthetic data; Section 8 shows how CLOVER can be used to improve the accuracy of virial stability analyses of structures with multiple velocity components; and Section 9 summarizes the paper. In addition, Appendix A provides an overview of the installation and usage instructions for the \texttt{astroclover} Python package.

\section{Data}

\subsection{Training Set: Generating Synthetic Spectra}
All machine learning classification projects require a training set composed of input feature vectors (a.k.a., ``samples'' or ``examples'') that belong to one of the possible output classes the model will be trained to predict.  In this paper, we train a network that has three distinct output classes: ``one-component'' spectra with only one velocity component along the line of sight, ``two-component'' spectra with two velocity components along the line of sight, and ``noise-only" spectra with negligible emission.

To generate the training set, synthetic spectral cubes on a 3$\times$3 pixel grid with 500 spectral channels were created.  For the ``one-component'' class, a single Gaussian spectrum was injected into the grid's central pixel with peak intensity ($T_{peak}$) set to 1 K and values of velocity dispersion ($\sigma$) and centroid velocity ($V_{LSR}$) chosen at random from a uniform distribution with the following limits:

\begin{itemize}
\item $\sigma$: $2-11$ channels, which produces both narrow and broad Gaussians similar to real emission lines. 
\item $V_{LSR}$: channel 112 to channel 388 of the 500 channel spectrum. This range is equivalent to $-0.55$ km s$^{-1}$ to 0.55 km s$^{-1}$ when the spectral axis has been normalized to $-1.0$ km s$^{-1}$ for the lowest velocity channel and $1.0$ km s$^{-1}$ for the highest velocity channel.  This range provides a variety of centroid velocities while ensuring the emission line edges do not spill off the edges of the spectrum.
\end{itemize} 

The Gaussians for the surrounding pixels in the 3$\times$3 grid are determined by applying a perturbation to the central pixel's Gaussian parameters.  This step was done by drawing values from three normal distributions (one for each parameter) with mean of zero and variance of 0.05.  The randomly drawn values were then added to the central pixel's parameter values to generate new Gaussians with slight offsets in $\sigma$, $V_{LSR}$, and $T_{peak}$.  Finally, noise with an RMS drawn from a uniform distribution between 0.05 K and 0.25 K was injected into each spectral cube, creating both low-, mid-, and high- signal-to-noise ratio (SNR) training examples.  

For the ``two-component'' class, two Gaussians were injected into the central pixel of the 3$\times$3 grid.  The values of $\sigma$ for the two Gaussians were both drawn at random as described above for the one-component class. The value of $T_{peak}$ for the second Gaussian was drawn randomly from a uniform distribution between $2\times$RMS and 1 K, where RMS is the noise level selected for the cube.  Similarly, the $V_{LSR}$ for the second Gaussian was randomly drawn from a uniform distribution between $V_{LSR,1}\pm1.5 \times \sigma_{max}$ and $V_{LSR,1}\pm5 \times \sigma_{max}$, where $\sigma_{max}$ is the value of $\sigma$ for the wider of the two Gaussians, $V_{LSR,1}$ is the centroid of the first component, and the sign of the offset ($\pm$) is chosen at random.  Thus, the second component can be on either the left or right of the first component along the spectral axis.  

This two-component sample generation approach created variations in the relative heights, velocity dispersions, and centroids of each velocity component.  Moreover, the velocity centroid separation threshold for each velocity component minimized the number of two-component samples that are indistinguishable from one-component samples.  This characteristic of the training set was necessary to prevent the CNN from overfitting (a tendency to predict the two-component class when it was clear the one-component class was more appropriate).  Such separation thresholds are also often implemented in traditional line-fitting methods \citep[see, e.g.,][]{Lindner_2015, Henshaw_2016, Riener_2019} when deciding whether or not a multiple-component fit is appropriate.  The outer pixels in the 3$\times$3 grid were filled by adding perturbations to both of the Gaussian components, as described above for the one-component class.



For each training example cube, only two spectra are used as input into the CNN: 1) the spectrum of the central pixel in the 3$\times$3 grid and 2) the averaged spectrum over all nine spectra in the 3$\times$3 grid.  The first spectrum provides a ``local'' view of the pixel for which the class prediction is being made, while the second spectrum provides a ``global'' view of neighboring pixels that can provide insight into whether the central pixel is a one-component, two-component, or noise-only spectrum.  Both spectra are normalized by dividing by the value of the brightest channel.   This ``local+global'' setup also provides for a simple way to make predictions on real observations.  In that case, a sliding window of size 3$\times$3 pixels is moved across the position-position plane of a spectral cube and a class prediction is made on the central pixel after feeding its ``local'' and ``global'' spectra into the trained network. 

\begin{figure}[htb]
\epsscale{0.84}
\plottwo{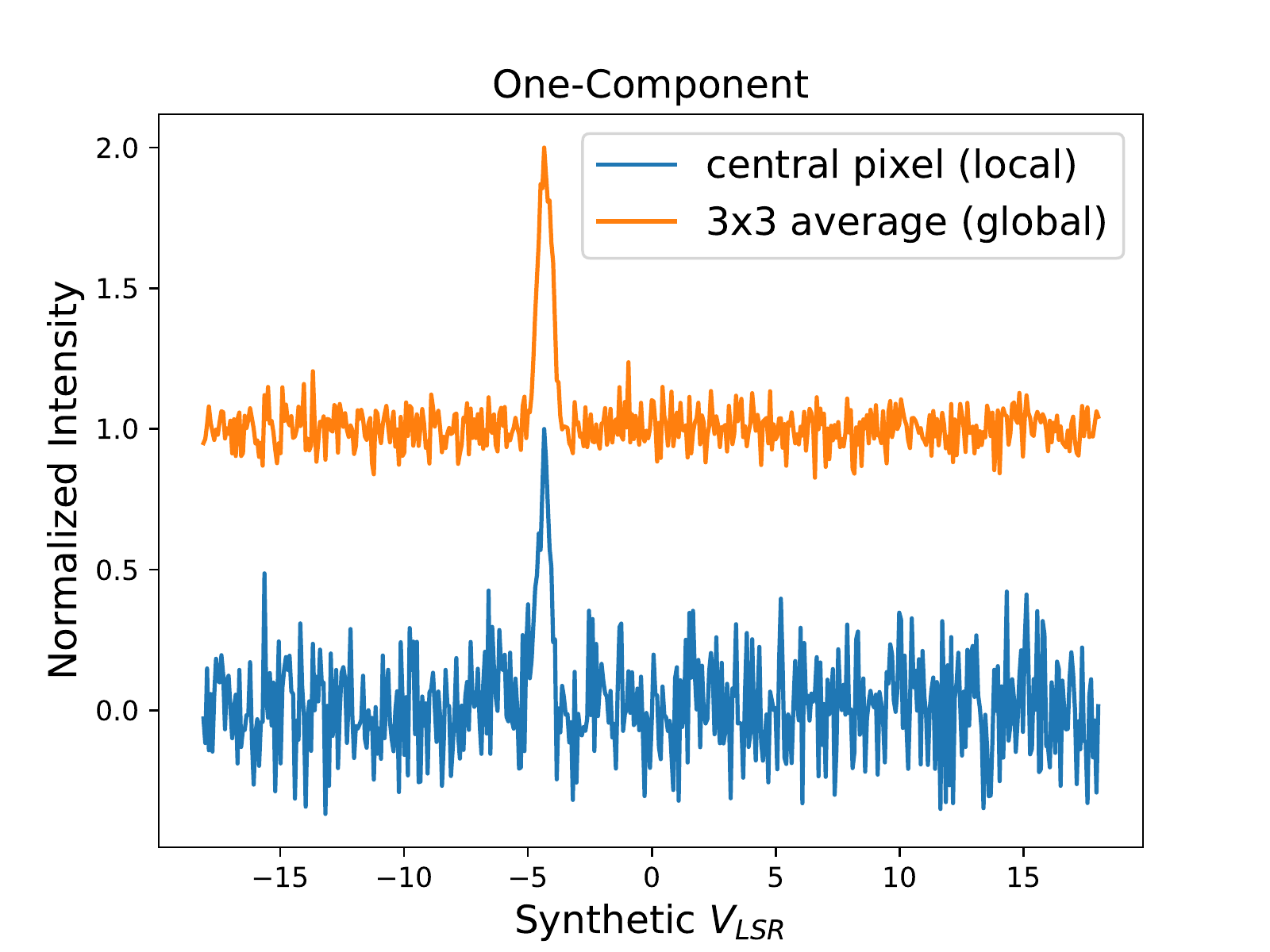}{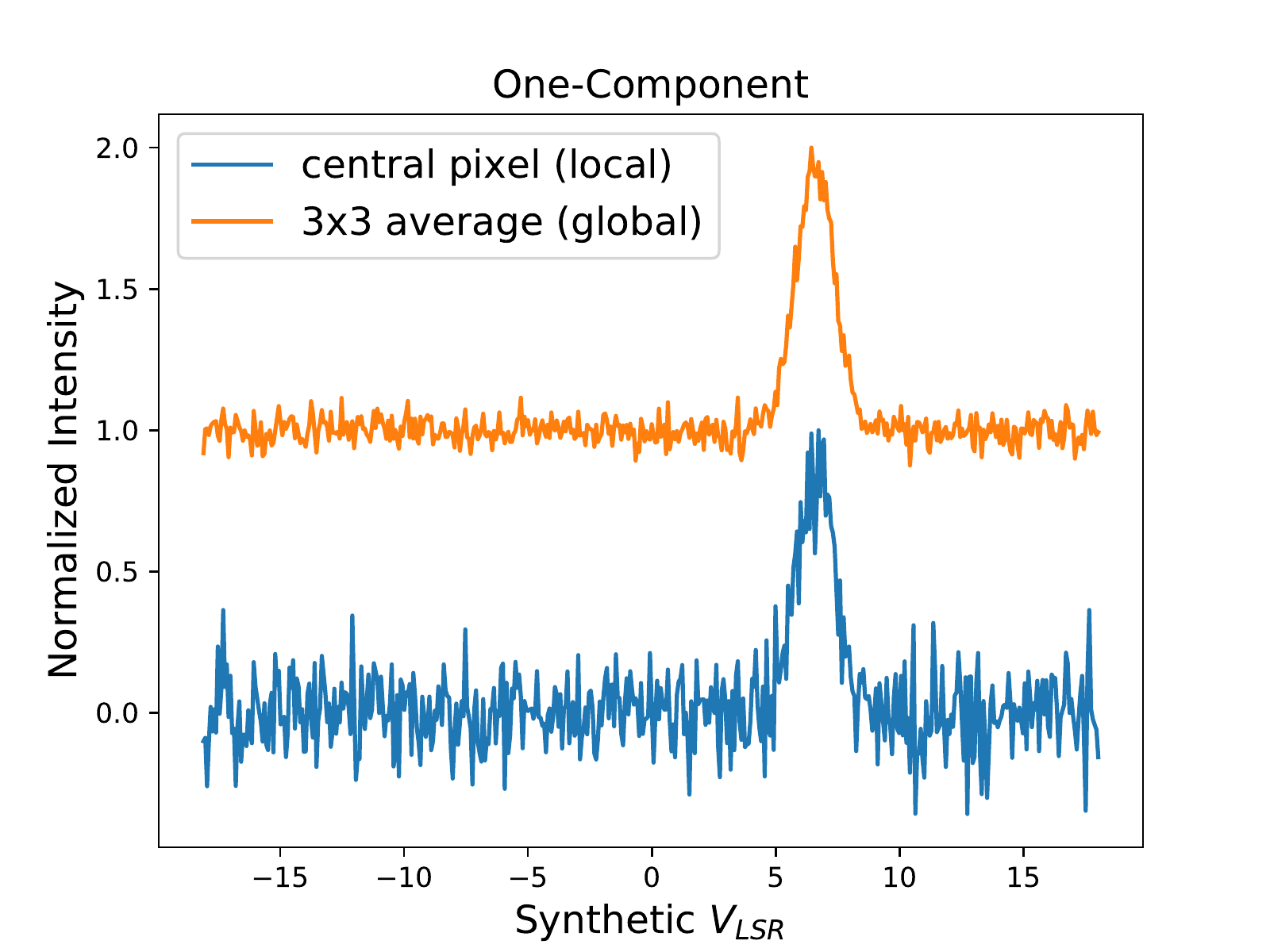}
\plottwo{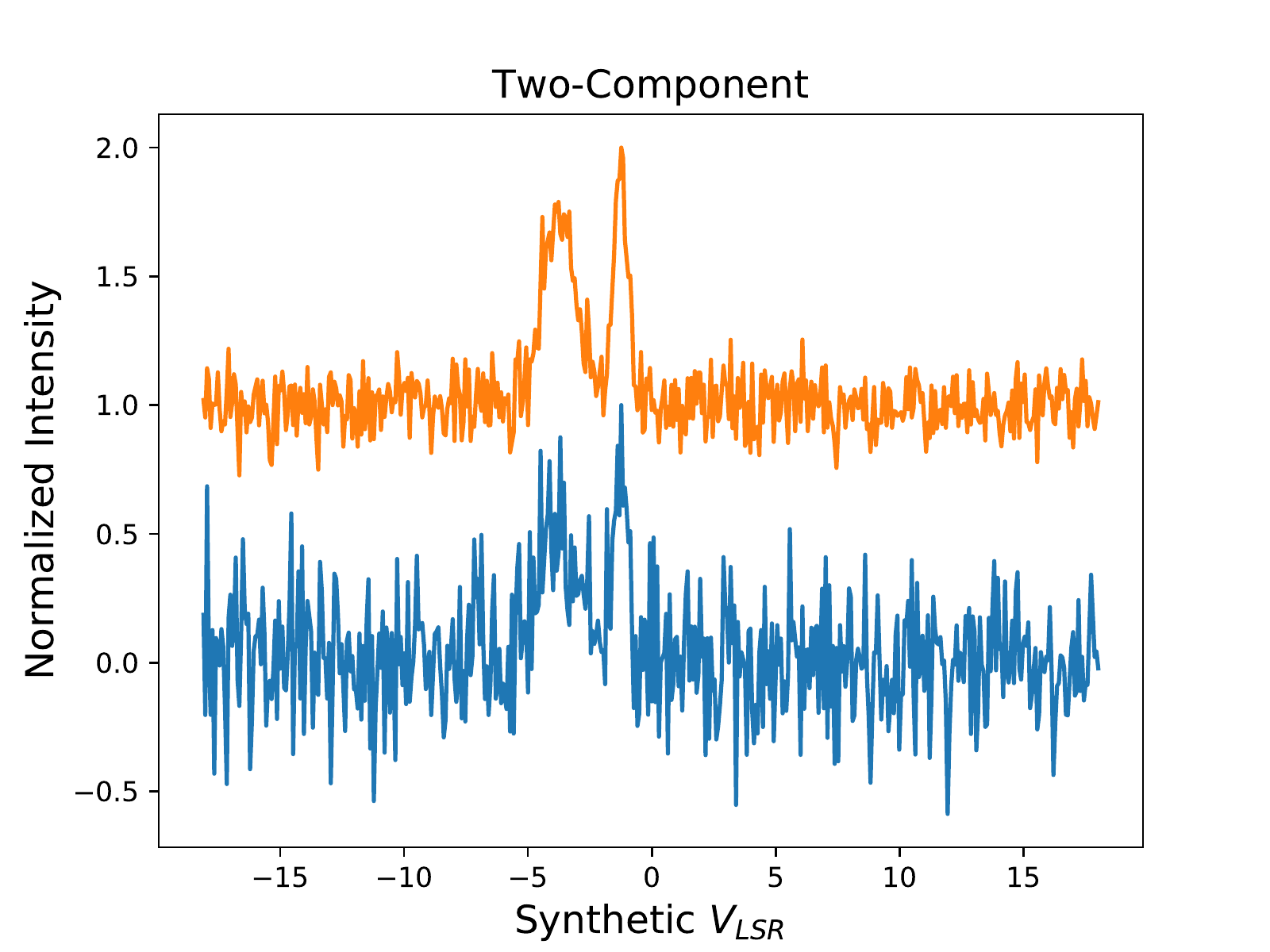}{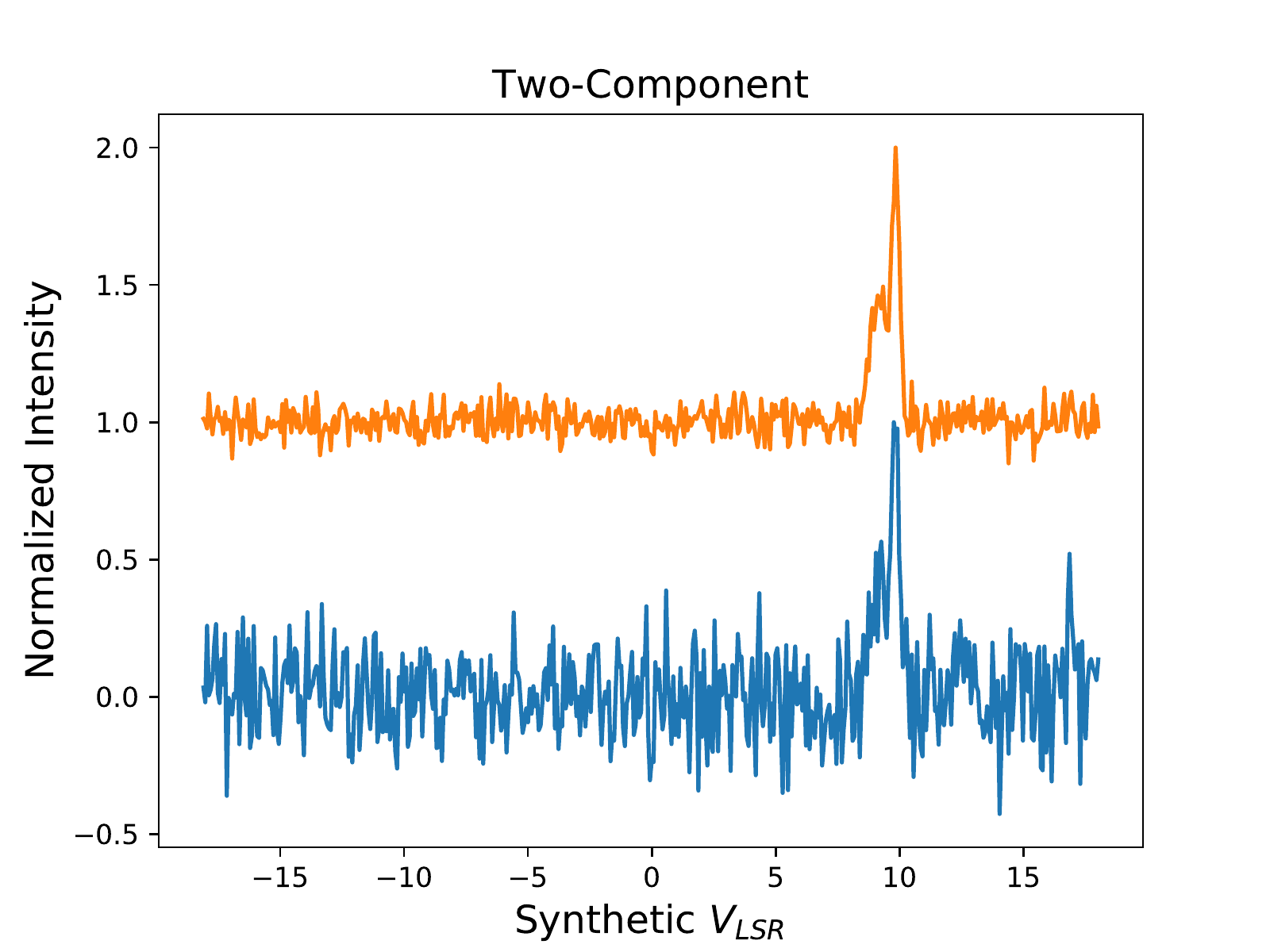}
\centering
\plottwo{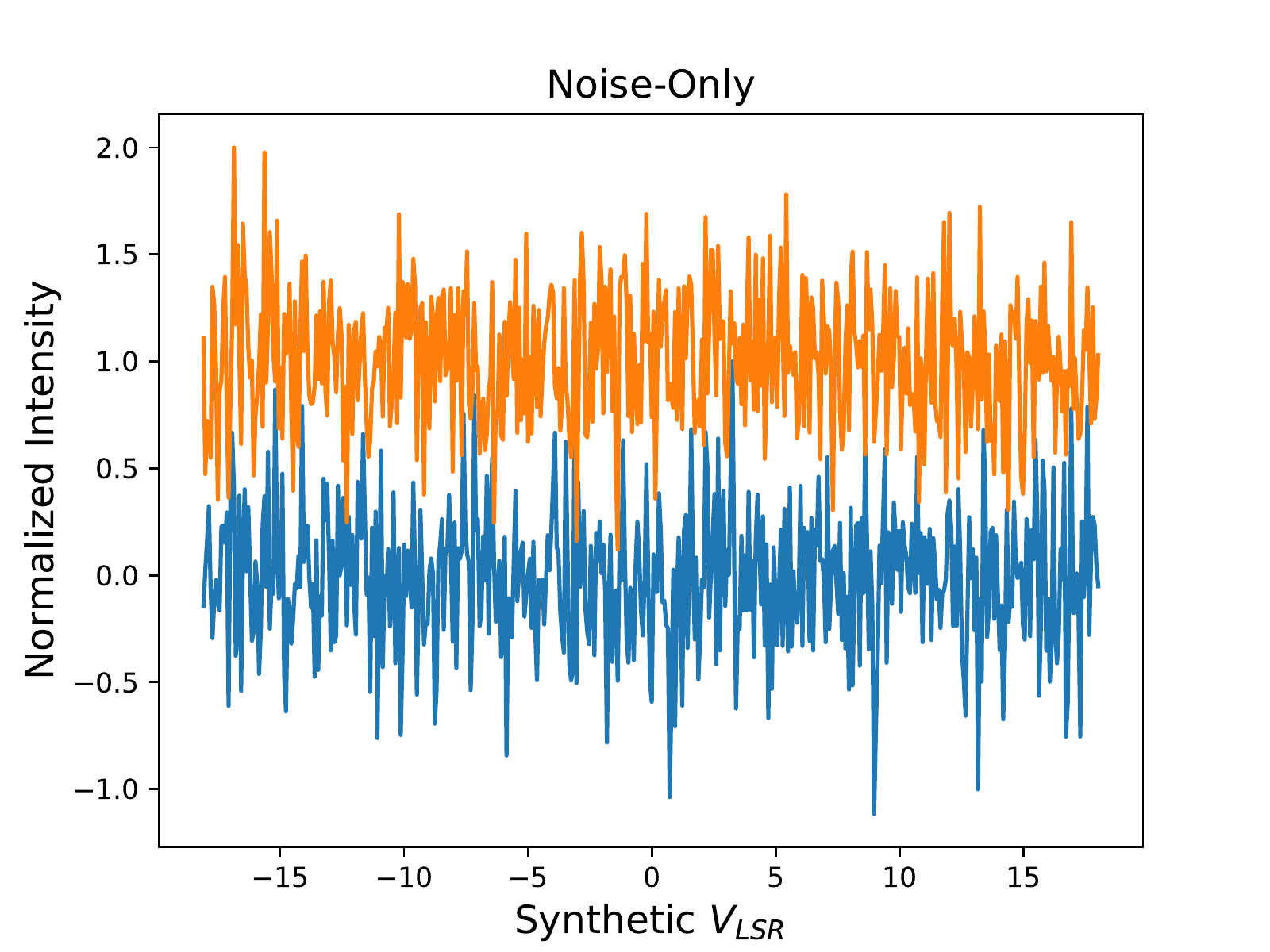}{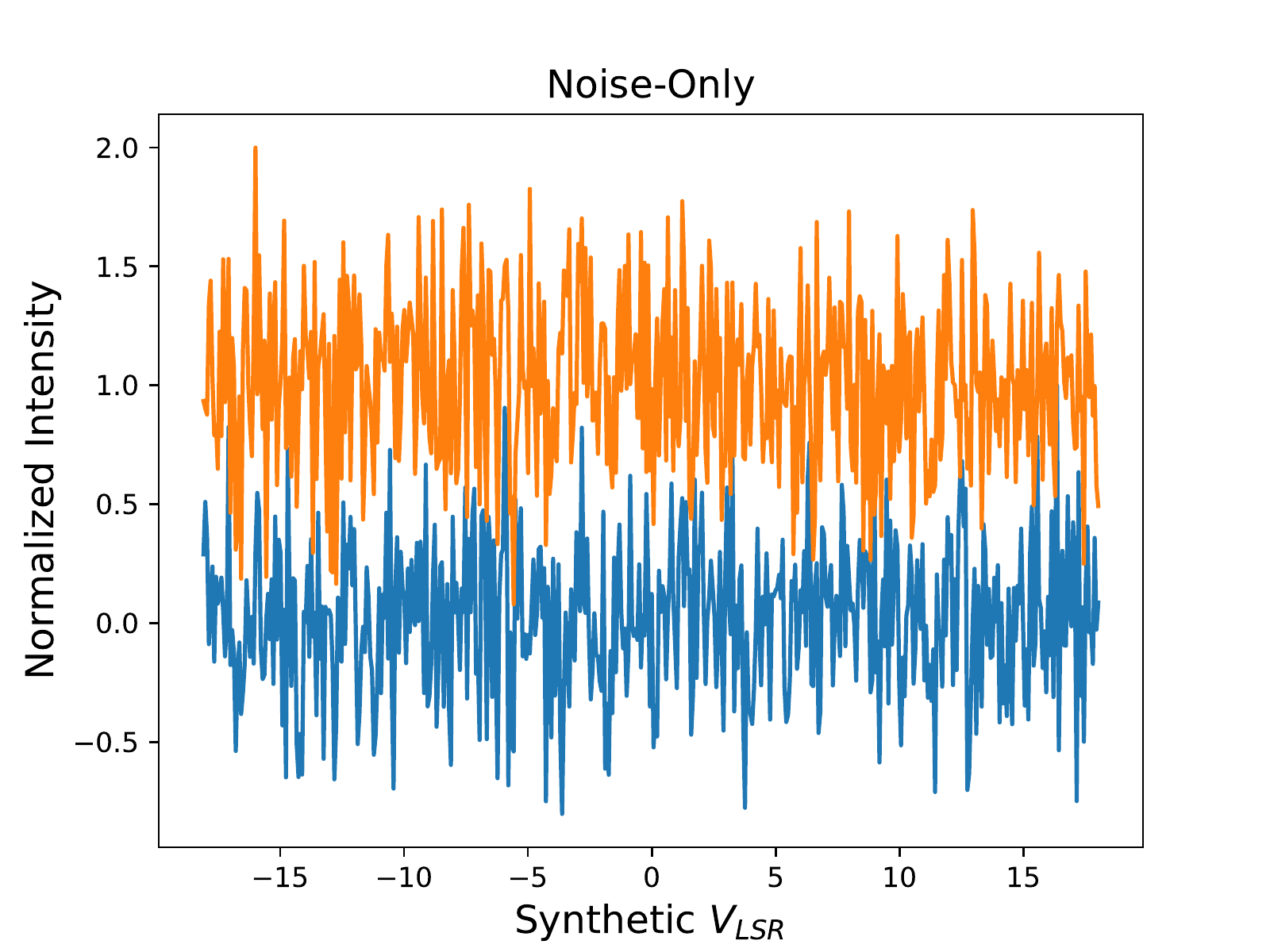}
\caption{Example synthetic spectra included in the one-component (top row), two-component (middle row), and noise-only (bottom row) training set classes.  Blue spectra show the central pixel in the spectral cube window (i.e., the ``local'' spectrum) and the orange spectra represent the $3\times3$ pixel average around the central pixel (i.e., the ``global'' spectrum).}
\label{training_set}
\end{figure}

Following the aforementioned method, 300,000 synthetic samples (100,000 for each training set class, i.e., a ``balanced'' training set) were generated.  Figure \ref{training_set} shows example local and global spectra for training set samples in the one-component, two-component, and noise-only classes.  A validation set of 90,000 additional synthetic spectra (30,000 in each class) was also generated for monitoring performance during training (see Section 3).  After training, the network's performance is tested on ten additional collections of 30,000 synthetic spectra (10,000 in each class).

\subsection{Test Set: Real $^{13}$CO, C$^{18}$O, \& HC$_5$N Spectral Cubes}
The test sets used to gauge the trained CNN's performance included three real spectral cubes observed from three different surveys of three distinct star-forming regions.  The first cube was a $^{13}$CO ($1-0$) observation of L1689 in the Ophiuchus molecular cloud from the COMPLETE survey\footnote{available at \url{https://www.cfa.harvard.edu/COMPLETE/}} \citep{Ridge_2006} on the Five College Radio Astronomy Observatory (FCRAO).  This cube has a spectral resolution of $\sim0.07$ km s$^{-1}$, the pixel size is $23\arcsec$, and the FCRAO has an angular resolution of $\sim46\arcsec$ at the rest frequency of $^{13}$CO ($1-0$).  

The second cube was a C$^{18}$O ($3-2$) observation of DR21\footnote{available at \url{http://www.cadc-ccda.hia-iha.nrc-cnrc.gc.ca/en/} with proposal ID: M10BD01} in the Cygnus X giant molecular cloud complex observed by the James Clerk Maxwell Telescope (JCMT).  The observations were accessed from the JCMT archive and have not been previously published, but were originally observed as part of a $^{12}$CO ($3-2$) survey by \cite{Gottschalk_2012}.  The native spectral resolution of the cube was $\sim$ 0.056 km s$^{-1}$, but to improve the spectral SNR, we smooth spectrally with a Gaussian kernel to half the original spectral resolution: 0.11 km s$^{-1}$. The native angular resolution of the JCMT at the rest wavelength of C$^{18}$O $(3-2)$ is $\sim15\arcsec$, which we convolve to 32$\arcsec$ to improve SNRs further.  The pixel scale of the cube is 7.2$\arcsec$.  

The third cube was a HC$_5$N ($9-8$) observation of B18\footnote{available at \url{https://dataverse.harvard.edu/dataverse.xhtml?alias=GAS_Project}} in the Taurus molecular cloud observed by the Green Bank Ammonia Survey \citep{Friesen_2017} on the 100m Green Bank Telescope.  This cube has a spectral resolution of $\sim0.07$ km s$^{-1}$, a pixel scale of $\sim11\arcsec$, and an angular resolution of $\sim31\arcsec$.  

\begin{figure}[htb]
\epsscale{0.7}
\plotone{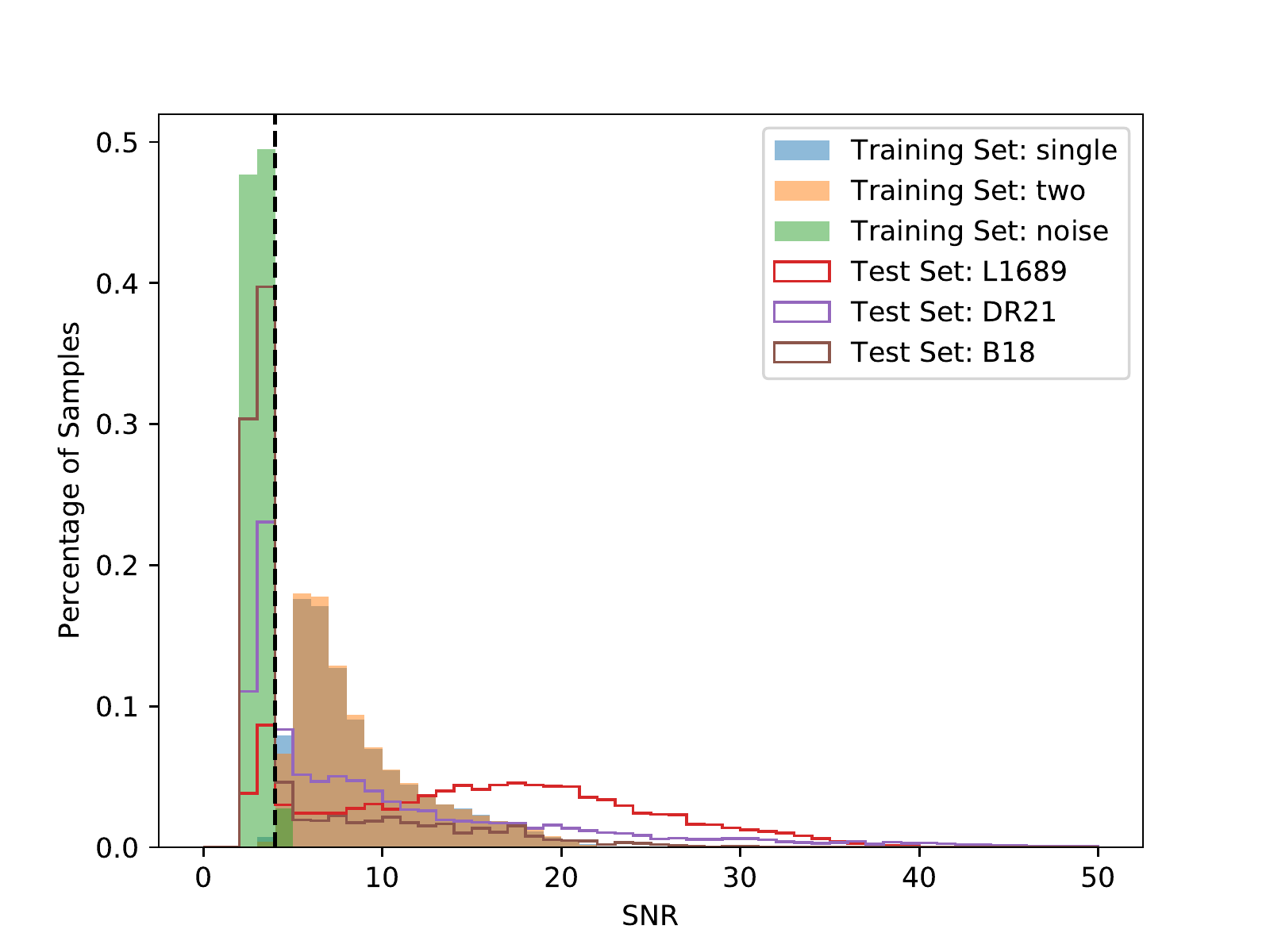}
\caption{Histograms of signal-to-noise ratio for the ``local'' spectra in the training and test sets.  Blue, orange, and green represent the one-component, two-component, and noise-only classes of the synthetic training set, respectively.  Red, purple, and brown show the distributions for the real observations of L1689, DR21, and B18, respectively. The black dashed line shows SNR=4.}
\label{SNR}
\end{figure}

Table 1 outlines the characteristics of each spectral cube and star-forming region in the test set.  These three regions were chosen due to their differing levels of star formation activity.  B18 is a fairly quiescent, nearby ($d\sim135$ pc), star-forming region, while L1689 is a more active nearby cloud ($d\sim119$ pc) and DR21 is a distant ($d\sim1700$ pc) high-mass star-forming region producing O- and B-type stars.  Thus, these spectral cubes provide a thorough test of the CNN performance across a variety of star-forming environments, instruments, and emission line transitions.

\begin{deluxetable}{ccccccccc}
\rotate
\tablewidth{0pt}
\tablecolumns{4}
\tablecaption{Test Set Spectral Cubes}
\tablehead{\colhead{Region} & \colhead{Cloud} & \colhead{Distance}  & \colhead{Transition} & \colhead{Telescope} & \colhead{Rest Freq.\tablenotemark{e}} & \colhead{Spectral Res.} & \colhead{Spatial Res.} & \colhead{Pixel Scale}\\
   & & (pc) & & & (MHz) & (km s$^{-1}$) & ($\arcsec$)  & ($\arcsec$)}
\startdata
L1689 & Ophiuchus & 119$\pm$6\tablenotemark{a} & $^{13}$CO ($1-0$) & FCRAO & 110201.354 & 0.07 & 46 & 23\\
DR21 & Cygnus X & 1700\tablenotemark{b} & C$^{18}$O ($3-2$) & JCMT & 329330.552 & 0.11 & 32 & 7.2\\
B18 & Taurus & 135$\pm$20\tablenotemark{c} & HC$_5$N ($9-8$) & GBT & 23963.9010 & 0.07 & 31& 11\\
M17SW & M17 & 1700\tablenotemark{d} & NH$_3$ (1,1) & GBT & 23694.4955 & 0.07 & 32& 8.8\\
MonR2 & Orion-Monoceros & 900\tablenotemark{c} & NH$_3$ (1,1) & GBT & 23694.4955 & 0.07 & 32 & 8.8\\

\enddata
\tablenotetext{a}{\cite{Lombardi_2008}}
\tablenotetext{b}{\cite{Schneider_2006}}
\tablenotetext{c}{\cite{Schlafly_2014}}
\tablenotetext{d}{\cite{Xu_2011}}
\tablenotetext{e}{Accessed from \cite{Lovas_2004}.}

\label{Table_regions}
\end{deluxetable}

To be consistent with the synthetic spectra used to train the CNN, the spectral axis on all cubes was clipped to 500 channels centered on the line-of-sight velocity to each cloud.  The ``local'' and ``global'' view spectra were extracted from each cube by sliding a $3\times3$ pixel window across the position-position plane of the cube.  The CNN then makes a prediction on the class of the central pixel in the window using the ``local'' and ``global'' view spectra as input. 

Figure \ref{SNR} shows the distributions of SNR, defined as the ratio of the peak emission line channel to the standard deviation of the off-line channels, for the ``local'' spectra in the test set cubes.  Figure \ref{SNR} also displays the SNR distributions for the one-component, two-component, and noise-only training set classes.  The one- and two-component training set classes have similar SNR distributions, which ensures that the network must use morphological differences rather than SNR differences to distinguish those classes.  In addition, the one- and two-component training set distributions have a similar range in SNR as the majority of the real data.  This similarity suggests the training set is representative of real data, which is necessary for the trained network to generalize its predictions for handling real observations with a wide range of SNR.  Although the high SNR end of the training set is less populated than the real data, this difference is acceptable since the high SNR examples are typically much easier to classify than low SNR examples.  We also note that the one- and two-component classes in the training set have a significant drop-off below SNR=4, which is effectively the SNR threshold where the ``noise'' class begins.  An SNR of 4 is similar to typical minimum thresholds set for traditional line-fitting pipelines, however, making it a reasonable signal versus noise threshold for the network's predictions.

In Section 5.2, we also use NH$_3$ (1,1) cubes observed by the KFPA Examinations of Young STellar Object Natal Environments (KEYSTONE) survey (PI: James Di Francesco; Keown et al. 2019, submitted) using the 100m Green Bank Telescope.  Our analysis is focused on the KEYSTONE observations of M17 and MonR2 at distances of 900 pc and 2000 pc, respectively.  These cubes have a spectral resolution of 0.07 km s$^{-1}$, beam size of 32$\arcsec$, and pixel width of 8.8$\arcsec$.  

\section{Methods: CNN Architecture}
\begin{figure}[htb]
\epsscale{0.85}
\plotone{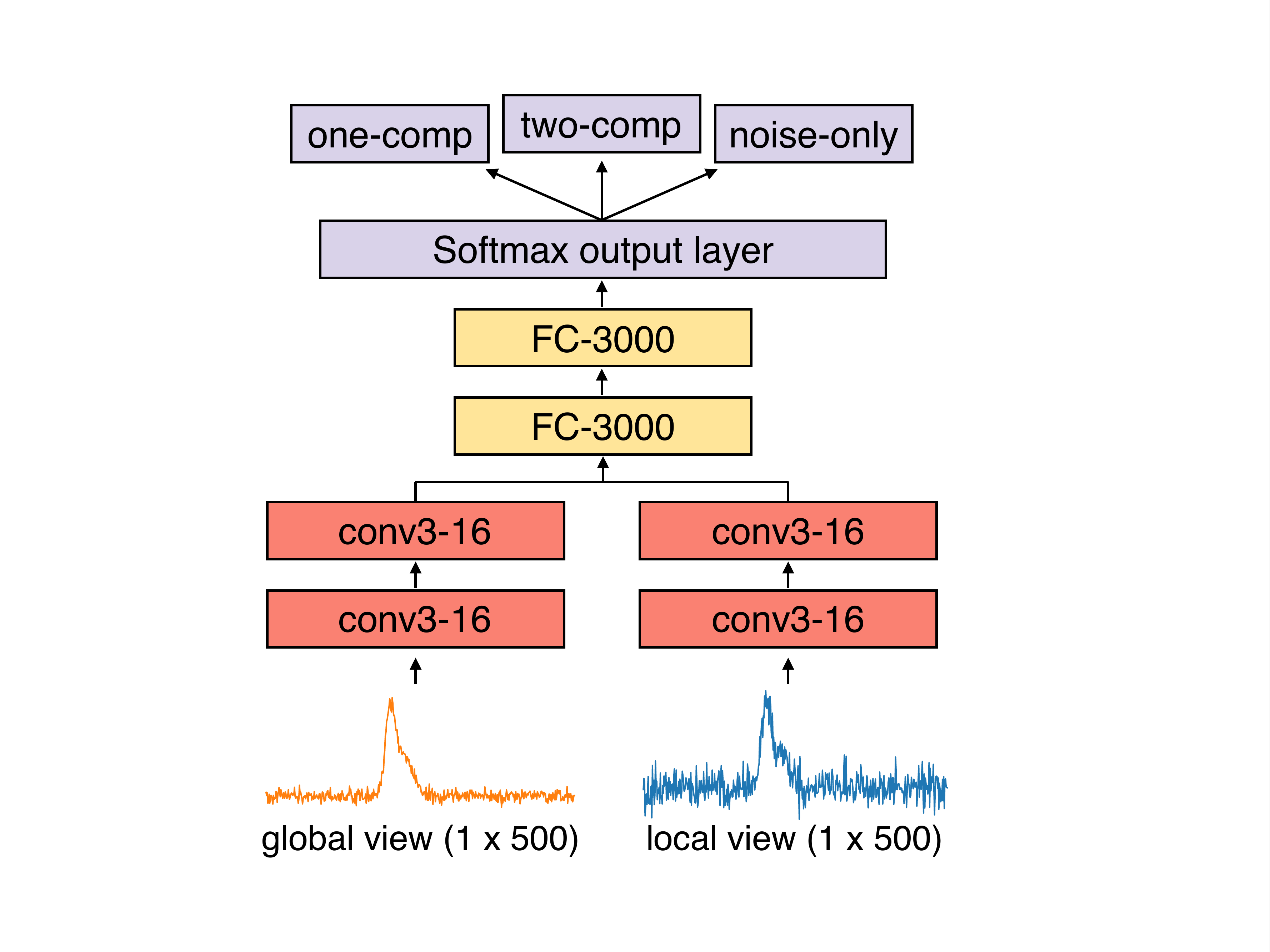}
\caption{Architecture of the CNN chosen for this paper.  The ``local'' and ``global'' view spectra for each sample are fed into individual columns of convolutional layers with 16, three-channel width kernels.  The convolved features maps from each column are then joined into a single column of fully-connected layers with 3000 artificial neurons.  The final layer predicts the class of the object using the softmax activation function. All other layers use a rectified-linear (relu) activation function.}
\label{architecture}
\end{figure}

The architecture of the 1D CNN adopted in this paper is shown in Figure \ref{architecture}.  The network's hyper-parameters were set based on the success of previously published 1D CNNs featured in \cite{Fabbro_2018} and \cite{Shallue_2018}, which used spectra and light curves, respectively, as input.  Following those papers, we also use a 1D CNN because each sample in our training set consists of two 1D spectra with 500 channels.  The ``local'' and ``global'' view spectra are fed into individual convolutional columns before being reconnected into a joint fully-connected layer.  The convolutional columns consist of two convolutional layers, each with 16 kernels with a width of 3 spectral channels.  The ``convolution'' in the convolutional layers involves taking the dot product between the input spectra and the kernels, which are moved across each channel of the input spectra. The output are convolved feature maps (one for each kernel) that are used as input into the next layer of the network. 

The weights on the convolutional kernels are learned during training and attempt to create convolved feature maps that highlight spectral features that can be used to make a decision about the class of the sample.  The resulting convolved features from the two convolutional columns are then combined as inputs into a joint column of two fully-connected layers with 3000 artificial neurons each.  All of the neurons in these layers have a rectified-linear (`relu') activation function, which transforms the inputs it receives into an output that is sent to the next layer in the network.  The rectified-linear activation function is commonly used in deep neural networks because it solves the ``vanishing gradients problem,'' wherein large networks fail to train properly because the error of neurons deep in the network go to zero and can't be properly updated by gradient descent methods \citep{Hochreiter_2001}. 


The final output layer has three artificial neurons with a `softmax' activation function.  Each of these neurons has its own weight vector ($w$) that is the length of the output vector ($x$) from the previous network layer.  The neurons first apply a weighted sum of $x$ by performing the dot product of $x$ and $w$.  The output of the three dot products performed by each individual neuron form a new vector ($y$) of length three that is then passed to the softmax function, which is given by $e^{y_i}/\Sigma_je^{y_j}$ where $y_i$ is every element of $y$ and $\Sigma_je^{y_j}$ is the sum of the exponential of each element in $y$.  Thus, the softmax activation function output is a length three vector that always sums to one and is interpreted as the probability of the input sample being in each of the three input classes.  

The weights on the artificial neurons and convolutional kernels are optimized by minimizing the categorical cross-entropy loss function, using the `Adam' gradient descent optimization method \citep{Kingma_2014}.  Since the categorical cross-entropy loss function increases as the predicted probabilities of the training set samples diverge from their ground-truth values, the model prediction accuracy is maximized if the cross-entropy function is minimized.  For instance, an input training set sample that is a one-component class member has a label of [1, 0, 0].  If the softmax output of the network is [0.1, 0.5, 0.4] for that sample, the loss function output is high and thus the weights of the network need to be adjusted to minimize the loss.  The gradient of the loss function is then calculated to determine in which directions the model weights should be updated to get closer to the minimum loss.  

The weights of the CNN are updated by iteratively moving through the training set in batches of 100 samples (a.k.a. `mini-batch gradient descent').  To prevent over-fitting, `early-stopping' is implemented by monitoring the model's performance on a validation set of 90,000 additional synthetic spectra (30,000 in each class) during training.  After each epoch in the training process, where an epoch represents using all samples in the training set to update the weights of the CNN, the validation set loss is measured.  If the validation set loss does not improve for five epochs in a row, training is stopped and the model from the epoch with the best validation set loss is saved.  

As an additional comparison to the architecture using the local and global spectra as inputs, we also train two additional networks using only the local and only the global spectra as input.  The results for these architectures are presented in Section 4.  Tests of more complex architectures with additional layers, neurons, and larger kernel sizes produced models that overfit the training data.  As such, the simpler architecture presented in Figure \ref{architecture} was chosen for CLOVER.  


  
\section{Results}
\subsection{Testing on Synthetic Data}
\begin{figure}[htb]
\epsscale{1.0}
\plottwo{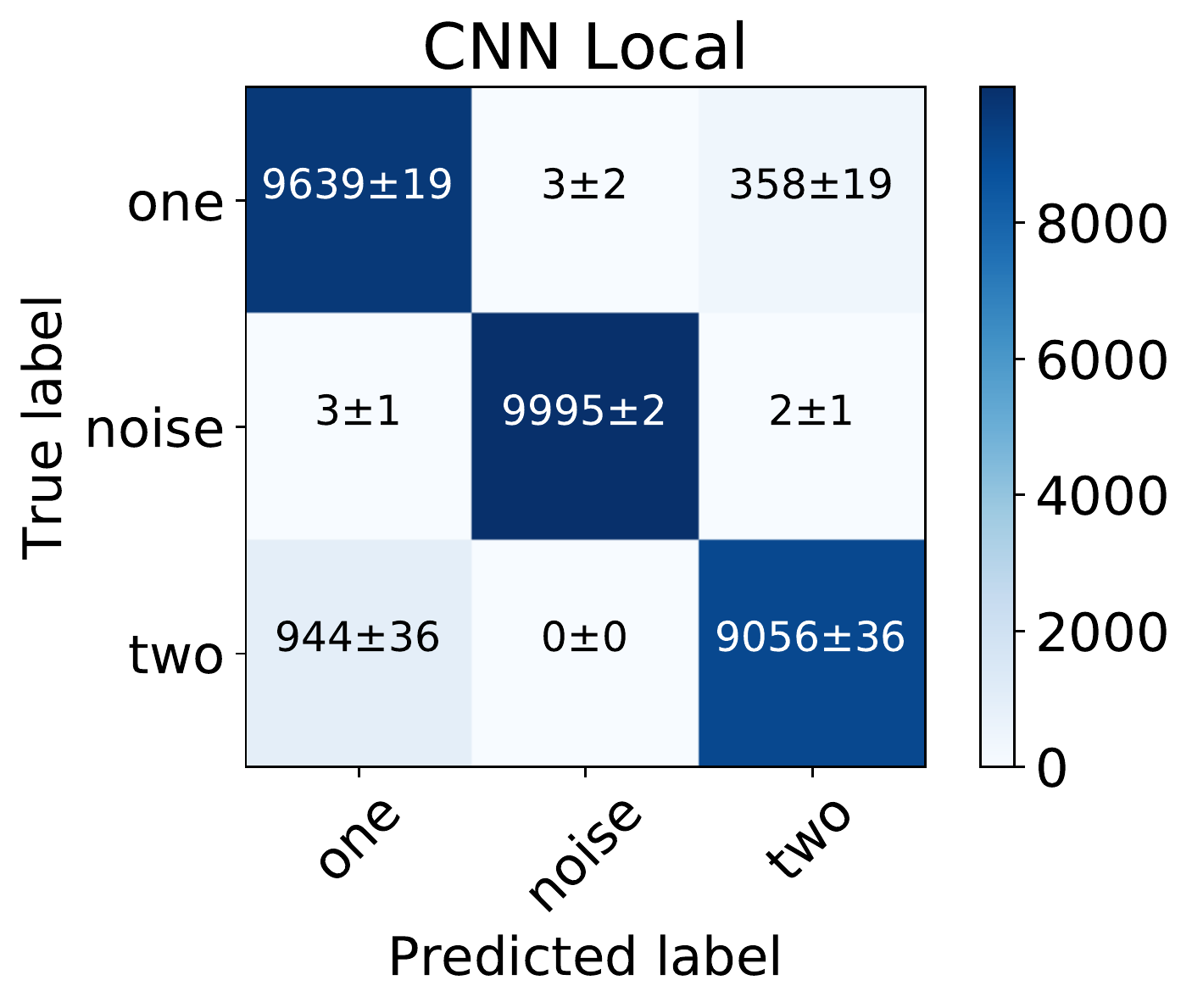}{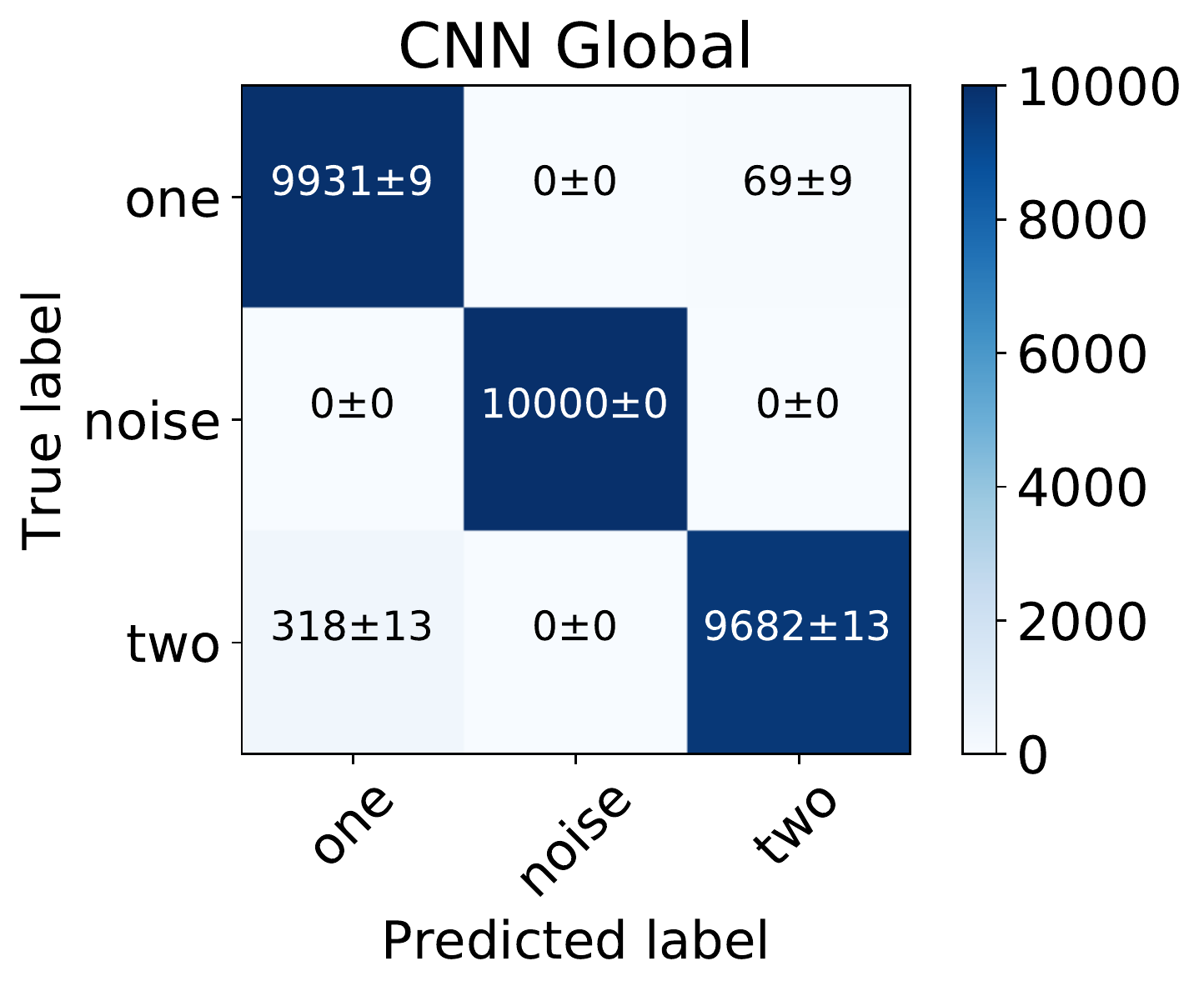}
\plottwo{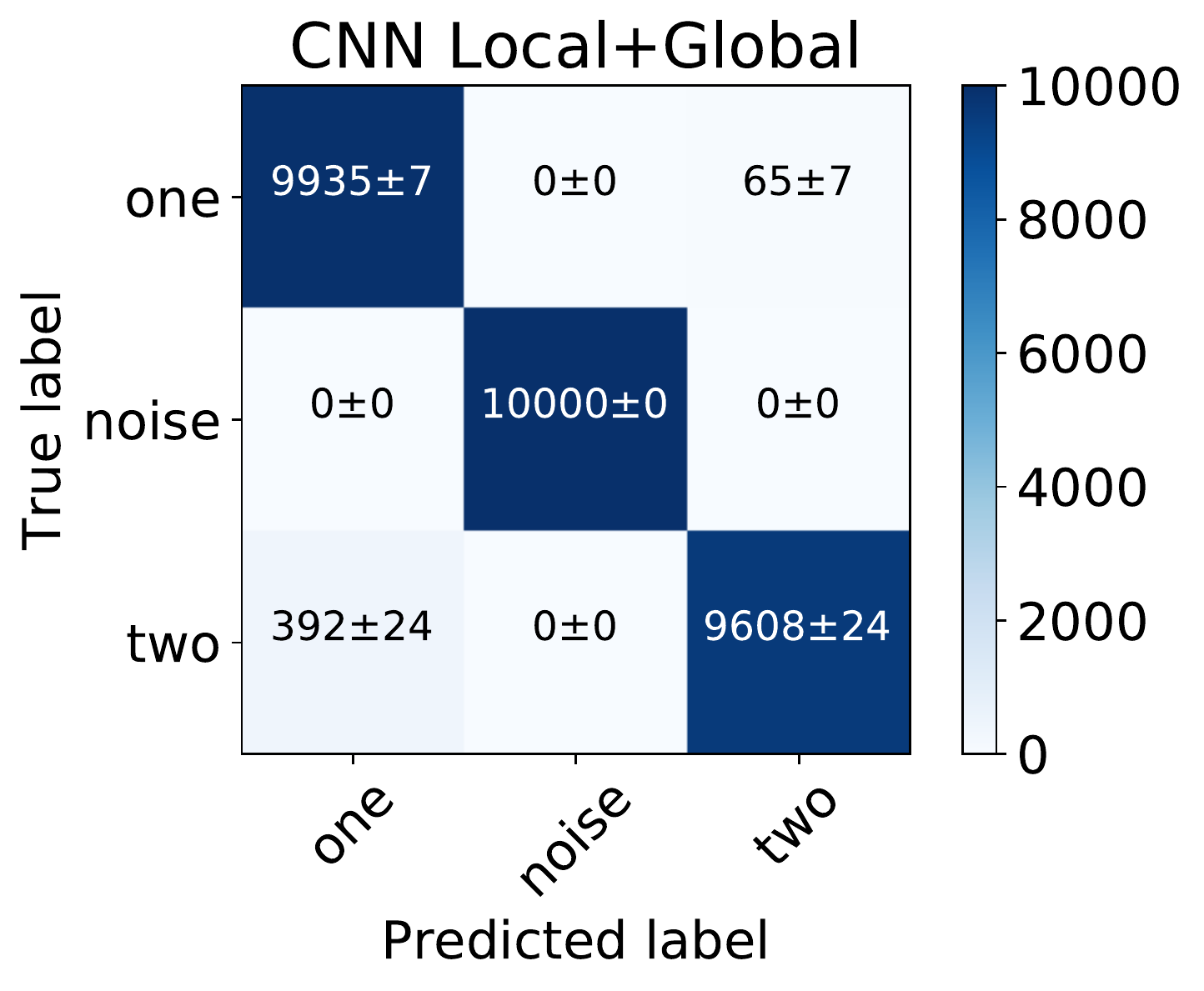}{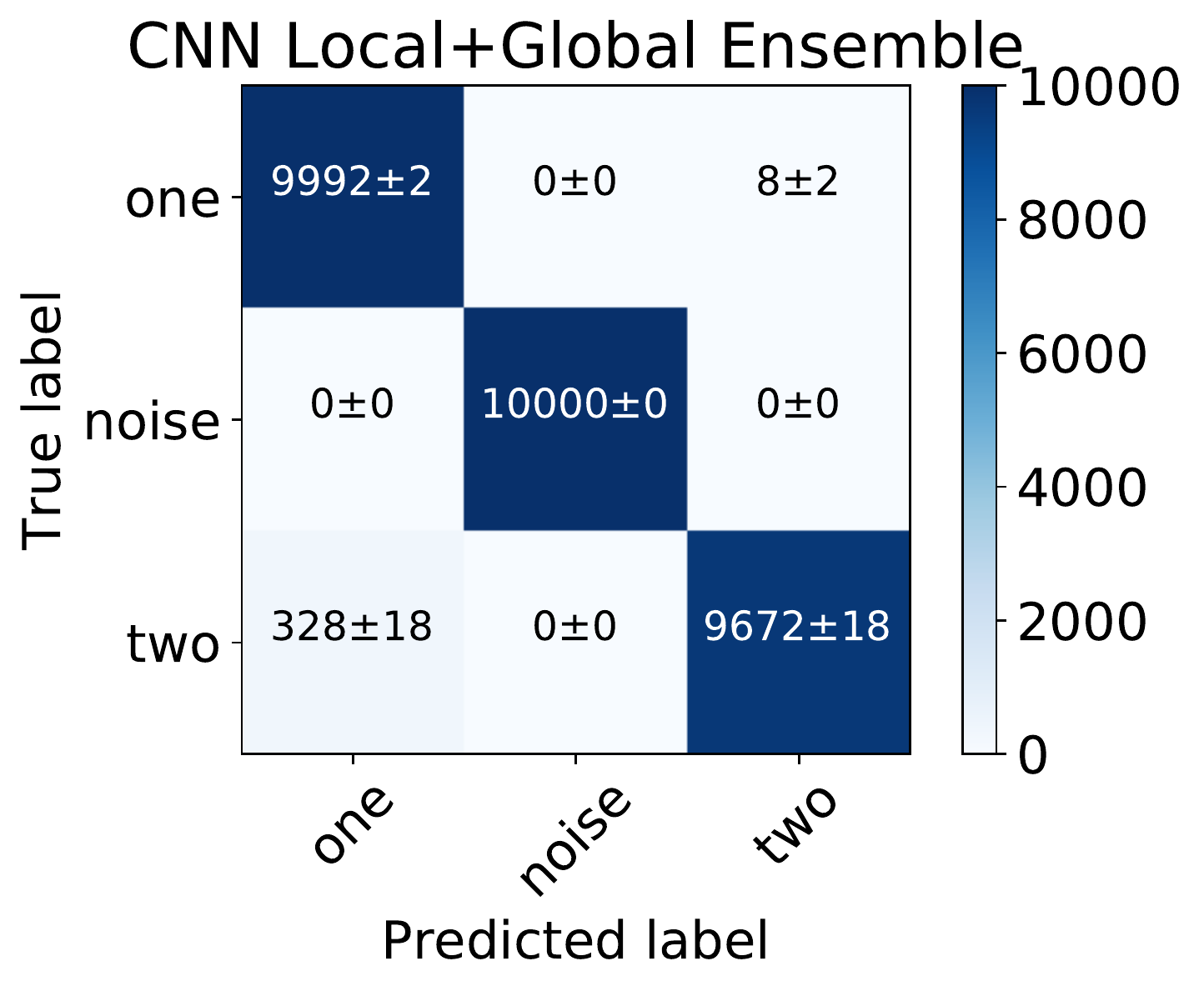}
\centering
\plottwo{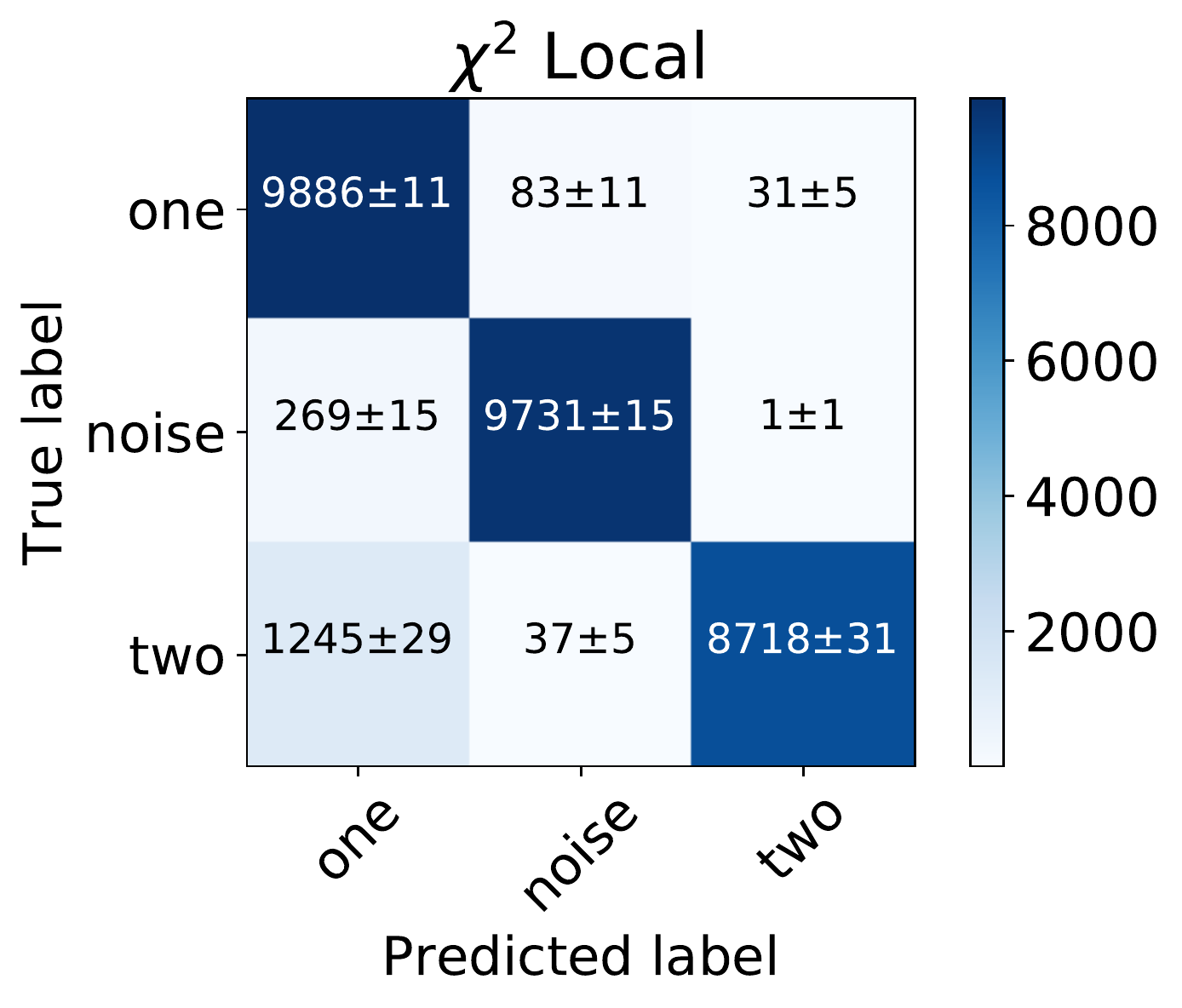}{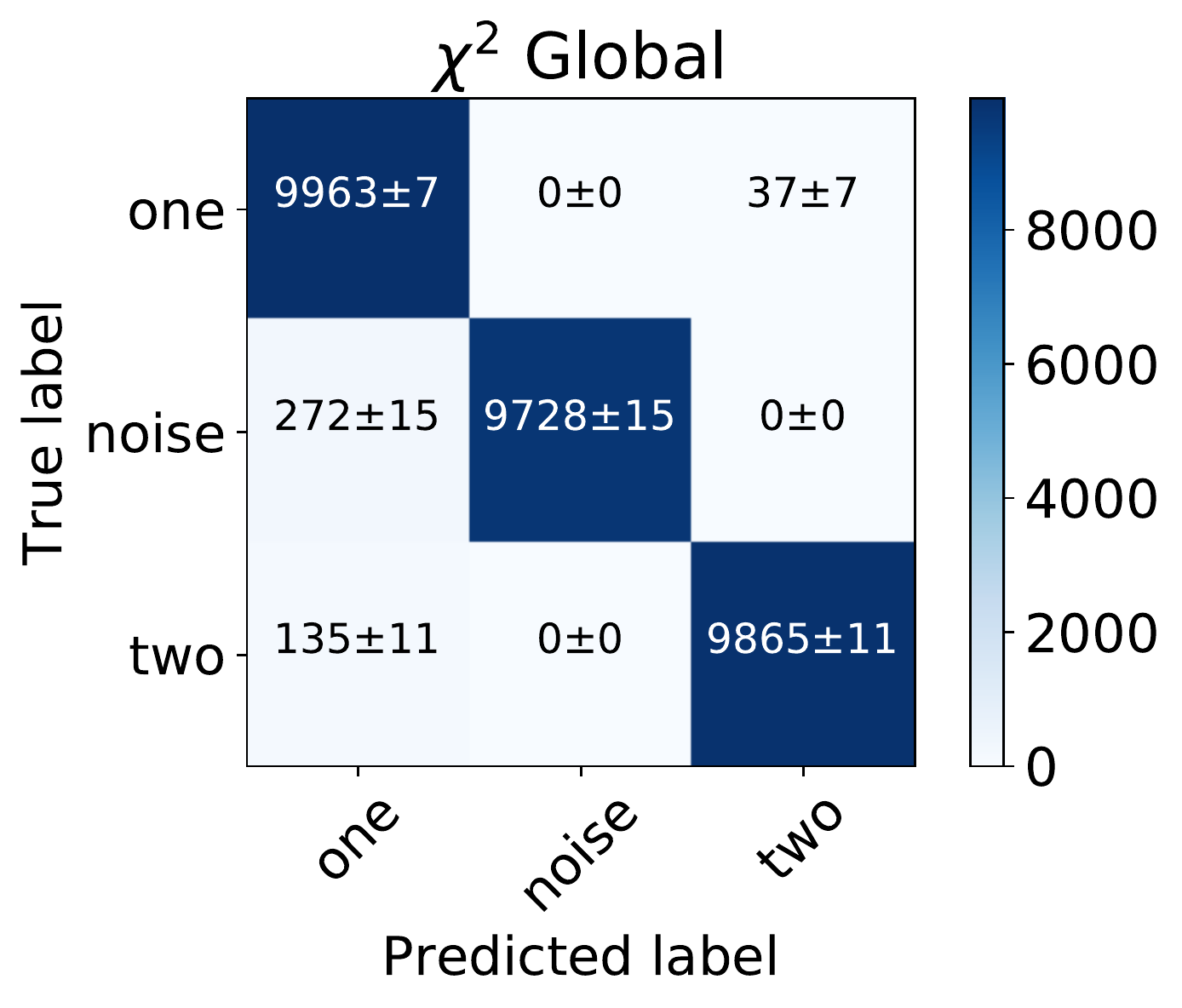}
\caption{Confusion matrices for ten validation sets of 30,000 synthetic spectra (10,000 in each class) classified by the CNN using only the ``local'' spectrum (top left), CNN using only the ``global'' spectrum (top right), CNN using both the local and global spectra (middle left), averaged ensemble of six CNNs (middle right), traditional $\chi^2$-minimization on the local spectrum (bottom left), and the traditional $\chi^2$-minimization on the global view spectrum (bottom right). The ``noise'' class for the $\chi^2$-minimization panels was selected based on a SNR threshold of 4.  Each panel in the confusion matrices shows the mean and standard deviation for the ten validation sets.}
\label{cm}
\end{figure}

After training the three CNN architectures (local-only, global-only, and local+global), we test their performance using ten independent validation sets of 30,000 synthetic spectra (10,000 in each class) described in Section 2.1.  Predictions are made on each of the ten validation sets by the trained CNNs.  The mean and standard deviation of the ten confusion matrices for each architecture are shown in Figure \ref{cm}.  Each square in the confusion matrix shows the amount of correct or incorrect classifications the CNN has made for each class.  For a perfect classification of all samples in the validation set, the upper left, central, and lower right squares in the matrices would each be 10,000.  The off-diagonal squares represent the number of misclassifications in each class.  

The mean classification accuracies and standard deviations for the CNN using only the local spectrum as input across the ten independent validation sets are $96.39 \pm 0.19 \%$, $99.95 \pm 0.02\%$, and $90.56 \pm 0.36\%$ for the one-component, noise, and two-component classes, respectively.  For the CNN using only the global spectrum as input, the classification accuracies improve to $99.31 \pm 0.03 \%$, $100\%$, and $96.82 \pm 0.13\%$ for the three classes.  This improvement can be attributed to the higher SNR of the global spectra, which makes it easier to classify the samples.  When using both the local and global spectra as input to the CNN, the classification accuracies become $99.35 \pm 0.07 \%$, $100\%$, and $96.08 \pm 0.24\%$, which are similar to those of the global-only CNN.  

Although the global-only and local+global CNNs show similar performance, we opt to use the local+global CNN for the remainder of our analysis since the local spectra can be useful for preventing overfitting on real data.  For instance, there are scenarios in real observations where the global spectrum may appear to have two velocity components, but the local spectrum shows only a single component.  For those cases, using the local spectrum as input prevents misclassification. 

The accuracy of the local+global CNN is improved further by averaging the outputs of six independently trained CNNs.  Since each CNN is trained with different random initializations for their parameter weights, there is a variance in their output predictions on a given test set.  Averaging their predictions, however, reduces this variance and often leads to improved overall performance since each CNN may perform better or worse on particular samples.  Known as `ensembling' or model `averaging,' this technique involves summing the three output class probabilities predicted by each CNN before selecting the class with the highest probability as the predicted class for a given sample.  The confusion matrix for this ensemble CNN is shown in the middle right panel of Figure \ref{cm}.  The accuracies for the one-component, noise, and two-component classes improves to $99.92 \pm 0.02 \%$, $100\%$, and $96.72 \pm 0.18\%$.  We refer to this ensemble of CNNs as the ``ensemble CNN'' for the remainder of the paper.

\clearpage

\begin{figure}[htb]
\epsscale{1.1}
\plottwo{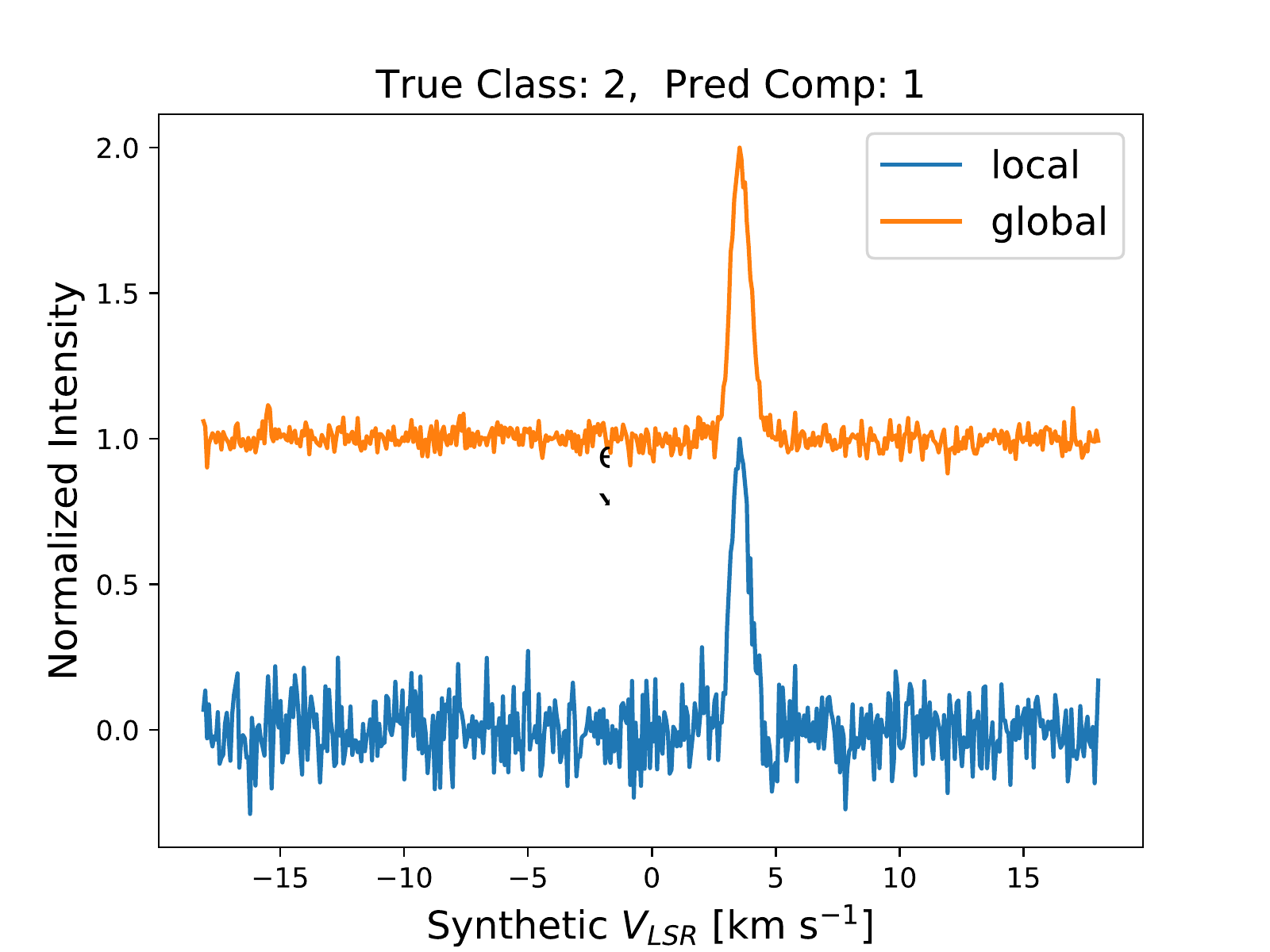}{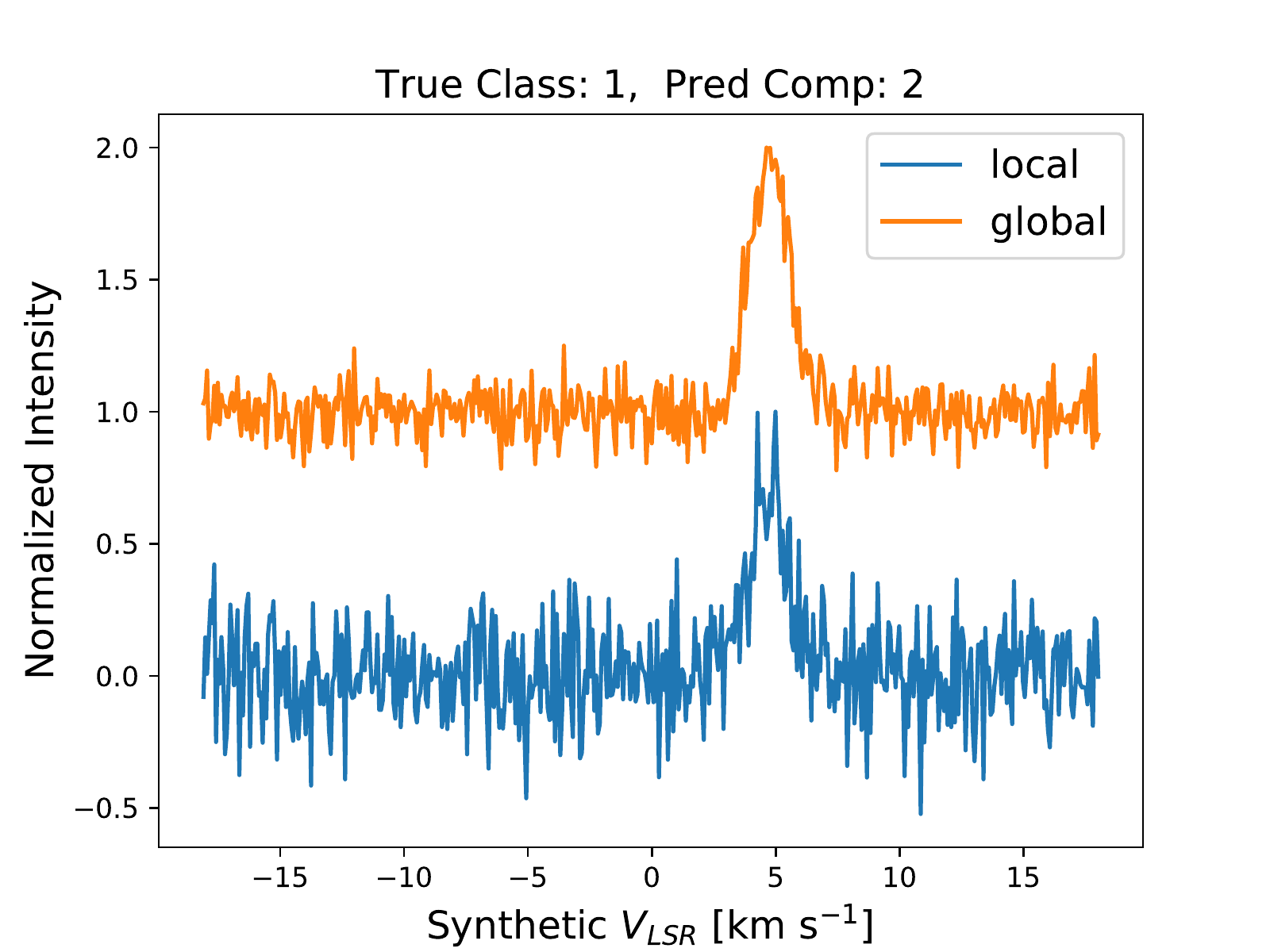}
\plottwo{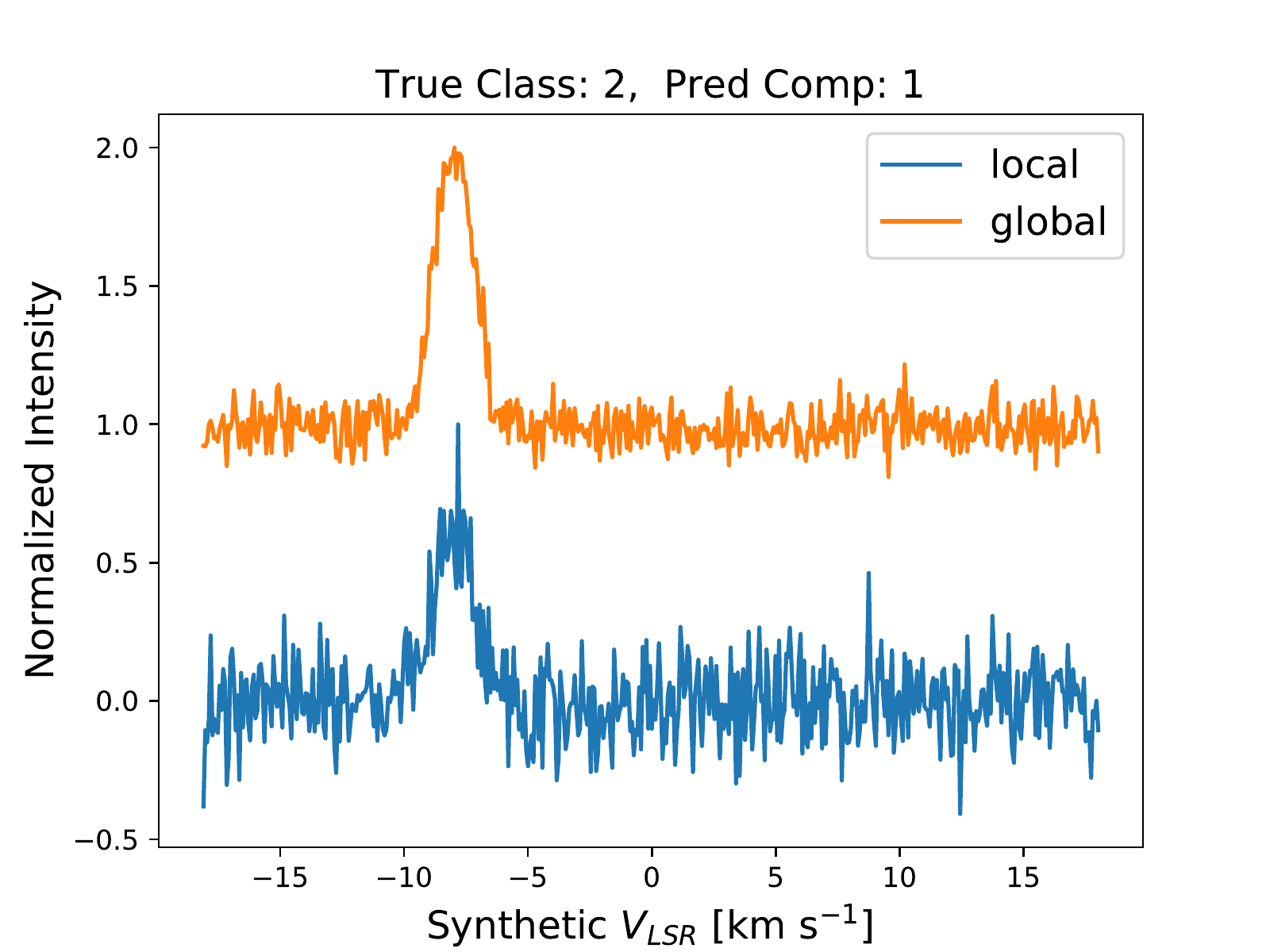}{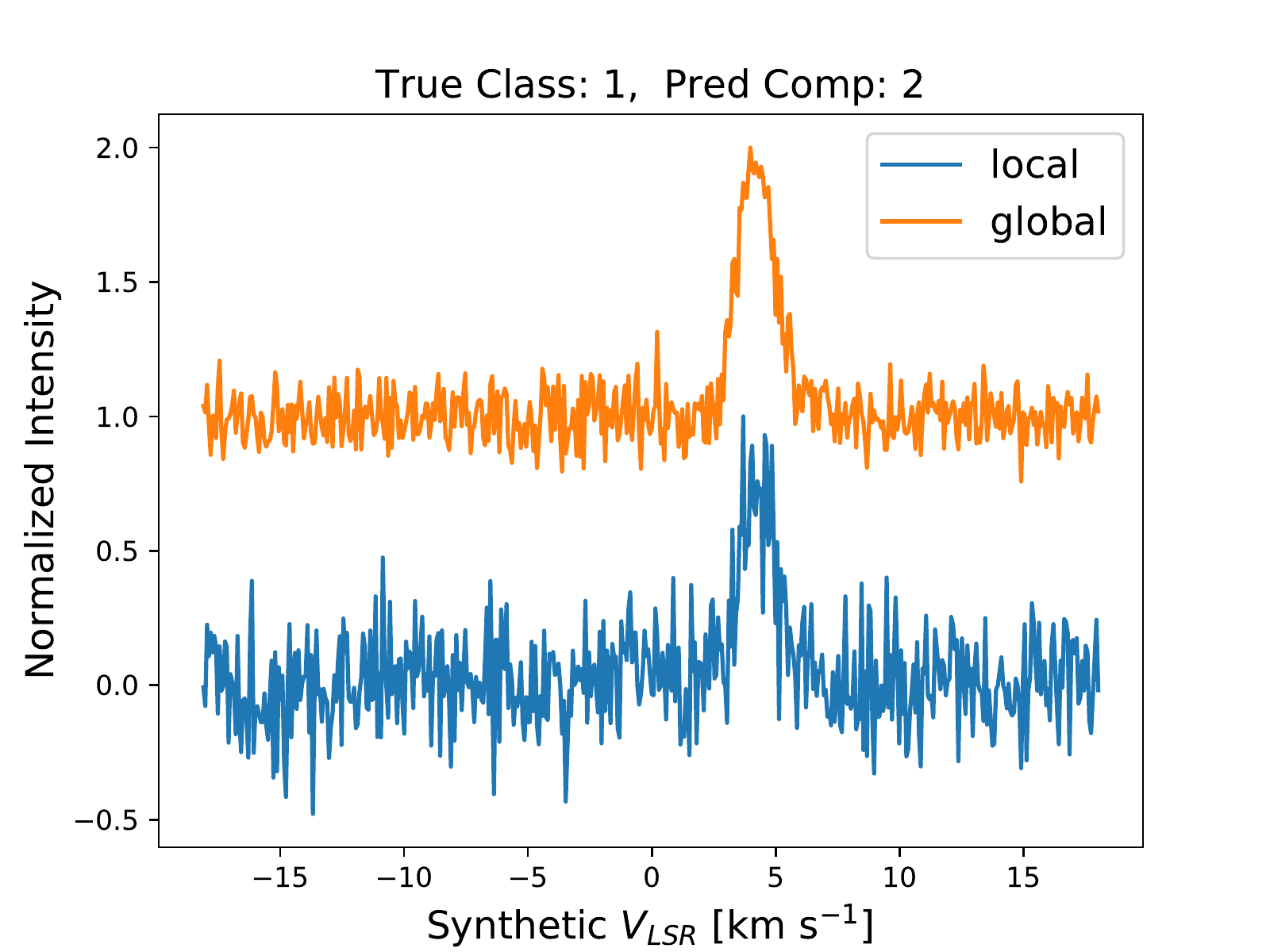}
\centering
\plottwo{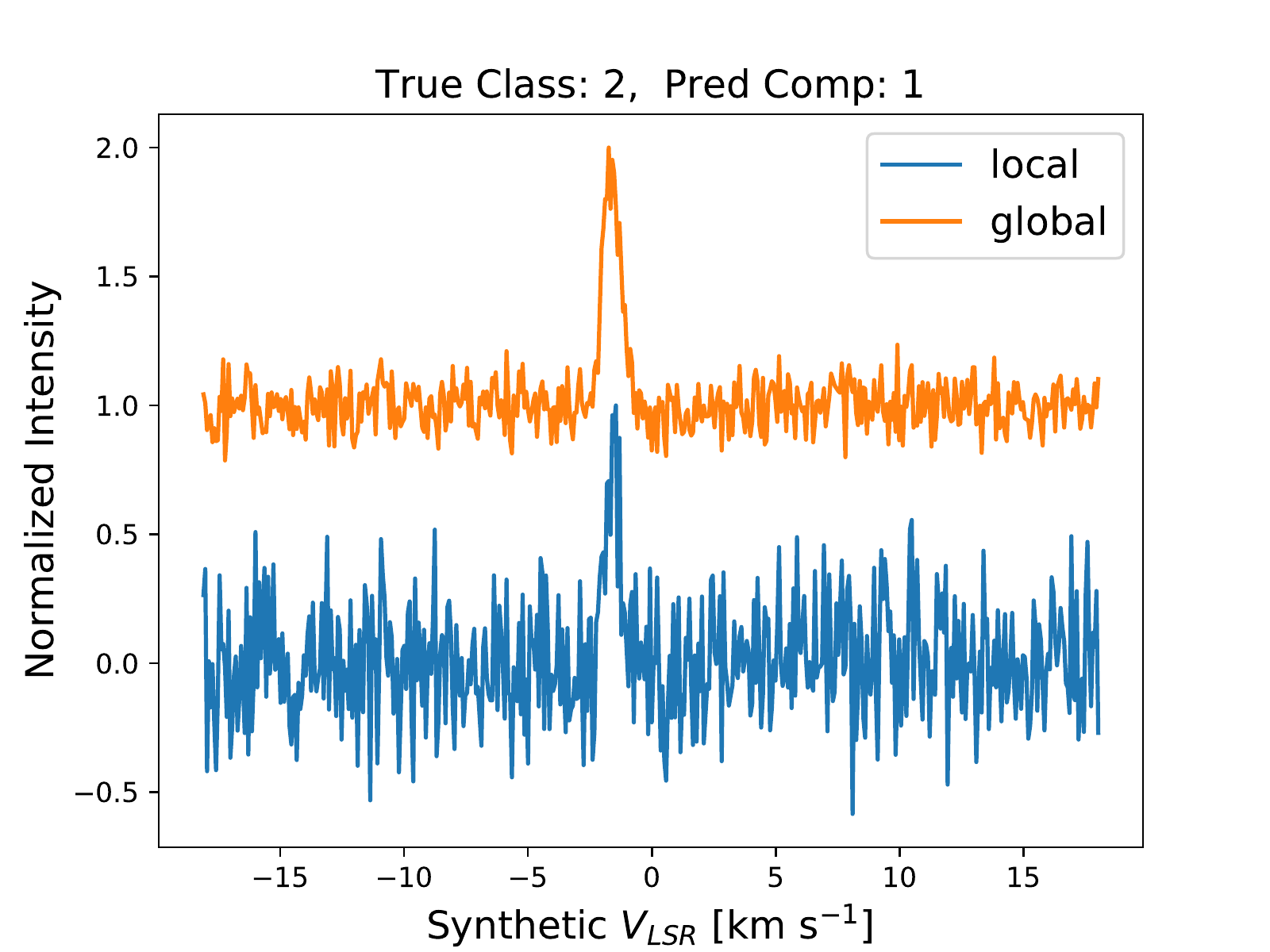}{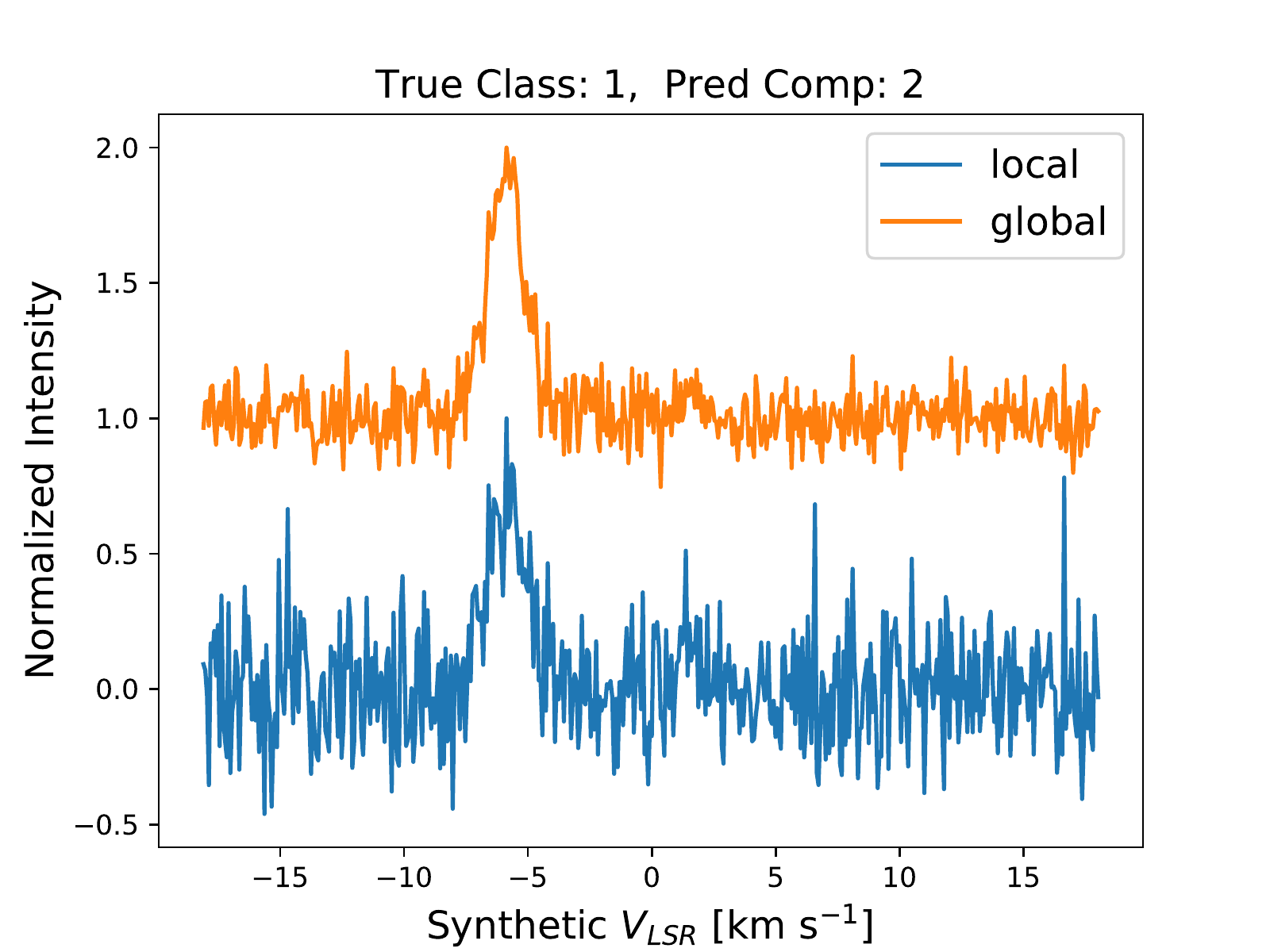}
\caption{Samples in the synthetic validation set misclassified by the ensemble CNN.  The left column shows true two-component samples (True Class: 2) classified as one-component (Pred Comp: 1) by the ensemble CNN. The right column shows true one-component samples (True Class: 1) classified as two-component (Pred Comp: 2) by the ensemble CNN.}
\label{misclass}
\end{figure}
\clearpage

Visual inspection of the ensemble CNN misclassifications in the validation set also reveals that many of those samples indeed exhibit characteristics of the misclassified class.  Several examples of these misclassified samples are shown in Figure \ref{misclass}.  For instance, the true one-component samples that the ensemble CNN incorrectly identified as two-components often have a visible, but subtle, second peak due to the randomness of the noise injection.  Similarly, the true two-component samples identified as one-components by the ensemble CNN are often indistinguishable from a true one-component sample.  As such, these misclassifications are actually a positive sign that the ensemble CNN has ``learned'' the subtle differences between the one- and two-component classes rather than simply memorizing the samples in the training set. 

\subsection{Performance Versus Two-Component Gaussian Line Fitting}
To gauge the ensemble CNN's performance against traditional line fitting methods, we also use a $\chi^2$-minimization model selection technique to make class predictions on the validation set.  Both a single- and two-component Gaussian model are fit to the ``local'' spectrum for each validation set sample using the Levenberg-Marquardt $\chi^2$-minimization method in the \texttt{scipy.optimize.curve$\_$fit} Python package.  For the one-component fit, we use the peak channel in the spectrum as the initial guess for centroid velocity, a set value of 1.0 for the peak intensity guess (spectra are scaled to a max value of 1.0), and a set value of 10 channels for the velocity dispersion guess. To find the optimal solution for the two-component fit, which is more susceptible to falling into local minima solutions rather than global minima, we perform the line-fitting using a grid of initial parameter guesses.  The model with the lowest $\chi^2$ value was selected as the best-fitting two-component model.  The initial guess for the first velocity component in the two-component model was set in the same way as the one-component model, while the second velocity component guesses were set as follows:


\begin{itemize}
\item $T_{peak}$: 0.1 less than the solution found for the one-component fit.
\item $V_{LSR}$: [$\pm$ 10, $\pm$ 30, $\pm$ 50, $\pm$ 70, $\pm$ 90, $\pm$ 110, $\pm$ 130, $\pm$ 150, $\pm$ 170, $\pm$ 190] channels from the solution found for the one-component fit.  Thus, we search for centroids to the left and right of the one-component fit.
\item $\sigma$: 0.1 channel larger than the solution found for the one-component fit.
\end{itemize}

The $\chi^2$ values for the best-fitting single- and two-component models are then compared to select the ``better'' model for the spectrum.  To penalize the larger number of model parameters in the two-component model and consider the number of data points being fit, we apply the Bayesian Information Criterion \citep[BIC; ][]{Schwarz_1978} to each model's $\chi^2$ value.  This approach is similar to the Akaike Information Criterion used in other traditional two-component line fitting methods \citep[e.g.,][]{Henshaw_2016}, but has a built-in penalty for the number of data points in the models being compared.  Namely, the model with the lowest BIC value is selected as the preferred model, where the BIC is given by the following expression:
\begin{equation}
BIC = N\ln(\chi^2) + p\ln(N)    ~~~,
\end{equation} where $p$ is the number of model parameters and $N$ is the number of fitted data points.  The BIC attempts to balance goodness-of-fit (e.g, $\chi^2$ value) against model complexity (i.e., the number of model parameters in relation to data set size) when comparing models.  This approach tries to avoid overfitting (selecting a model that is too complex simply because it fits the data better), but at the same time limit underfitting (selecting a simpler model when a more complex model is more appropriate for the data).  

The results of the BIC comparisons for the ``local'' spectra in the validation set are also shown in the bottom left panel of Figure \ref{cm}.  The ``noise'' class in this case is defined by any spectrum below SNR=4.0.  For this traditional model selection approach, the classification accuracies for the one-component, noise-only, and two-component classes are $98.86 \pm 0.11\%$, $97.31 \pm 0.15\%$, and $87.18 \pm 0.31\%$, respectively.  These accuracies are similar to those from the CNN local-only classifications, with slightly lower accuracies for the noise-only and two-component classes, but a slight increase in accuracy for the one-component class.  



We also repeat the traditional line fitting and model selection method using the ``global'' spectrum for each training sample.  The lower right panel in Figure \ref{cm} shows the accuracy for those classifications.  The classification accuracies in this case are $99.63 \pm 0.07\%$, $97.28 \pm 0.15\%$, and $98.65 \pm 0.11\%$ for the one-component, noise-only, and two-component classes, respectively.  The higher SNR of the ``global'' spectra are likely contributing to this accuracy improvement. 



\begin{figure}[htb]
\epsscale{1.0}
\plottwo{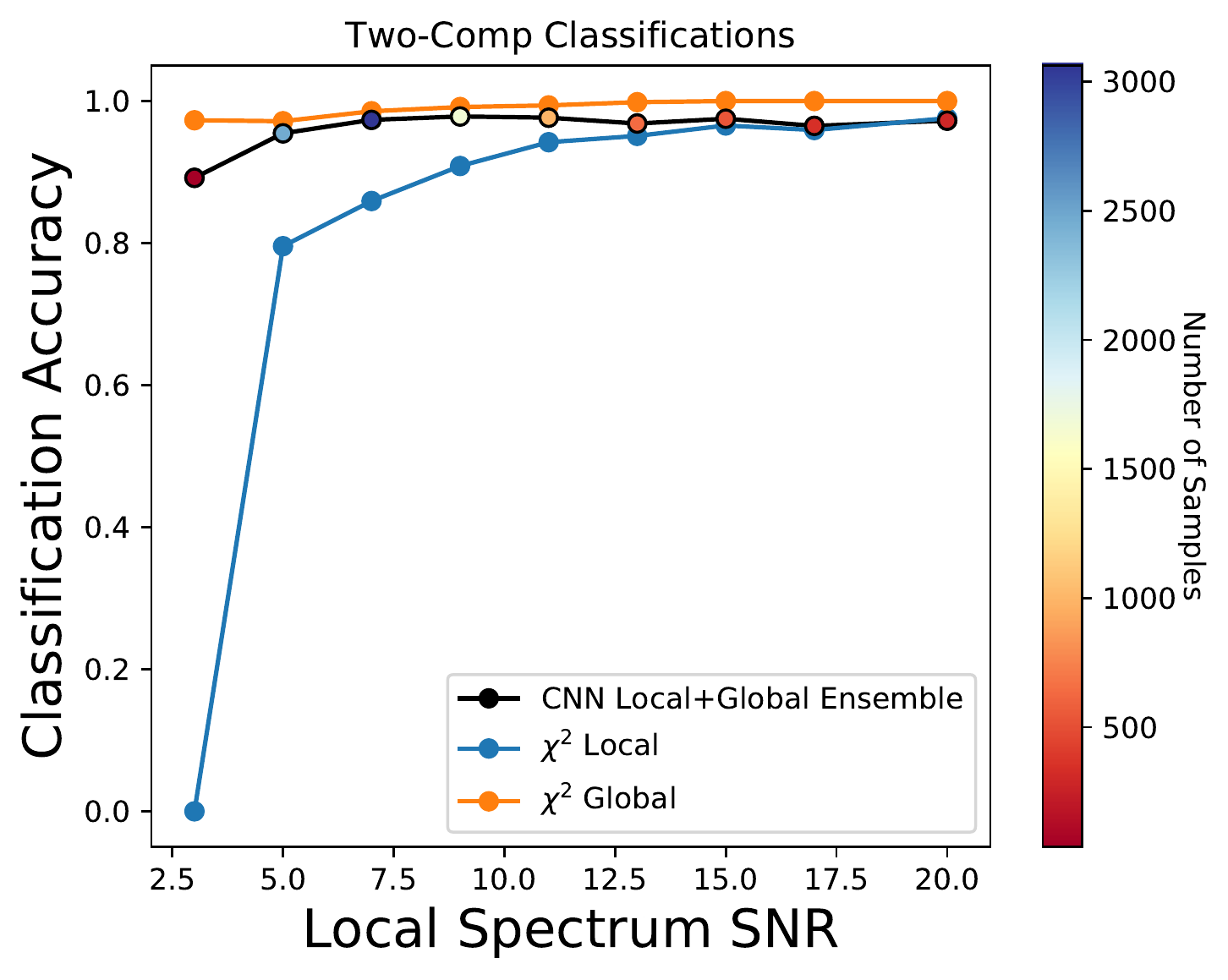}{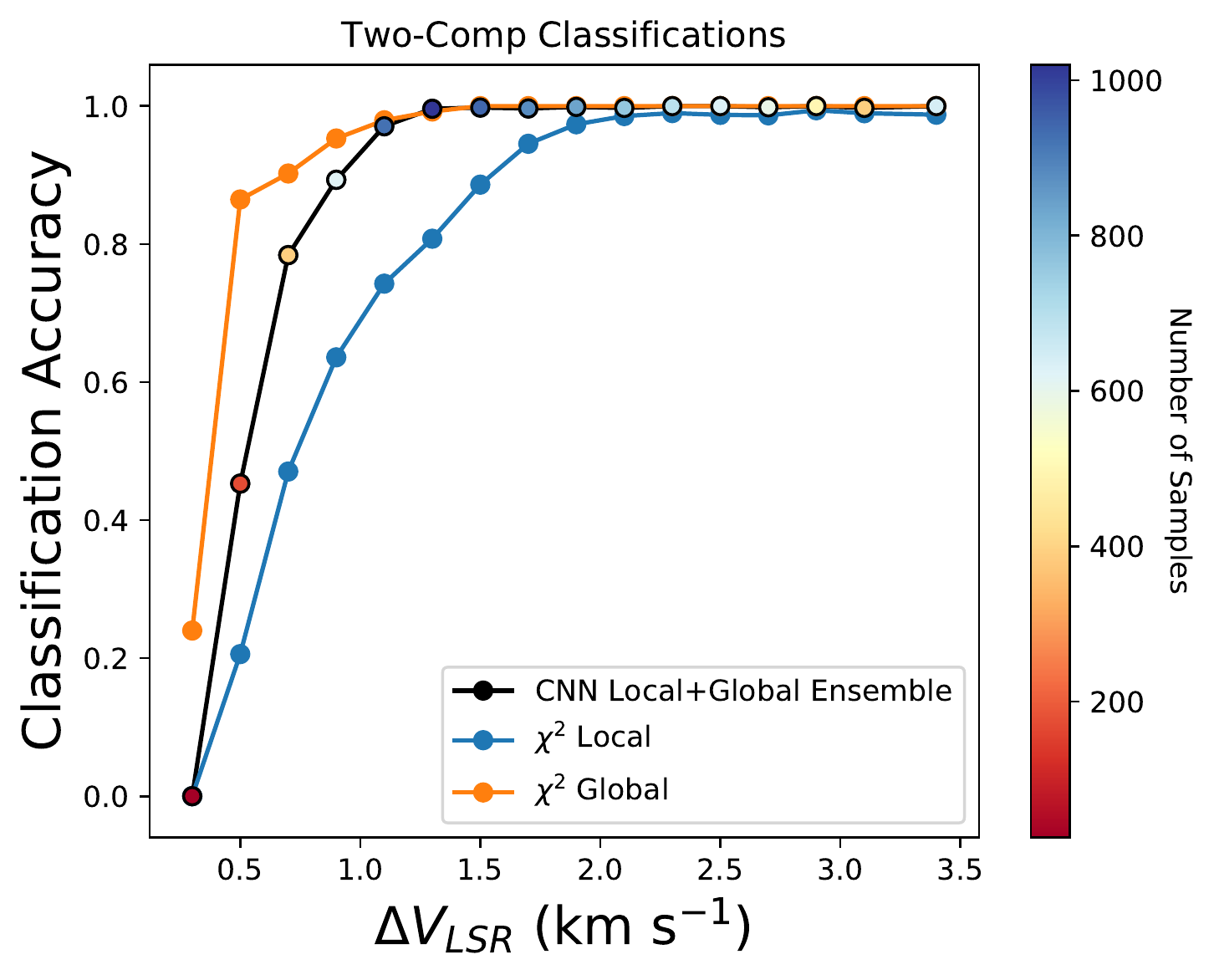}
\caption{Model classification accuracy versus SNR (left) and centroid velocity separation (right) for two-component samples in the synthetic test set.  Each data point represents the classification accuracy for samples within a bin centered on the data point's x-axis position.  The classifications for the traditional $\chi^2$-minimization methods on the ``local'' and ``global'' spectra are shown in blue and orange, respectively.  The color of the CNN Local+Global Ensemble data points (outlined in black) show the amount of test set samples within each bin.  The centroid velocity separation calculation assumes each spectral channel is separated by $\sim0.07$ km s$^{-1}$.}
\label{acc}
\end{figure}  

As an additional comparison between the ensemble CNN and $\chi^2$-minimization predictions, we also show in Figure \ref{acc} each method's classification accuracy for the two-component samples in the synthetic test set versus SNR and centroid velocity separation ($\Delta V_{LSR}$).  Figure \ref{acc} shows that the classification accuracy for the $\chi^2$-global and ensemble CNN methods is stable between an SNR range of 4-20, with variations less than a few percent.  In contrast, the $\chi^2$-local method has a severe drop-off in accuracy for the lowest SNR bin.  Since the $\chi^2$-global and ensemble CNN methods incorporate the higher SNR global spectra into their classifications, they are less affected by the lower SNRs of the local spectra.  

In terms of centroid velocity separation, all three methods show a significant drop-off in classification accuracy below $\Delta V_{LSR} \sim 1$ km s$^{-1}$.  This effect is due to many of the low velocity separation two-component samples being indistinguishable in appearance to one-component class members.  At $\Delta V_{LSR} > 1$ km s$^{-1}$, the components are distinct and easy to identify, causing accuracies to be near $100\%$ for all three methods.  We also see that the $\chi^2$-local method's accuracy begins to decrease at higher velocity separations $\Delta V_{LSR} \sim 2$ km s$^{-1}$ than the other two methods.  This behavior is once again likely related to the lower SNR of the local spectrum, which makes classifying close velocity components more difficult.  

We also note that using solely an averaged spectrum (e.g., the global spectrum) to make classifications is not common for other traditional line-fitting methods \citep{Henshaw_2016, Sokolov_2017, Clarke_2018}.  Typically, a fit to an averaged spectrum is used to set the initial parameter guesses for a second fit to an individual spectrum.  For this reason, all further comparisons will be between the $\chi^2$-local and the ensemble CNN methods.  

 
\subsection{Testing On Real Observations}

\subsubsection{L1689 - $\mathit{^{13}}$CO ($\mathit{1-0}$)}

Although the ensemble CNN has demonstrated high classification accuracy on synthetic data, it is only useful if it can accurately classify real emission-line spectra.  To test the model's performance on real observations, we have collected a $^{13}$CO ($1-0$) spectral cube from L1689 in the Ophiuchus molecular cloud observed by the COMPLETE survey \citep{Ridge_2006}.  This cube provides an excellent test for the ensemble CNN since it displays spectra that belong in all three of the classes in our training set.  For this test, we implicitly assume that the line emission observed is optically thin everywhere.  The use of CLOVER or traditional two-component line-fitting techniques on data with self-absorbed single-component lines will likely result in erroneous conclusions about the nature of the emission.  Nevertheless, even if the observed emission is optically thick and self-absorbed, it still provides an adequate test set since self-absorption features mimic the appearance of optically thin emission with two velocity components along the line of sight.  

\begin{figure}[htb]
\epsscale{0.75}
\plotone{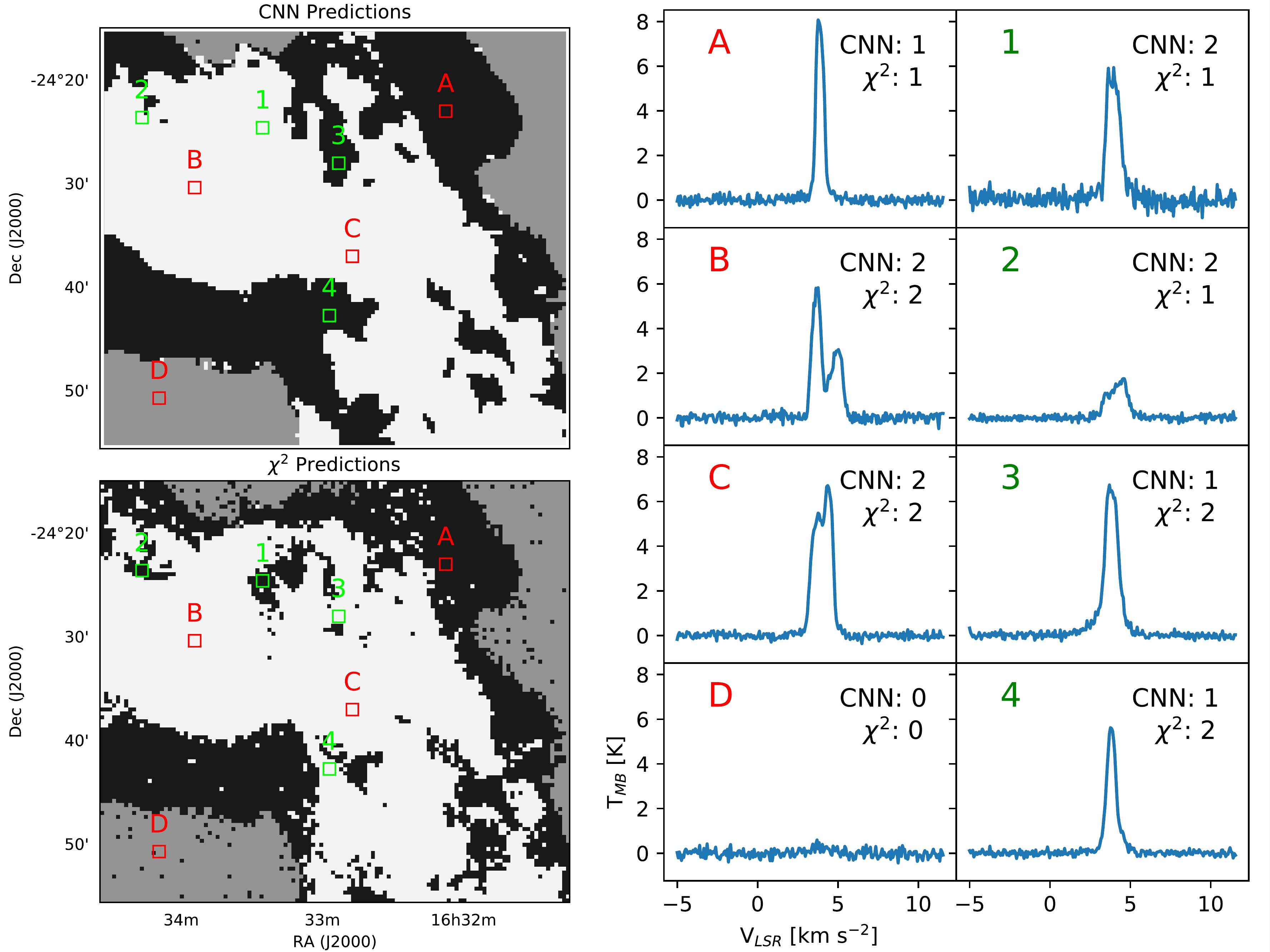}
\caption{Left panels: example segmentations of a $^{13}$CO ($1-0$) spectral cube observation of L1689 into three classes: single velocity component spectrum (black), multiple velocity component spectrum (white), and noise (grey) using CLOVER's ensemble CNN (top) and traditional $\chi^{2}$-minimization model fitting (bottom).  Right panels: The ``global'' view spectra extracted from the observed spectral cube at the positions of the 3$\times$3 pixel windows overlaid onto the left panels.  Red letters denote positions where CLOVER and the $\chi^{2}$ technique agree in their class predictions, while the green numbers show positions where they disagree.  The text in the upper right corner of each panel shows the class predicted by CLOVER and the $\chi^{2}$ technique for that spectrum, where 2=two-component, 1=one-component, and 0=noise.}
\label{multi_comps_segmentation}
\end{figure}

Figure \ref{multi_comps_segmentation} shows the output predictions after a sliding window of size 3$\times$3 pixels has been moved across the position-position plane of the cube, the ``local'' and ``global'' spectra are extracted, and fed into the ensemble CNN.  Gray pixels in Figure \ref{multi_comps_segmentation} denote those that were predicted to be in the noise class, while black represents pixels predicted to be in the one-component class and white shows those predicted to be two-component class members.  The red-lettered panels in Figure \ref{multi_comps_segmentation} show the ``global'' spectra at different locations on the data cube.  Since we have no a priori knowledge of the physical processes that created any apparent two-component features in these real spectra, we must rely only on the appearance of the spectra when determining the success of the CNN's predictions.  Nevertheless, comparing the spectra highlighted in Figure \ref{multi_comps_segmentation} to the ensemble CNN prediction map reveals that the CNN can distinguish the spectral differences between each class.  Even two-component spectra with closely separated velocity components are correctly identified by the model (see, e.g., spectrum C in Figure \ref{multi_comps_segmentation}).

Figure \ref{multi_comps_segmentation} also displays the class predictions of each pixel obtained from the traditional BIC model selection method using the ``local'' spectrum.  For this method, any pixel with SNR $< 4$ is deemed noise.  The red-lettered panels in Figure \ref{multi_comps_segmentation} show locations where the traditional model selection method agrees with the ensemble CNN predictions.  As can be seen, these tend to be high SNR spectra where the class of the object is obvious.  The green-numbered panels in Figure \ref{multi_comps_segmentation} show spectra from locations where the two methods disagree in their class predictions.  These disagreeing cases reveal clear examples of the $\chi^{2}$-minimization method underfitting (labeled panels 1 and 2 in Figure \ref{multi_comps_segmentation}) and overfitting (labeled panels 3 and 4 in Figure \ref{multi_comps_segmentation}) the spectra.  Visual inspection of the individual best-fit models for the $\chi^{2}$-minimization approach at those locations reveals that they are not cases in which the method fails to provide good fits to each spectrum, but rather they are failures of the BIC model selection technique. Conversely, the ensemble CNN is able to identify weak two-component features that are deemed to be one-component by the $\chi^{2}$-minimization approach (labeled panels 1 and 2 in Figure \ref{multi_comps_segmentation}), but is also resilient against predicting the two-component class when it is not warranted (labeled panels 3 and 4 in Figure \ref{multi_comps_segmentation}).  These examples serve as evidence for the advantage that the ensemble CNN can provide for identifying multiple velocity component spectra. 

Moreover, the ensemble CNN predictions on the entire spectral cube, which has dimensions of 118$\times$106 pixels (i.e., $\sim$ 12,500 individual predictions), take only 137 seconds ($\sim 23$ seconds for each of the six CNN predictions in the ensemble) on a single core of a 2.8 GHz Intel Core i7 CPU.  Although CLOVER's prediction speed could improve by utilizing multiple cores on a single CPU or GPU, the low computing power required for CLOVER to obtain fast performance is a marked advantage over traditional methods (see Section 5.1 for a comparison involving both classification and parameter predictions).  The number of pixels in typical spectral cubes is also growing with the advent of focal plane arrays that quickly map large areas of the sky \citep[e.g.,][]{Morgan_2008, Sieth_2014} and interferometers (e.g., ALMA, ngVLA, etc.) that map at high spatial resolutions.  As such, the quick prediction speeds provided by CLOVER make it an attractive tool for the next generation of large-scale spectroscopic surveys of star-forming regions.

\subsubsection{DR21 - C$\mathit{^{18}}$O ($\mathit{3-2}$)}
Although $^{13}$CO ($1-0$) emission is a common tracer of molecular gas in star-forming regions, it can become optically thick in some environments.  The high opacity emission sometimes leads to self-absorption dips, which can mimic the double-peaked structure of optically-thin two-component spectra \citep[e.g.,][]{Lee_1999, Sohn_2007, Schnee_2013, Keown_2016}.  To ensure that self-absorption is not affecting the CNN predictions, we also test the CNN on a C$^{18}$O ($3-2$) spectral cube of DR21 in the Cygnus X star-forming region.  Since C$^{18}$O is a much rarer isotopomer than $^{13}$CO, its emission is almost always optically thin and rarely suffers from self-absorption dips.  

We advise users of CLOVER to determine whether or not the emission they are inputting into the algorithm is optically thin or thick.  Since CLOVER makes its predictions under the assumption that the emission is optically thin, any significantly self-absorbed optically thick spectrum it receives as input will most likely be classified in the two-component class.

\begin{figure}[htb]
\epsscale{0.8}
\plotone{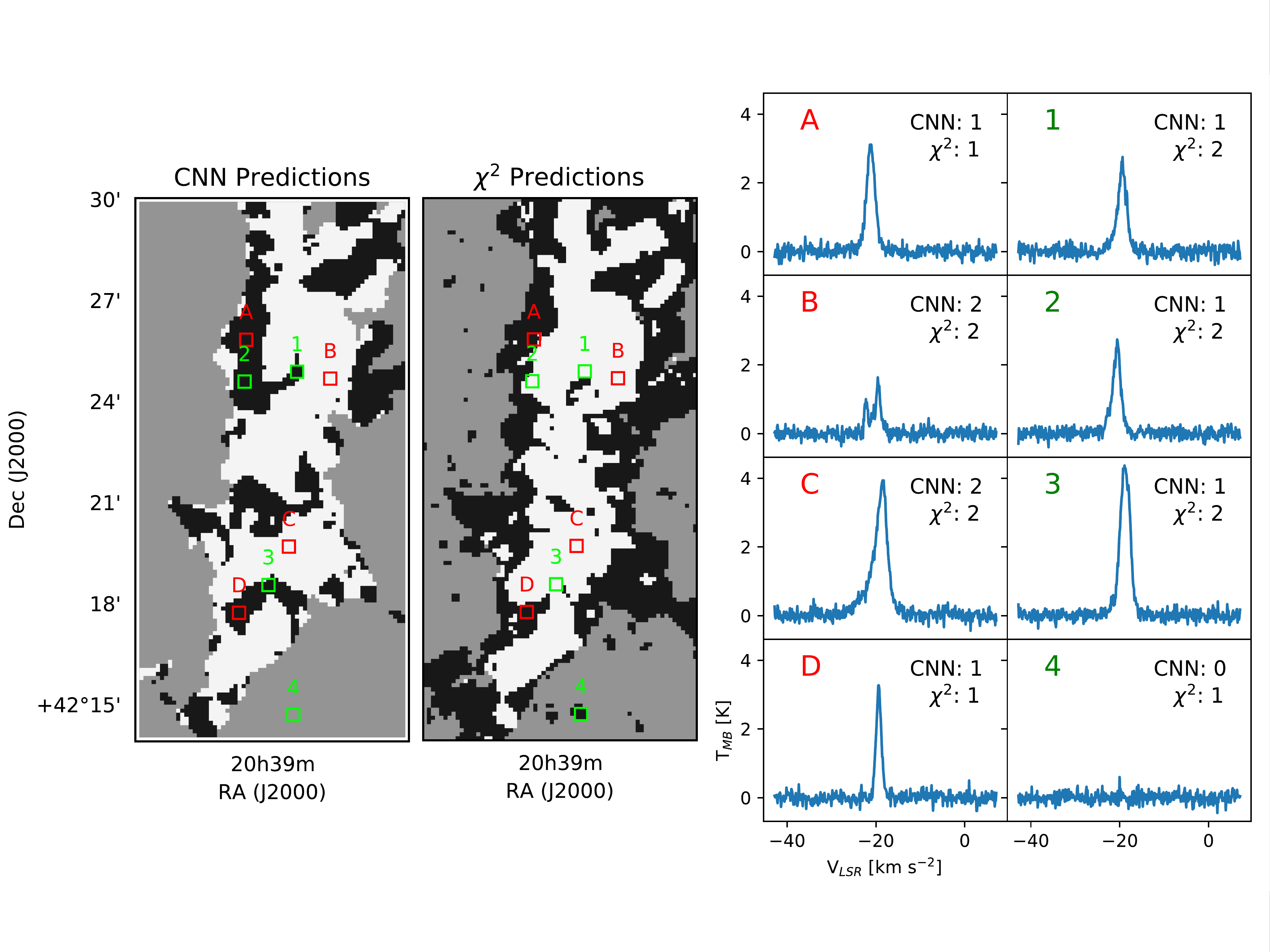}
\caption{Same as Figure \ref{multi_comps_segmentation}, but for the C$^{18}$O ($3-2$) observations of DR21.}
\label{multi_comps_segmentation_DR21}
\end{figure} 

\begin{figure}[htb]
\epsscale{0.8}
\plotone{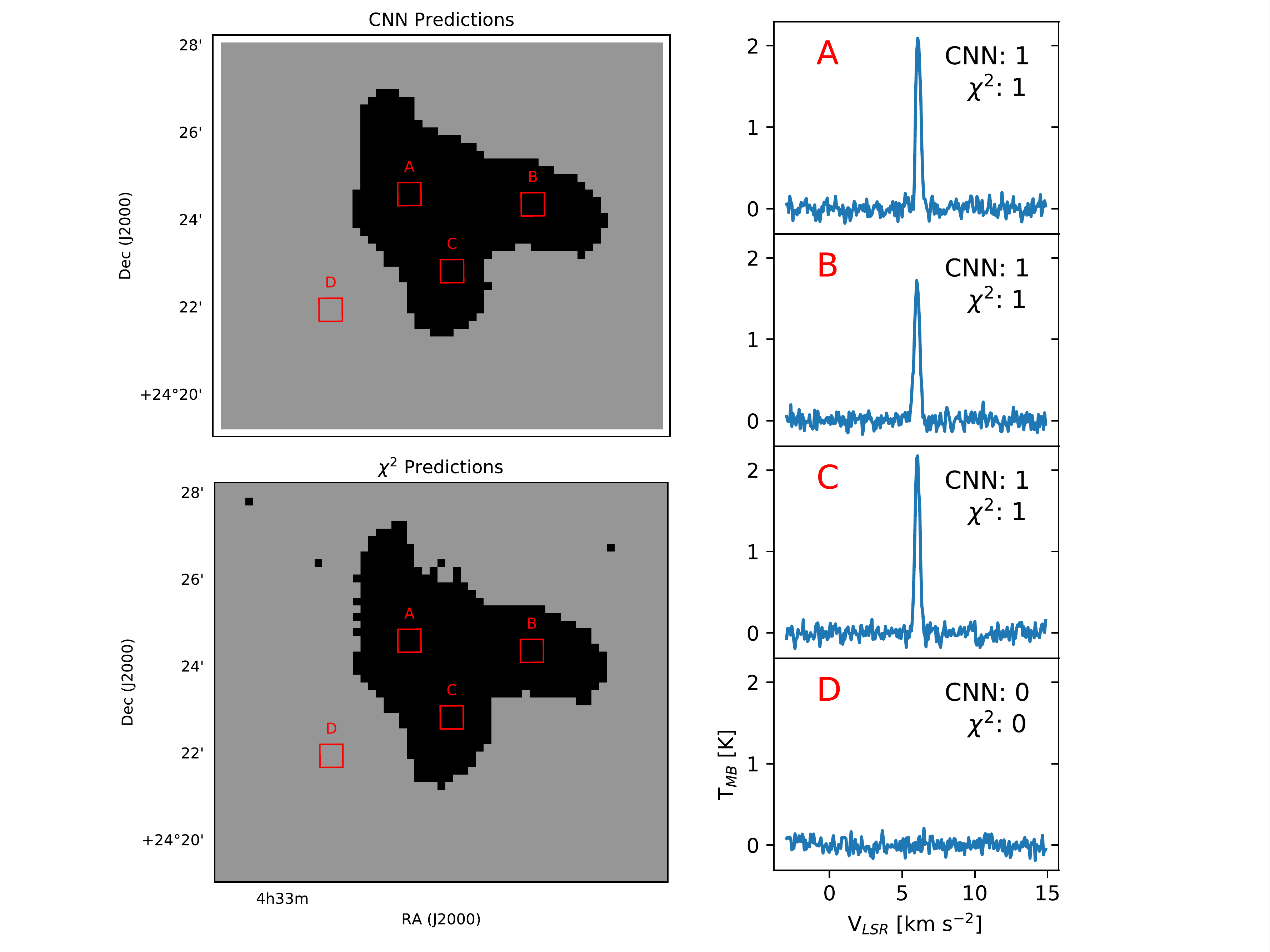}
\caption{Same as Figure \ref{multi_comps_segmentation}, but for the HC$_5$N ($9-8$) observations of B18.}
\label{multi_comps_segmentation_B18}
\end{figure} 

Figure \ref{multi_comps_segmentation_DR21} shows the results of both the ensemble CNN and $\chi^{2}$-minimization model selection technique on the C$^{18}$O ($3-2$) spectral cube.  Once again, we see that the ensemble CNN and $\chi^{2}$-minimization methods show overall agreement between their predictions.  As seen in the green numbered panels of Figure \ref{multi_comps_segmentation_DR21}, the ensemble CNN method is less susceptible to overfitting than the $\chi^{2}$-minimization method.  For instance, the $\chi^{2}$-minimization method frequently classifies spectra that appear to have a single velocity component (or very subtle wings) as two-components. 

In addition to showing the differences between one- and two-component class predictions for the ensemble CNN and $\chi^{2}$-minimization approach, Figure \ref{multi_comps_segmentation_DR21} also highlights the advantages of the CNN's noise class predictions over simple SNR thresholds.  For example, the $\chi^{2}$-minimization approach's SNR cutoff leads to many islands of one-component members that should instead be classified as noise (see, e.g., green spectrum 4 in Figure \ref{multi_comps_segmentation_DR21}).  Conversely, the ensemble CNN segmentation is much smoother, showing a clear distinction between the core of signal at the center of the cube and noise at the edges.  A similar distinction between the noise and signal can be seen in the L1689 CNN segmentation shown in Figure \ref{multi_comps_segmentation}.  This behavior provides further evidence of the advantages gained by incorporating CNNs into the line-fitting procedure. 



\subsubsection{B18 - HC$\mathit{_5}$N ($\mathit{9-8}$)}
As an additional comparison between the ensemble CNN and $\chi^{2}$-minimization approaches, we also test their performance on a HC$_5$N ($9-8$) spectral cube from B18 in the Taurus star-forming region observed by the Green Bank Ammonia Survey \citep{Friesen_2017}.  B18 is a much more quiescent region than L1689 and DR21, which means its emission tends to have only a single velocity component.  HC$_5$N ($9-8$) is also an optically thin transition, ensuring that self-absorption is not affecting the spectra.  Thus, this cube provides a test to see how robust the ensemble CNN is for cubes that lack two-component class members.  

Figure \ref{multi_comps_segmentation_B18} displays the segmentation results for the ensemble CNN and $\chi^{2}$-minimization approach applied to the B18 cube.  Overall, the two methods are in good agreement.  Both correctly identify that only one-component and noise-only spectra are within the cube.  As in the other test regions, we once again see that the ensemble CNN noise segmentation for B18 is superior to the SNR threshold of the $\chi^{2}$-minimization approach since there is a clear distinct between the noise and signal in the former.

\subsection{Testing on Three-Component Spectra}
One important assumption of CLOVER's ensemble CNN classifications are that they assume the input spectra belong to one of the three classes they were trained to predict (one-component, two-component, or noise-only).  In real observations, however, three or more velocity components may be present in a single spectrum \citep[e.g.,][]{Sokolov_2017, Clarke_2018, Chen_2019}.  As a simple test to see how CLOVER would classify spectra with three velocity components, we generate an additional synthetic test set of 30,000 three-component spectra.  The first and second velocity components for each sample in the test set were generated using the same steps described in Section 2.1 for generating the two-component samples.  A third velocity component was introduced by injecting a third Gaussian spectrum into each sample.  The velocity dispersion and centroid for this third Gaussian were randomly drawn from a uniform distribution with the following limits: $2$ channels $\leq \sigma \leq 11$ channels, and $-0.55$ km s$^{-1} \leq$ $V_{LSR} \leq 0.55$ km s$^{-1}$ (where the spectral axes has been normalized between $-1$ km s$^{-1}$ and 1 km s$^{-1}$).  $T_{peak}$ for the third Gaussian was drawn randomly from a uniform distribution between $2\times$RMS and 1 K, where RMS is the noise level selected for the cube.  The RMS level for each sample was set as described in Section 2.1.  

When predicting the class of the 30,000 three-component spectra, CLOVER assigns the two-component class to 29,827 ($\sim 99\%$) and the one-component class to only 173.  Thus, CLOVER's two-component classifications can be thought of as a ``multi-component'' class.  If presented with a sample containing more than two velocity components, the current implementation of CLOVER will likely place that sample in its two-component class. 

\section{Deriving Kinematics From Two-Component Spectra}
The quick and accurate classifications provided by CLOVER can be used to improve kinematics measurements in one of two ways: 1) As a preprocessing technique that will predict the class of each pixel, then a traditional line fitting method can be used to find the best-fitting parameters for that model. 2)  As a preprocessing technique into a second neural network that will predict the centroid velocity and line width directly from the spectra of the pixels identified as two-component class members.  In this section, we demonstrate the latter case - deriving kinematics directly from spectra.

\cite{Fabbro_2018} showed that CNNs have similar performance as traditional least-squares template fitting for deriving stellar parameters from APOGEE spectra.  More importantly, the \cite{Fabbro_2018} CNN made stellar parameter predictions significantly faster than least-squares template fitting, highlighting the advantages gained by utilizing neural network architectures.  With those results in mind, it is likely that neural networks can perform similarly to the traditional least-squares model fitting commonly used to derive kinematics from emission-line spectral data of star-forming regions.  

Using a similar neural network architecture described in Section 3 for CLOVER's spectral classification, we trained an additional network to use the local and global spectra for a two-component class member (i.e., a pixel predicted to be a two-component by the ensemble CNN) to predict the velocity centroid, dispersion, and peak intensity of each component.  There are two main changes to the architecture of this network from the spectral classification network: 1) Instead of the training set labels being classes, they are now the velocity centroid, dispersion, and peak intensity of each component (i.e., in the most general case, both the inputs and outputs of a machine learning problem can be multidimensional. Here, we have a multidimensional-output regression.).  The training set labels are a six-number array, with the first being the centroid of the lower-velocity component, the second being the centroid of the higher-velocity component, the third being the dispersion of the lower-velocity component, the fourth being the dispersion of the higher-velocity component, the fifth being the peak intensity of the lower-velocity component, and the sixth being the peak intensity of the higher-velocity component.  This setup ensures that the network always predicts the labels in the same order so that no label switching occurs.  2) The output layer consists of six output neurons (one for each label) with linear activation functions that predict continuous values rather than the probability of each class.  The centroid velocity labels are normalized between $-1$ and 1, with $-1$ being the left (lowest-velocity) edge of the spectrum and 1 being the right (highest-velocity) edge of the spectrum.  The velocity dispersion labels are represented in units of spectral channels.

\begin{figure}[htb]
\epsscale{0.95}
\plotone{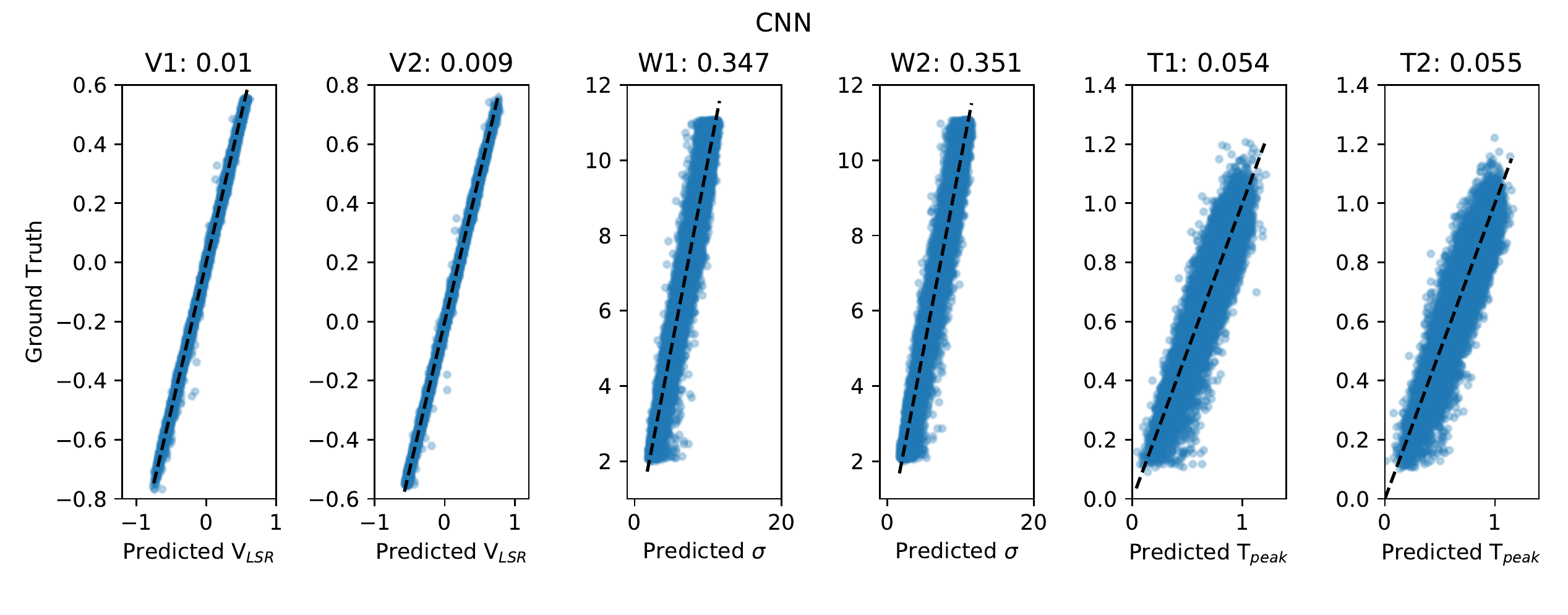}
\plotone{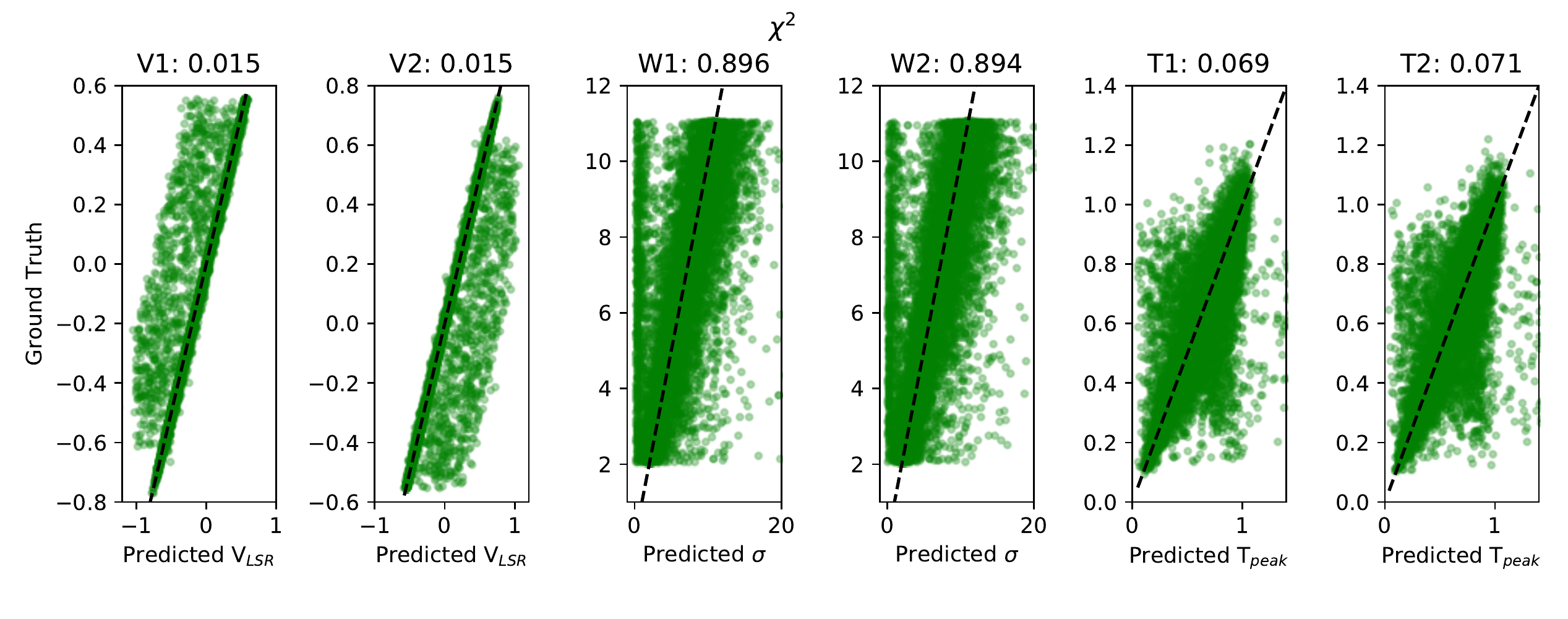}
\plotone{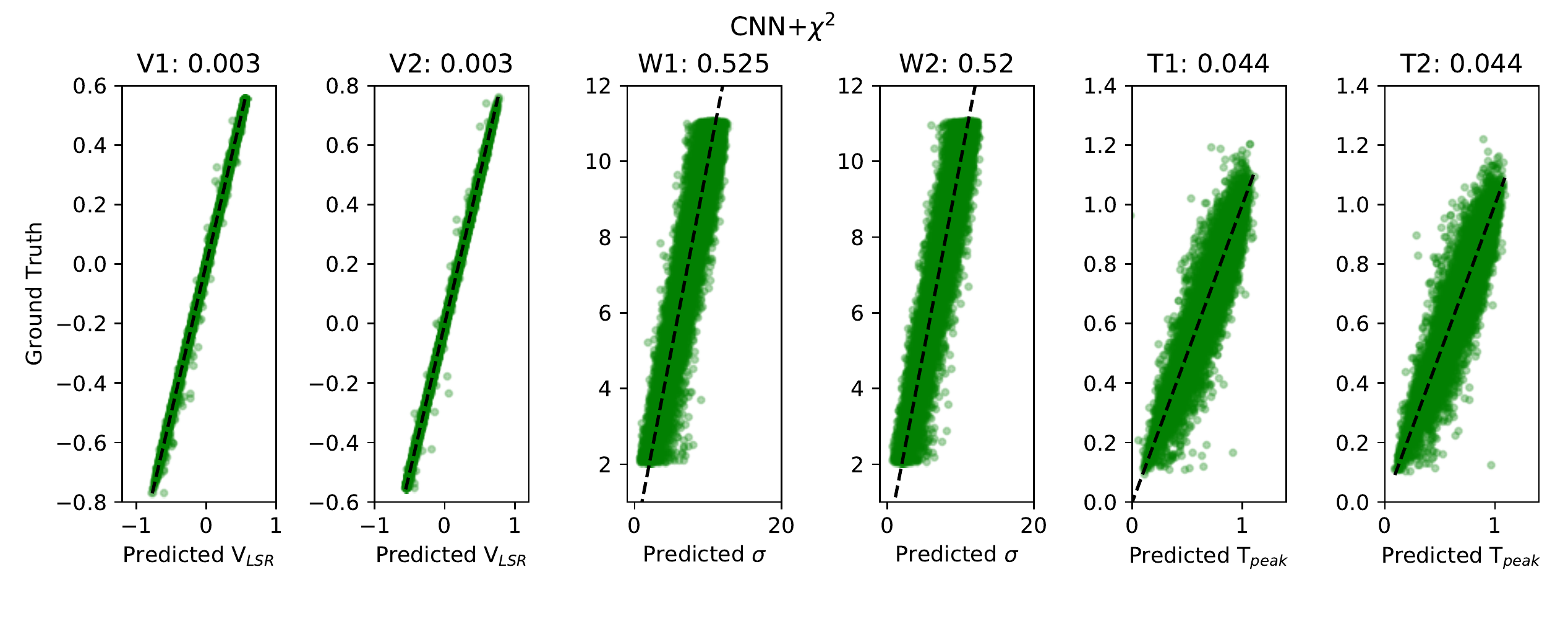}
\caption{Velocity centroid (two left columns), dispersion (middle two columns), and peak intensity (two right columns) predictions by CLOVER's trained regression CNN (top row), $\chi^2$-minimization grid search method (middle row), and $\chi^2$-minimization method with CNN initial guesses (bottom row) versus the ``ground-truth'' for the low-velocity component (V1, W1, T1) and high-velocity component (V2, W2, T1) for the 30,000 two-component spectra in the synthetic test set.  The dashed lines show a one-to-one correspondence. In all panels, the centroid velocities are normalized between $-1$ and 1.  The velocity dispersion units are the number of channels in the spectrum.  The subtitle above each panel shows the mean absolute error for that parameter.}
\label{multi_comps_regression}
\end{figure}

The training set for this regression network included 300,000 two-component spectra generated using the same method discussed in Section 2.1.  A validation set of an additional 90,000 spectra was also used to monitor the network's performance during training in order to apply early-stopping.  After training, a test set of 30,000 additional two-component samples were generated and used to gauge the network's performance.  The top row of Figure \ref{multi_comps_regression} displays the regression accuracy of the CNN predictions for the test set.  The model's predictions are accurate to mean absolute error (MAE = $\frac{1}{n}\sum_{t=1}^{n}|e_t|$, where $e_t$ is the error in the prediction of sample $t$) of $\sim 0.01$ for centroid velocity, $\sim 0.35$ for velocity dispersion, and $\sim 0.06$ for peak intensity.

\clearpage

\begin{figure}[htb]
\epsscale{1.0}
\plottwo{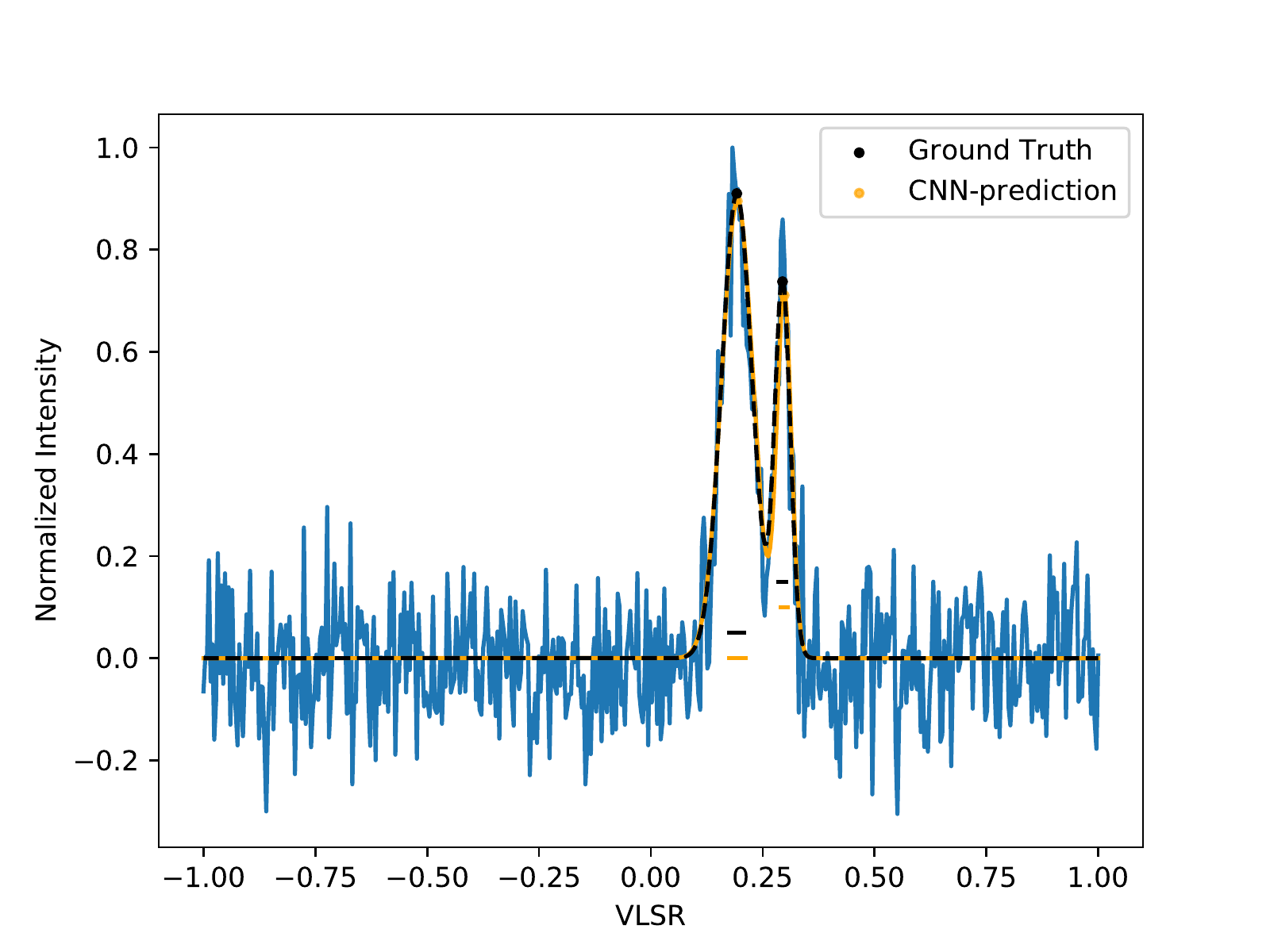}{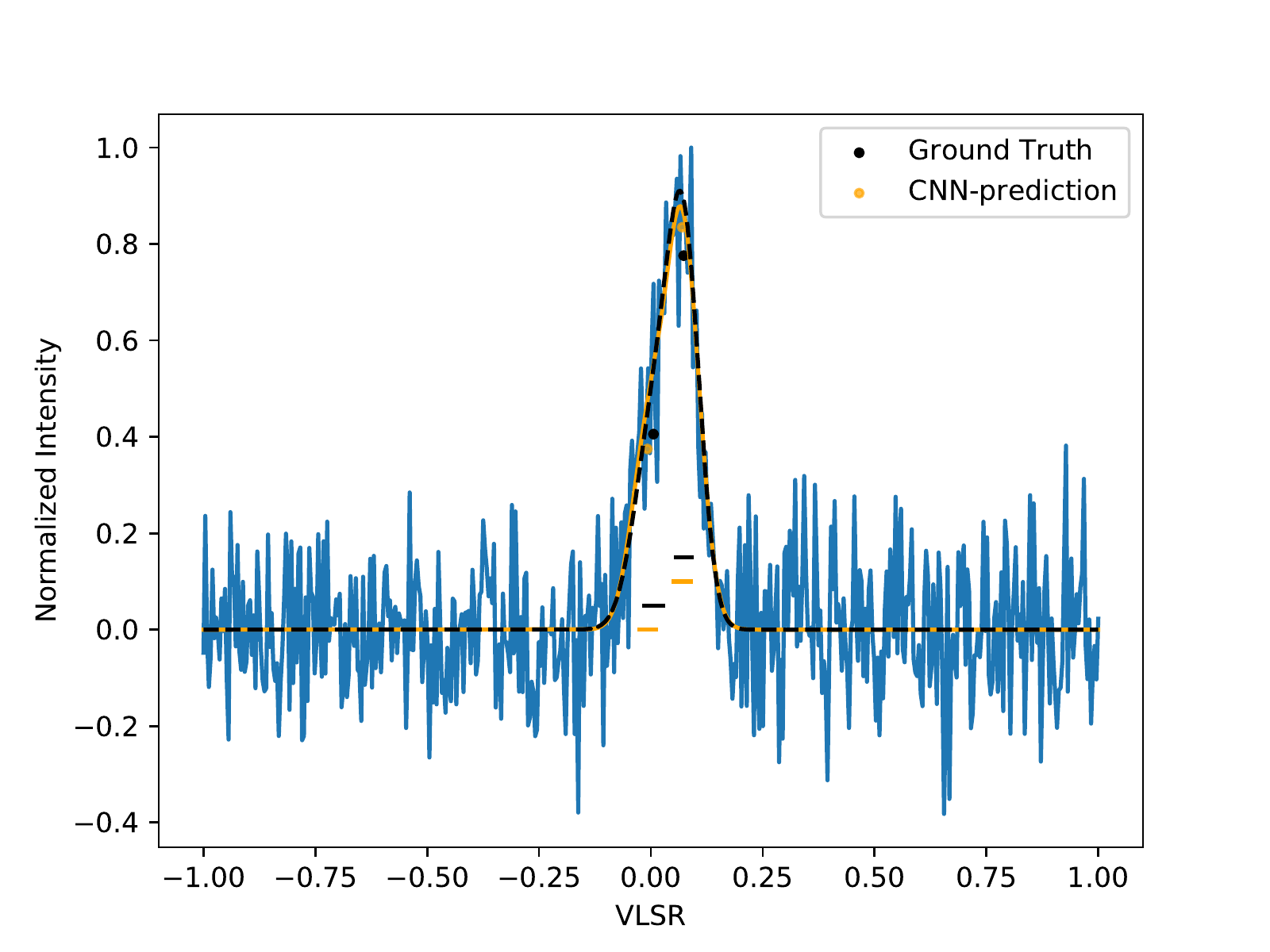}
\caption{Example predictions by CLOVER on previously unseen ``local-view'' spectra from the synthetic test set.  The black dots/bars show the positions of the ``ground-truth'' velocity centroids, peak intensity, and velocity dispersions used to generate the synthetic sample.  For comparison, the orange dots/bars show CLOVER's parameter predictions. The dashed black line shows the ground-truth model used to generate the synthetic sample, while the orange solid line shows the corresponding two-component model generated using CLOVER's parameter predictions.}
\label{test_reg}
\end{figure}

Figure \ref{test_reg} shows the trained network's predictions for two samples in the test set.  The model can accurately predict the kinematics of components that have large velocity separations (e.g., left panel of Figure \ref{test_reg}), but also those that are blended together (e.g., right panel of Figure \ref{test_reg}).

The middle row in Figure \ref{multi_comps_regression} shows the performance of the $\chi^2$-minimization method's best-fit two-component model for every sample's local spectrum in the test set.  Using this method, the mean absolute errors increase to $\sim 0.015$ for centroid velocity, $\sim 0.9$ for velocity dispersion, and $\sim 0.07$ for peak intensity, with a significant number of poor fits as shown by the abundance of outliers in each panel.  Disregarding the outliers, the spread of the $\chi^2$-minimization predictions about the one-to-one line is similar to the CNN predictions.  The lack of outliers in the CNN predictions, however, suggests that it is more resilient against fitting noise and/or falling into local minimum solutions compared to the $\chi^2$-minimization method. 

\begin{figure}[htb]
\epsscale{0.9}
\plottwo{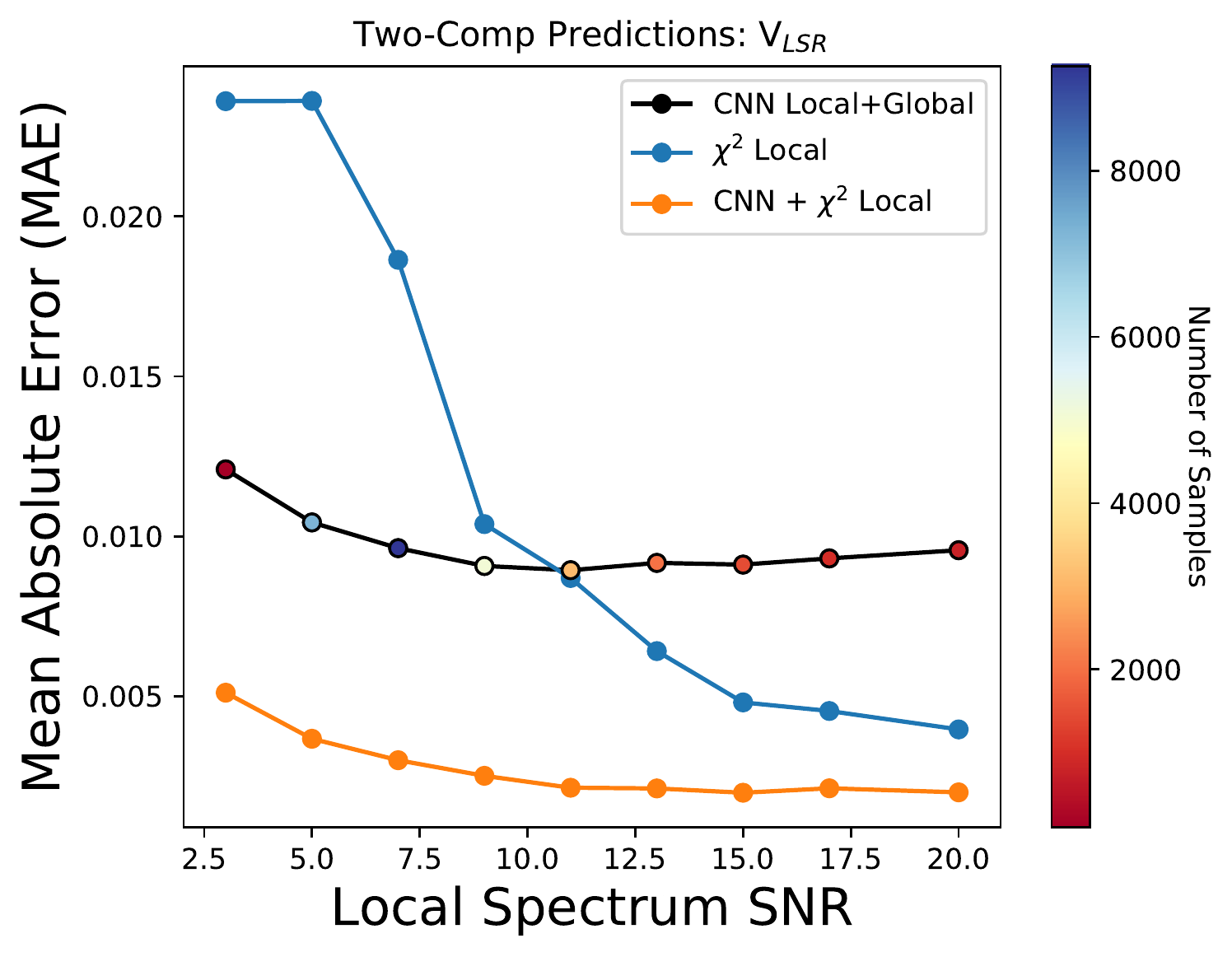}{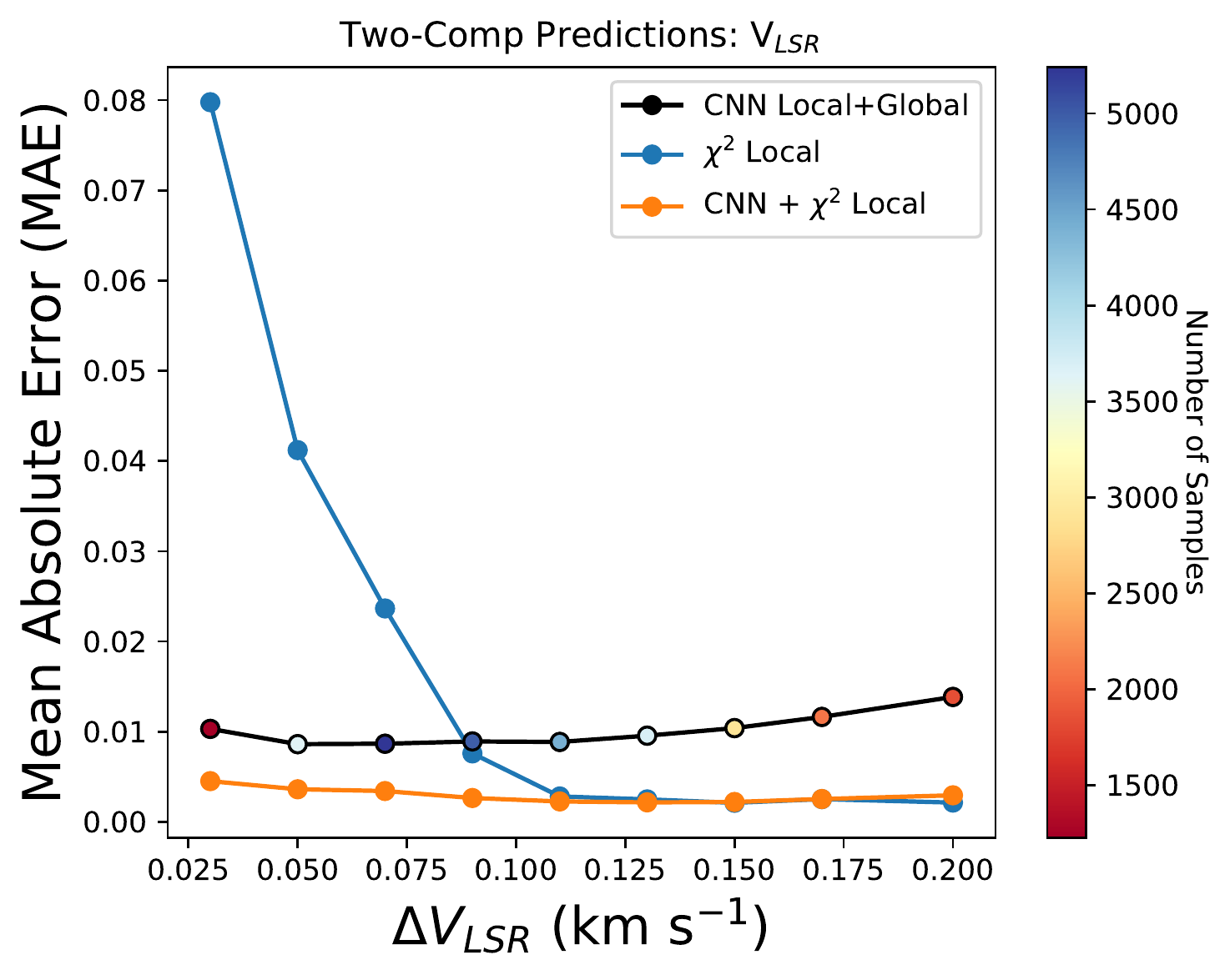}
\plottwo{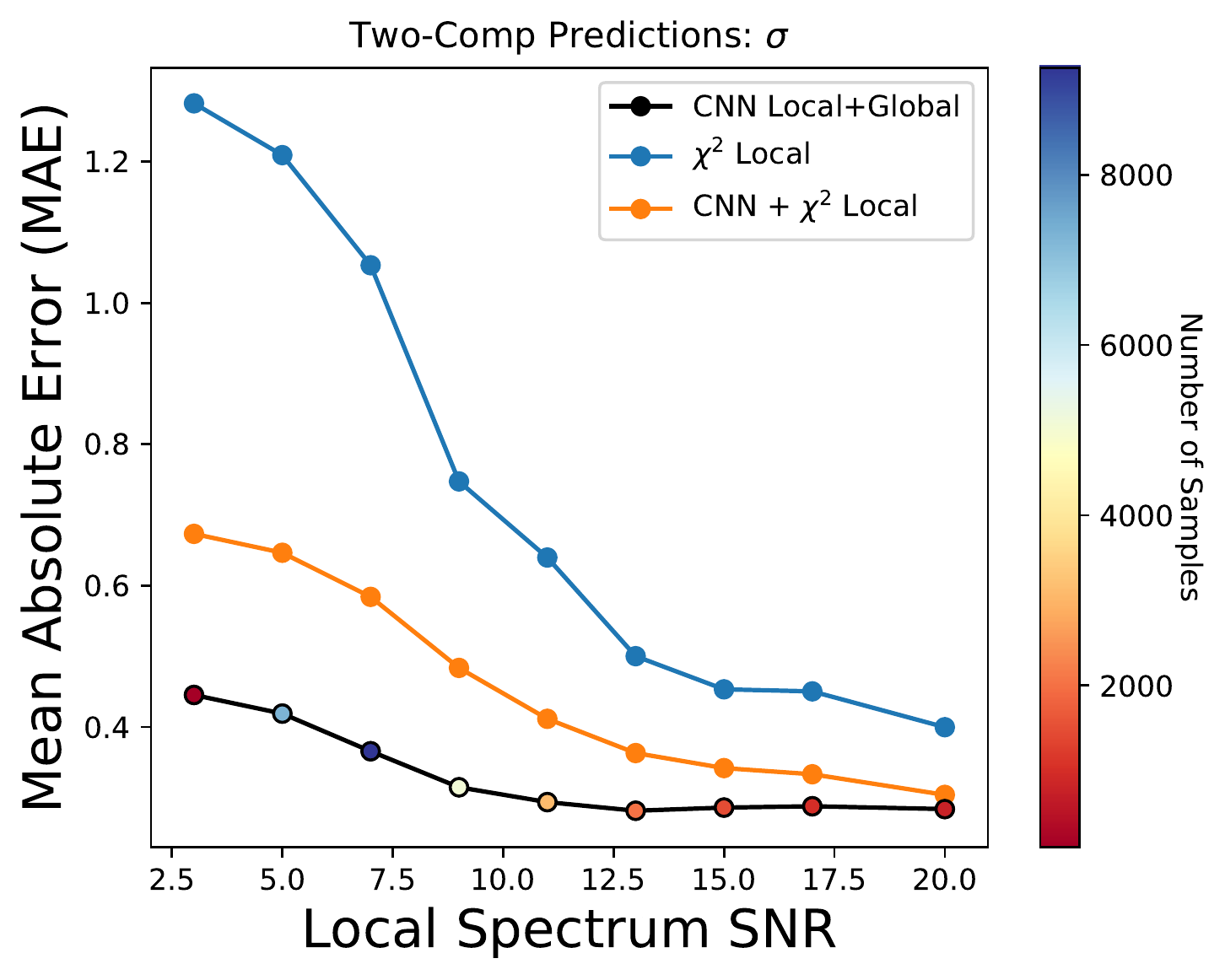}{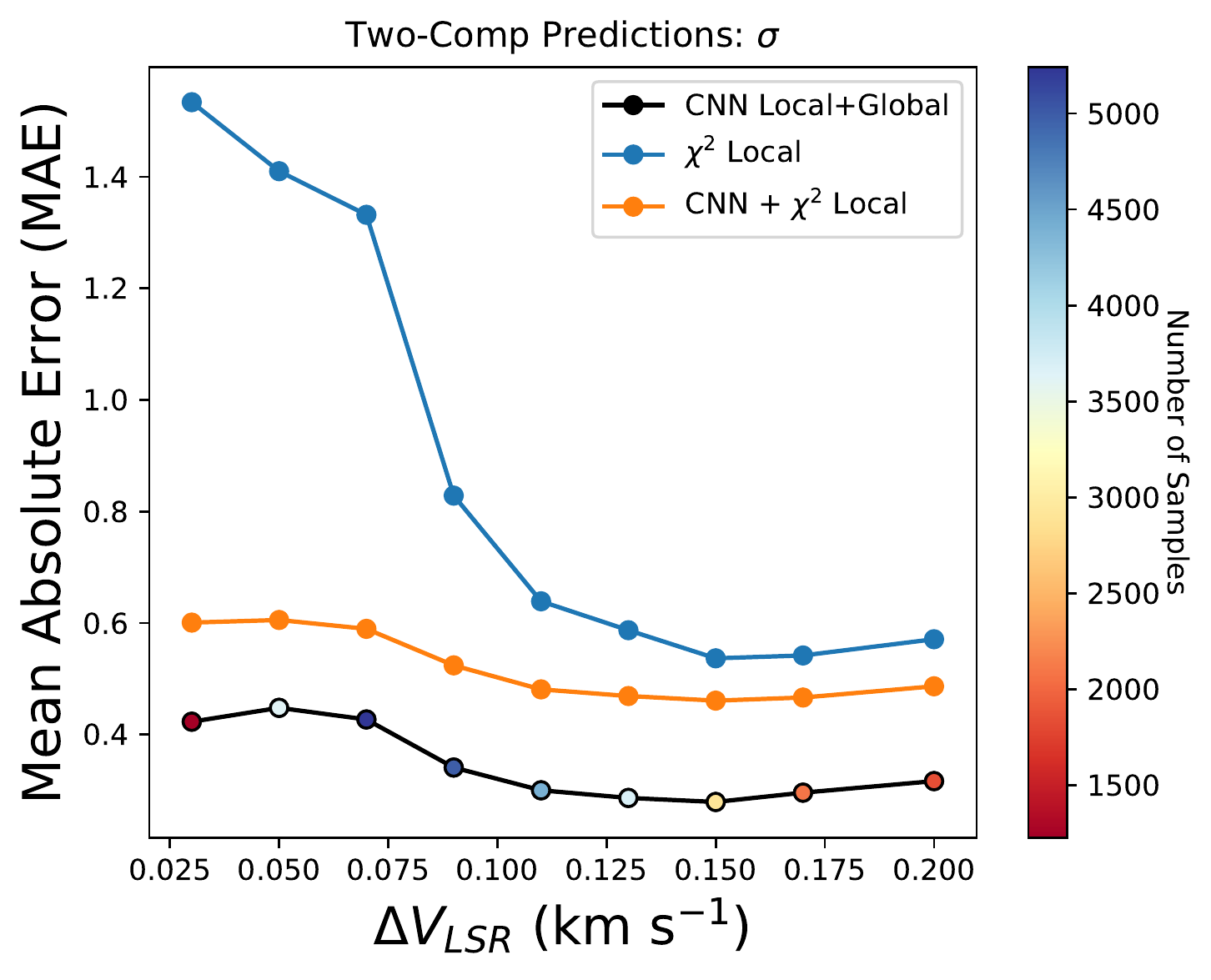}
\centering
\plottwo{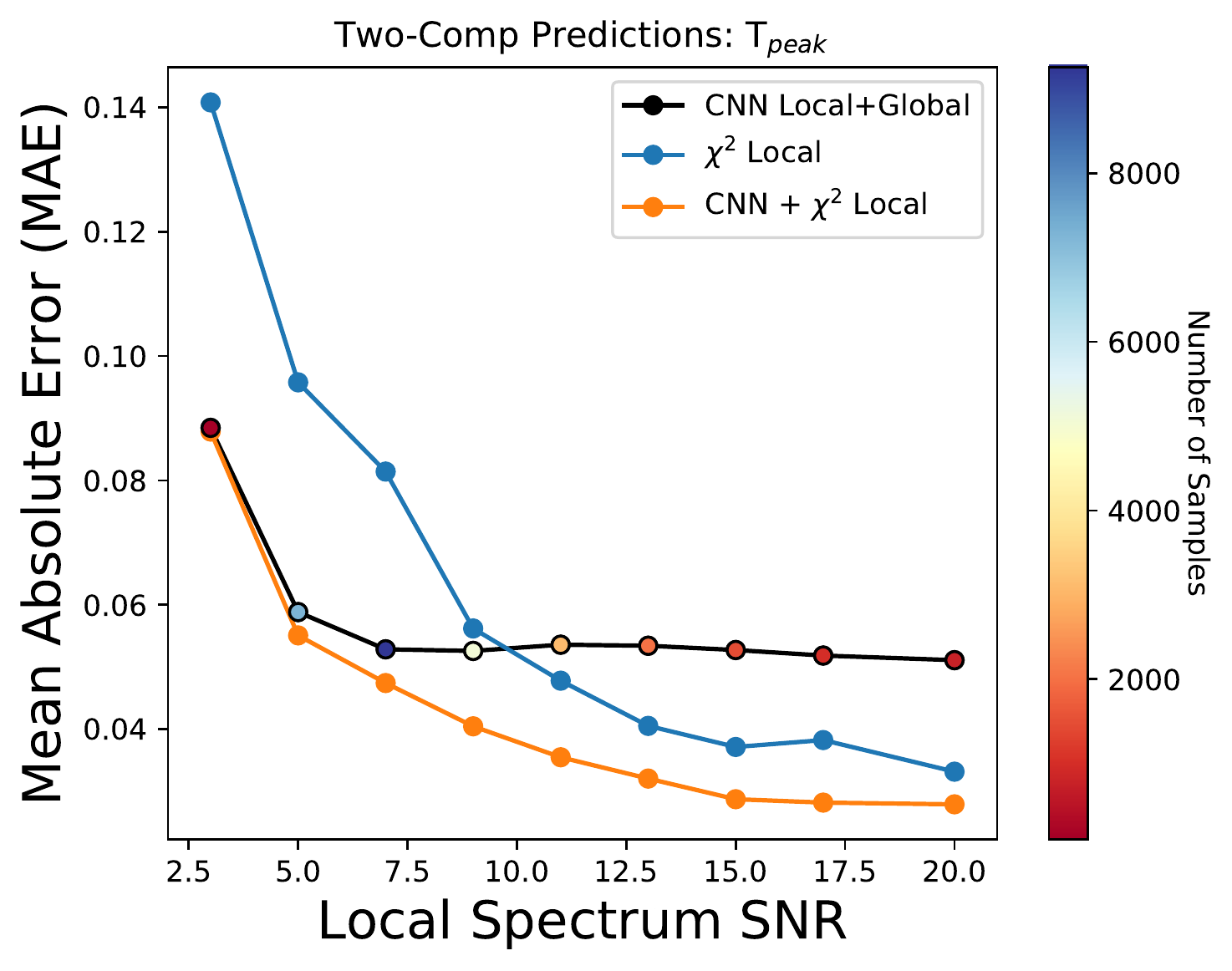}{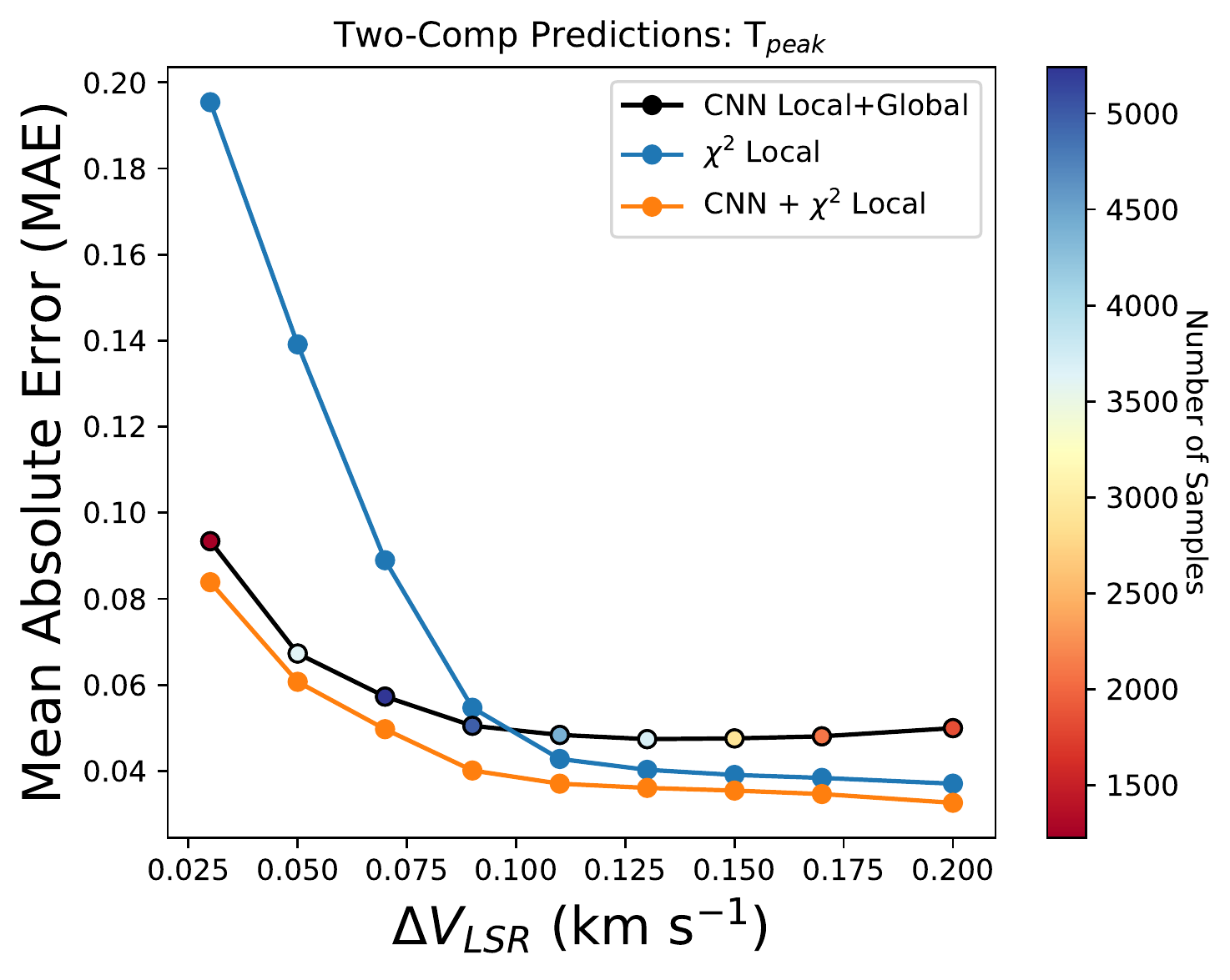}
\caption{Mean absolute error (MAE) versus SNR (left column) and centroid velocity separation (right column) for centroid velocity (top row), velocity dispersion (middle row), and peak intensity (bottom row) predictions.  Each data point represents the MAE for test set samples within a bin centered on the data point's x-axis position, averaged over both velocity components for each parameter (e.g., the average MAE for both V1 and V2 from Figure \ref{multi_comps_regression}).  The results for the traditional $\chi^2$-minimization method are shown for both the grid-search initial guess technique (blue, see Section 4.2) and when using CLOVER's regression CNN predictions to set initial guesses (orange, see Section 5). The color of the regression CNN Local+Global data points (outlined in black) show the amount of test set samples within each bin. The centroid velocity separation calculation uses the spectral axis after normalization between $-1$ km s$^{-1}$ and 1 km s$^{-1}$.}
\label{mae}
\end{figure}

Figure \ref{mae} shows the mean absolute error for each predicted parameter in bins of SNR and centroid velocity offset for the $\chi^2$-minimization method and CNN predictions.  It is clear from Figure \ref{mae} that the $\chi^2$-minimization method's outliers are caused by samples with low SNR and low centroid velocity offsets, which show higher values of MAE compared to samples with higher SNR and larger centroid velocity offsets.  Although the MAE values for the CNN predictions are more stable than those of the $\chi^2$-minimization method, the CNN still suffers from a moderate increase in MAE for samples with low SNR and low centroid velocity offsets.

\clearpage

Since the $\chi^2$-minimization method is susceptible to falling into local minimum solutions for samples with low SNR and low centroid velocity offsets, its performance can be improved by making better initial guesses and providing constraints on the parameter space explored.  One way to set these initial guesses is by using the CNN predictions, which are more resilient to low SNR and low centroid velocity offsets.  To demonstrate this use-case, we perform a second round of fitting on the test set using the $\chi^2$-minimization method.  Instead of using the initial guess grid-search method described in Section 4.2, the CNN predictions are used as the initial parameter guesses.  We also constrain the parameter space explored by the $\chi^2$-minimization method to be within the scatter of the CNN predictions, which also helps prevent fitting noise or falling into local minimum solutions.  In the bottom panel of Figure \ref{multi_comps_regression}, we show the results using this combined CNN and $\chi^2$-minimization method.  Using the CNN parameter constraints, the $\chi^2$-minimization method no longer falls into local minimum solutions.  The mean absolute errors improve to $\sim 0.003$ for centroid velocity, $\sim 0.5$ for velocity dispersion, and $\sim 0.04$ for peak intensity.  Moreover, Figure \ref{mae} also shows that using the CNN, rather than the grid-search technique, to set initial guesses for the $\chi^2$-minimization method reduces the MAE for samples with low SNR and low centroid velocity offsets.  As such, CLOVER provides a convenient way to improve existing line fitting pipelines that require initial guesses for centroid velocity, dispersion, and peak intensity.

In addition, Figures \ref{multi_comps_regression} and \ref{mae} show that the CNN+$\chi^2$-minimization approach has better performance than the CNN alone for both centroid velocity and peak intensity predictions.  Conversely, the accuracies of the velocity dispersion predictions for both $\chi^2$ methods are lower than those of the CNN.  The middle left panel of Figure \ref{mae}, however, shows that the velocity dispersion prediction accuracy of all three methods converges at high SNR.  The poorer performance of the $\chi^2$ methods are limited to the low SNR spectra, which make up the majority of the test set samples.  The better performance of the CNN is likely a result of its usage of the higher SNR ``global'' spectrum, which can provide better constraints on a given sample's velocity dispersions than the ``local'' spectrum alone. 


\subsection{Testing CLOVER's regression CNN on real data}
\begin{figure}[htb]
\epsscale{1.1}
\plottwo{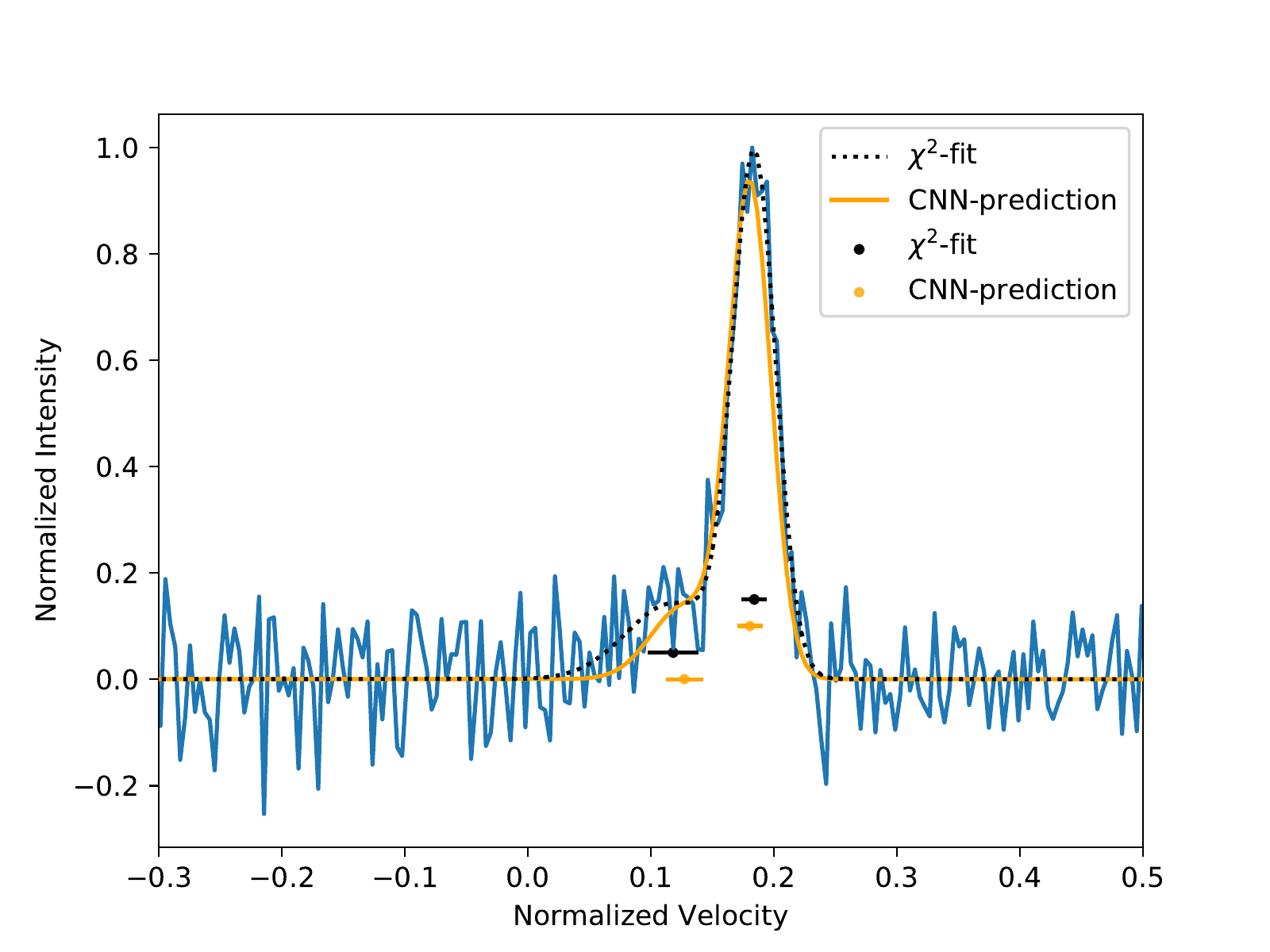}{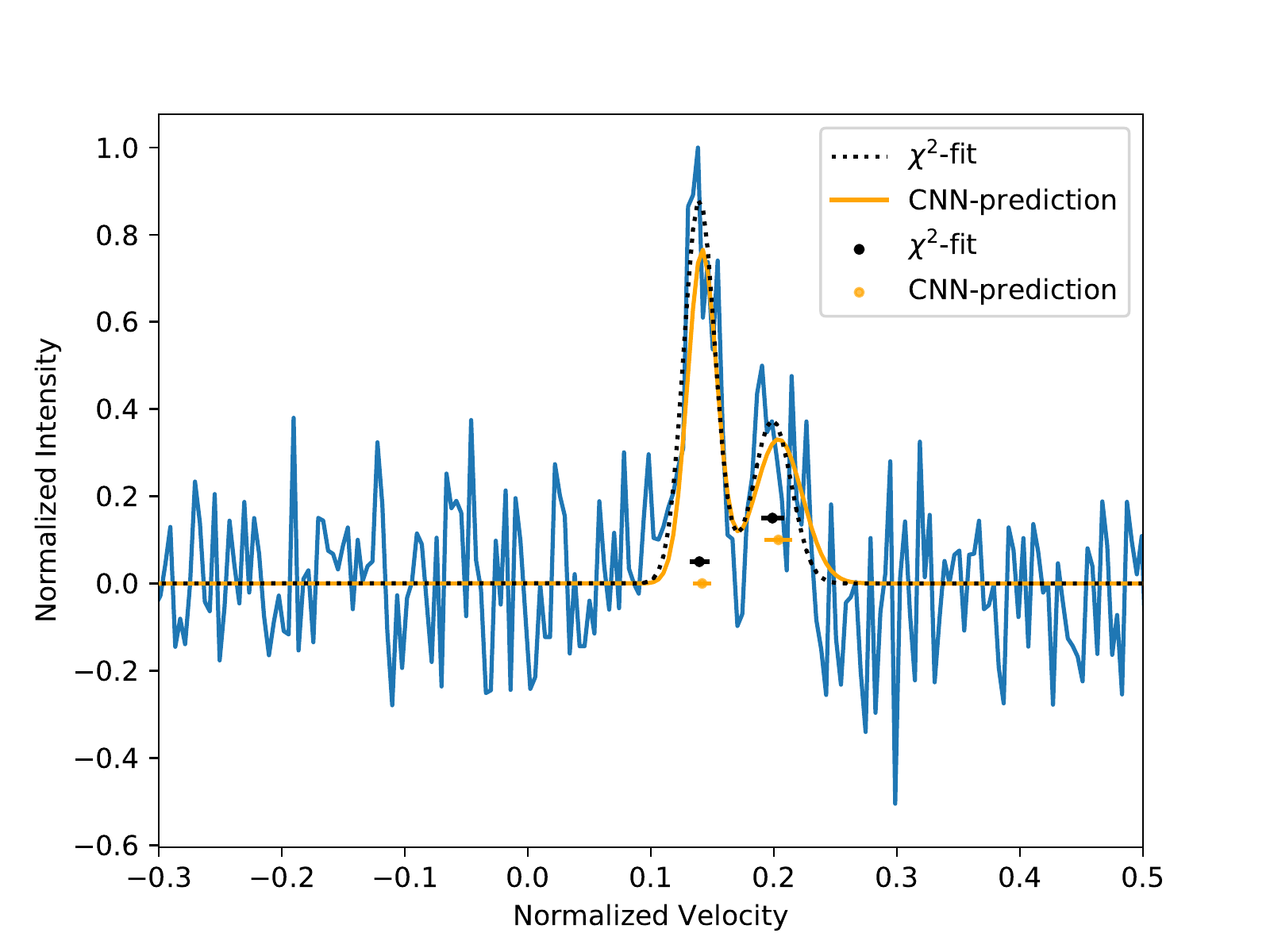}
\plottwo{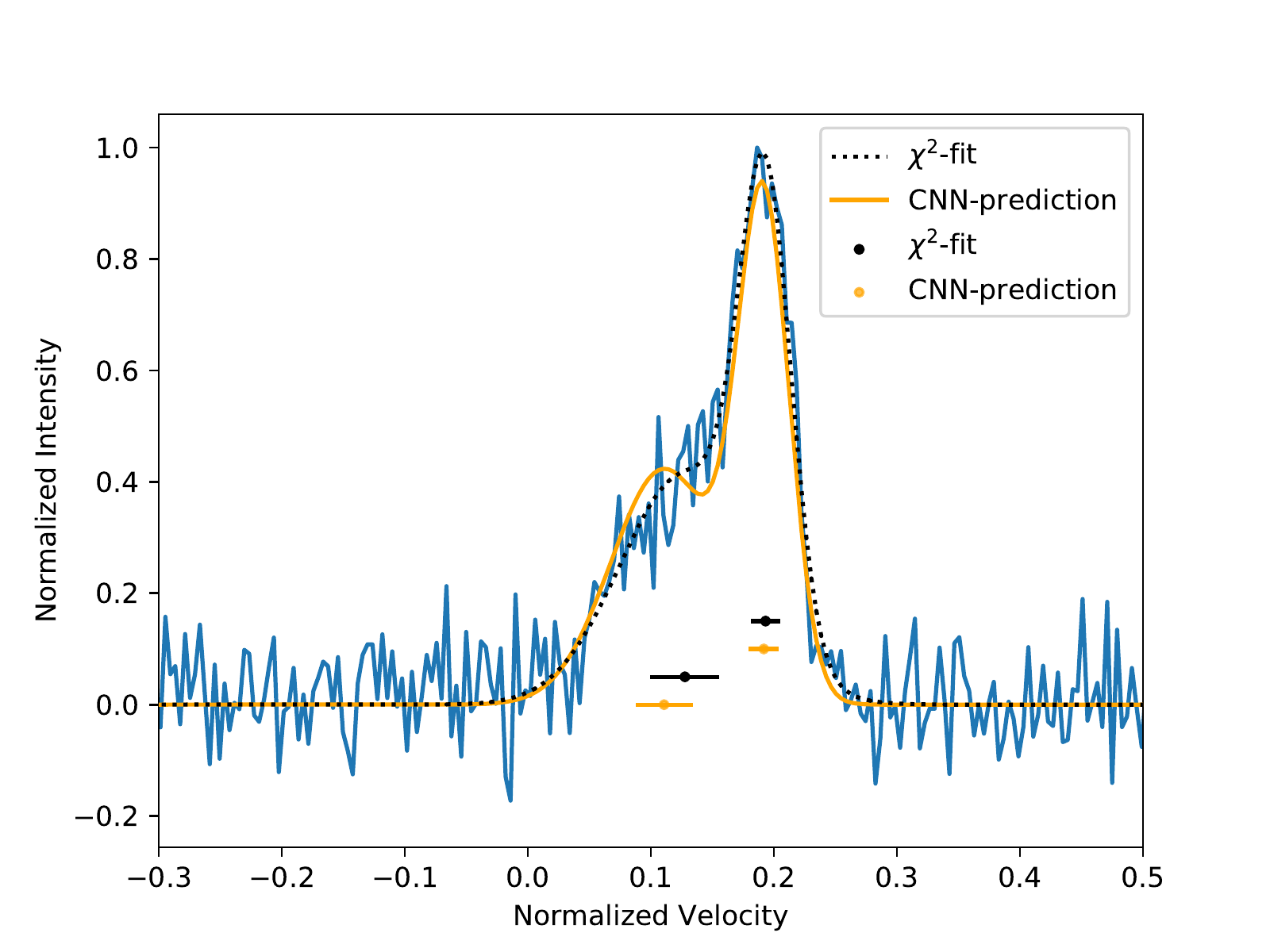}{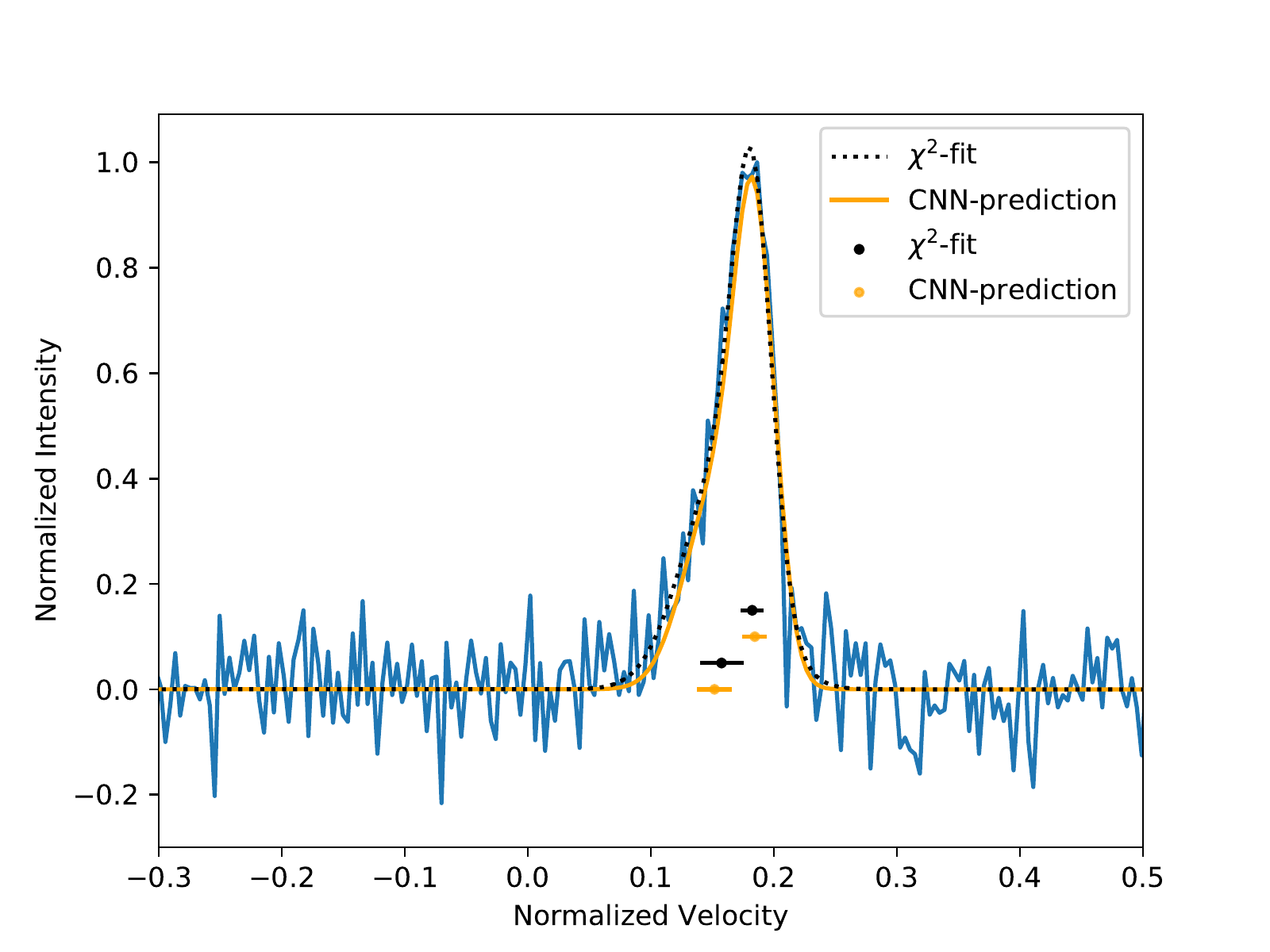}
\centering
\plottwo{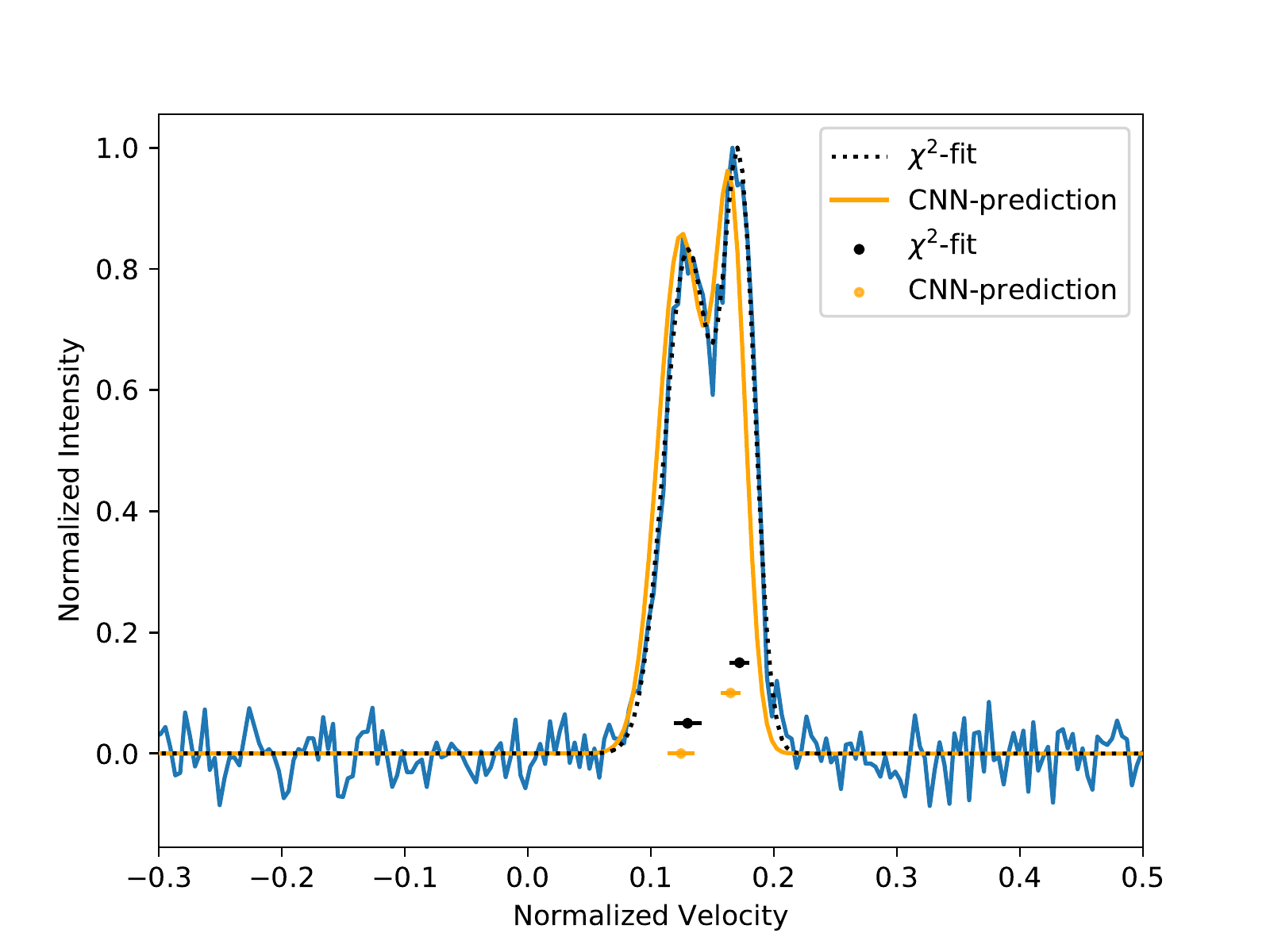}{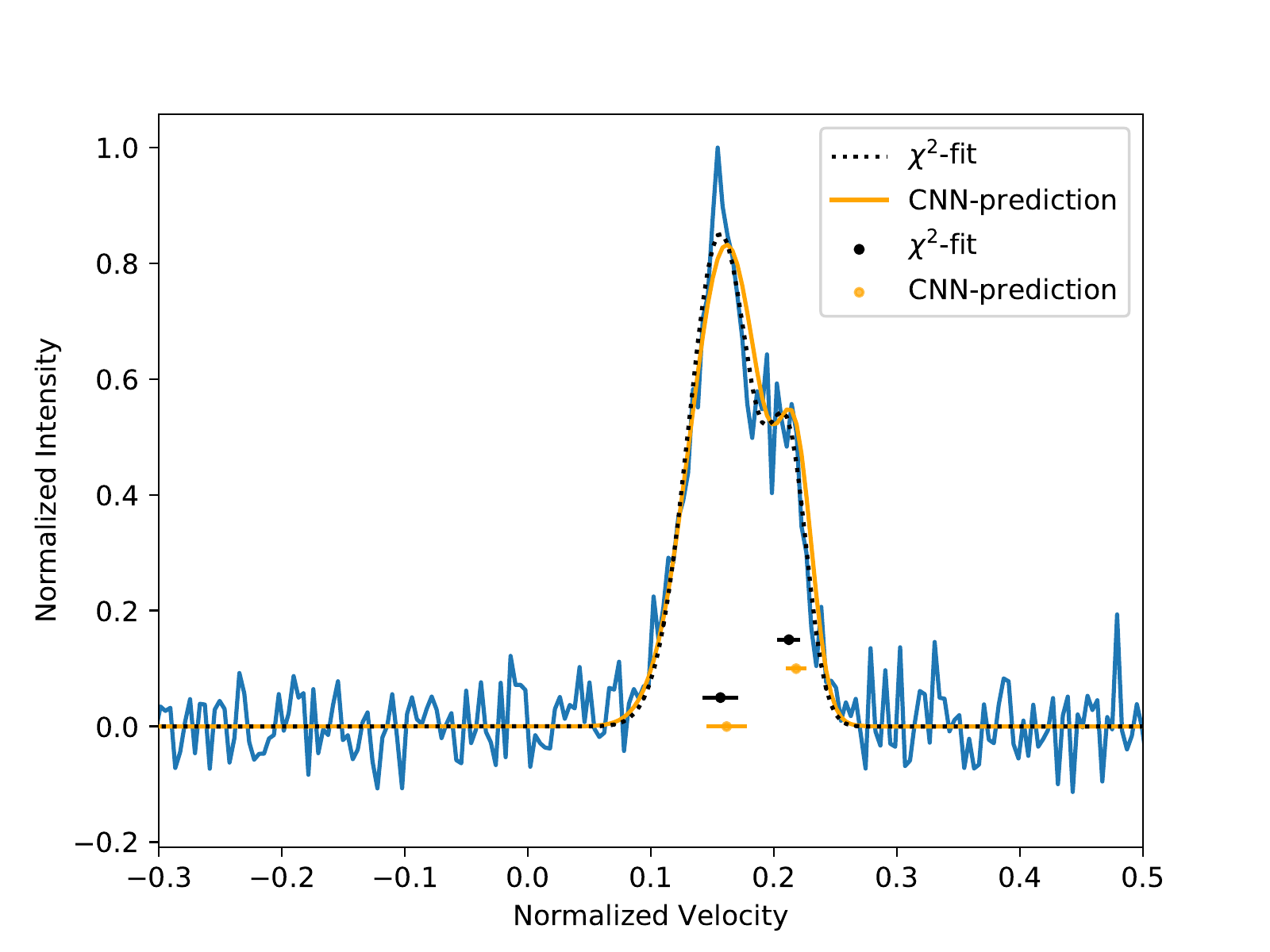}
\caption{Velocity centroid (orange dots) and dispersion (orange bars) predictions by CLOVER versus the centroids and dispersions obtained from a two-component Gaussian fit using $\chi^{2}$-minimization (black dots and bars) for six spectra observed in L1689. In all panels, the ``local'' view spectrum is shown in blue, while the best-fit two-component model from the $\chi^{2}$-minimization is displayed as a dotted black line.  The orange solid line shows the corresponding two-component model generated using CLOVER's parameter predictions.}
\label{multi_comps_regression2}
\end{figure}

To test the performance of CLOVER's regression CNN on real observations, we use the L1689 $^{13}$CO ($3-2$) spectra that were predicted to belong to the two-component class by CLOVER's classification CNN presented in Section 4.  Figure \ref{multi_comps_regression2} shows example predictions by the regression CNN on six of the L1689 spectra.  CLOVER is able to predict accurately the kinematics of blended two-component spectra with both broad and narrow components.  Figure \ref{multi_comps_regression2} also shows the kinematics and best-fit model determined by fitting a two-component Gaussian model to the data using the traditional $\chi^{2}$-minimization approach described in Section 4.2.  Overall, the predictions of the two methods are in good agreement across all of these test examples.  In most cases, however, the $\chi^{2}$-minimization approach provides a better fit visually to the data.  These better fits are likely related to the fact that the $\chi^{2}$-minimization approach uses only the local spectrum for fitting while CLOVER uses both the local and global spectra.  Since CLOVER also uses the global spectrum for its predictions, there is some bias in its predictions when viewed on the local spectrum.  Nevertheless, using the global spectrum allows CLOVER's predictions to have lower variance and suffer less from falling into the local minima solutions that plague the $\chi^{2}$-minimization approach (see, e.g., the discussion in Section 5).

Moreover, the design of the $\chi^{2}$-minimization approach provides  an advantage over the CNN for producing visually optimal fits.  The $\chi^{2}$-minimization approach is based on minimizing the residual between the spectrum and model.  When provided adequate initial model parameter guesses, the $\chi^{2}$-minimization approach will oftentimes produce a visually optimal fit.  Conversely, the CNN is attempting to ``guess'' the Gaussian model parameters based on similar examples it has seen during training.  The CNN knows nothing about the residual of the model it generates for the six parameters it predicts.  Rather, the CNN knows only that the new spectrum it receives has activated similar artificial neurons as examples it saw during training.  This approach is very different to residual minimization and does not always lead to the best possible visual fit.  The CNN approach can, however, get very close to the best possible visual fit (as shown in Figure \ref{multi_comps_regression2}).  To obtain the most robust parameter predictions possible, we therefore recommend users use CLOVER's regression predictions as initial guesses for a $\chi^{2}$-minimization approach.

\begin{figure}[htb]
\plottwo{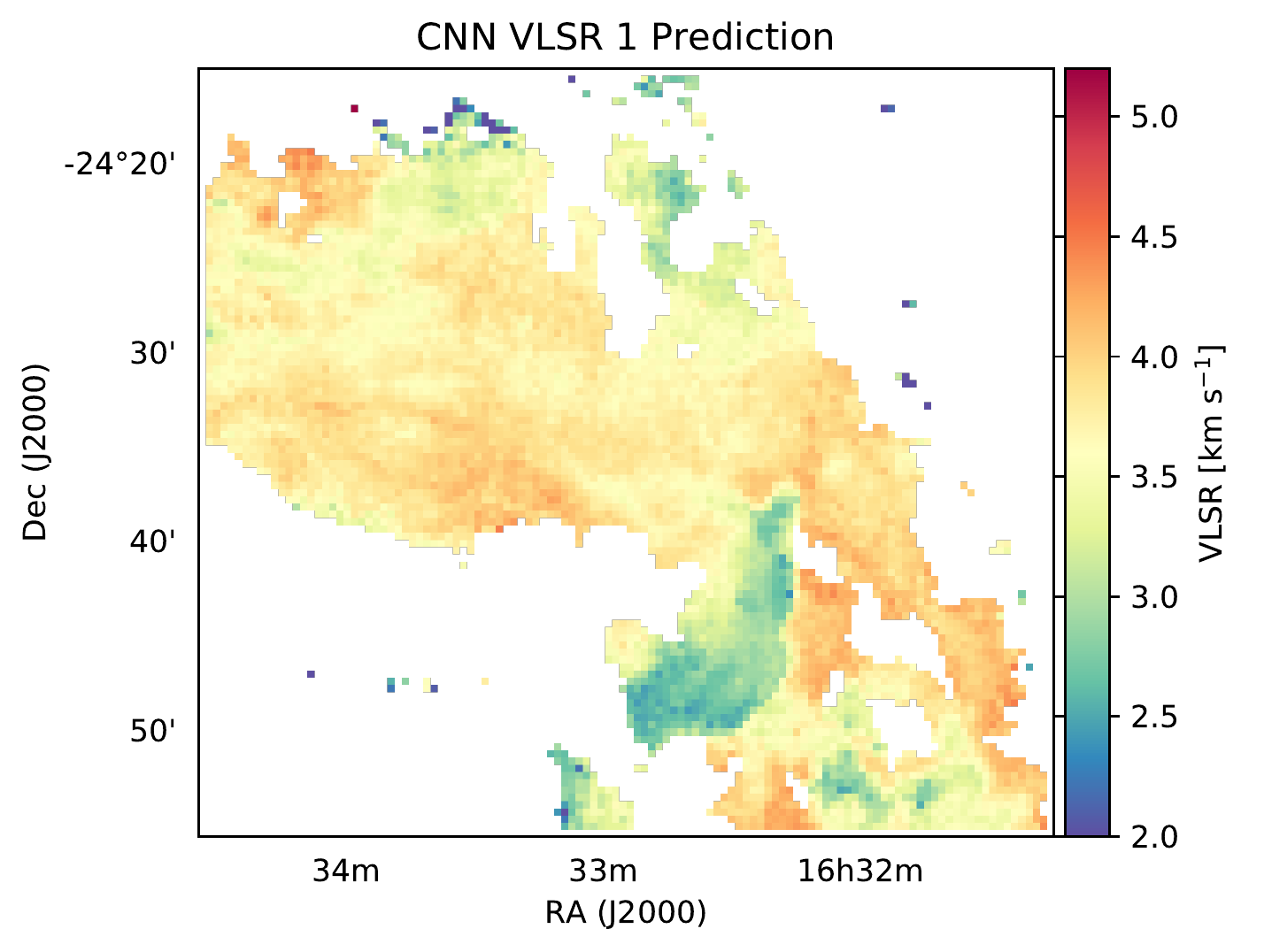}{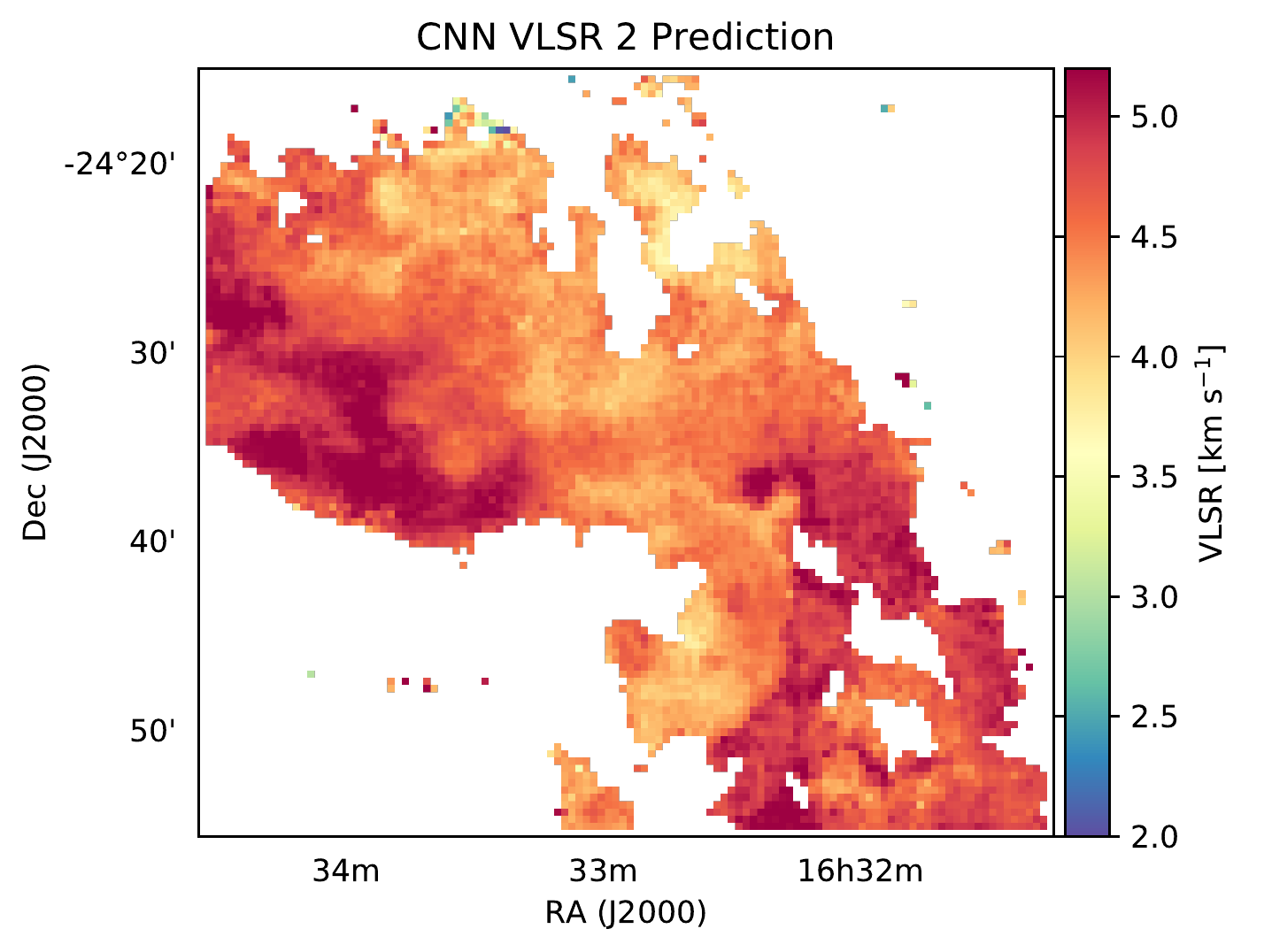}
\plottwo{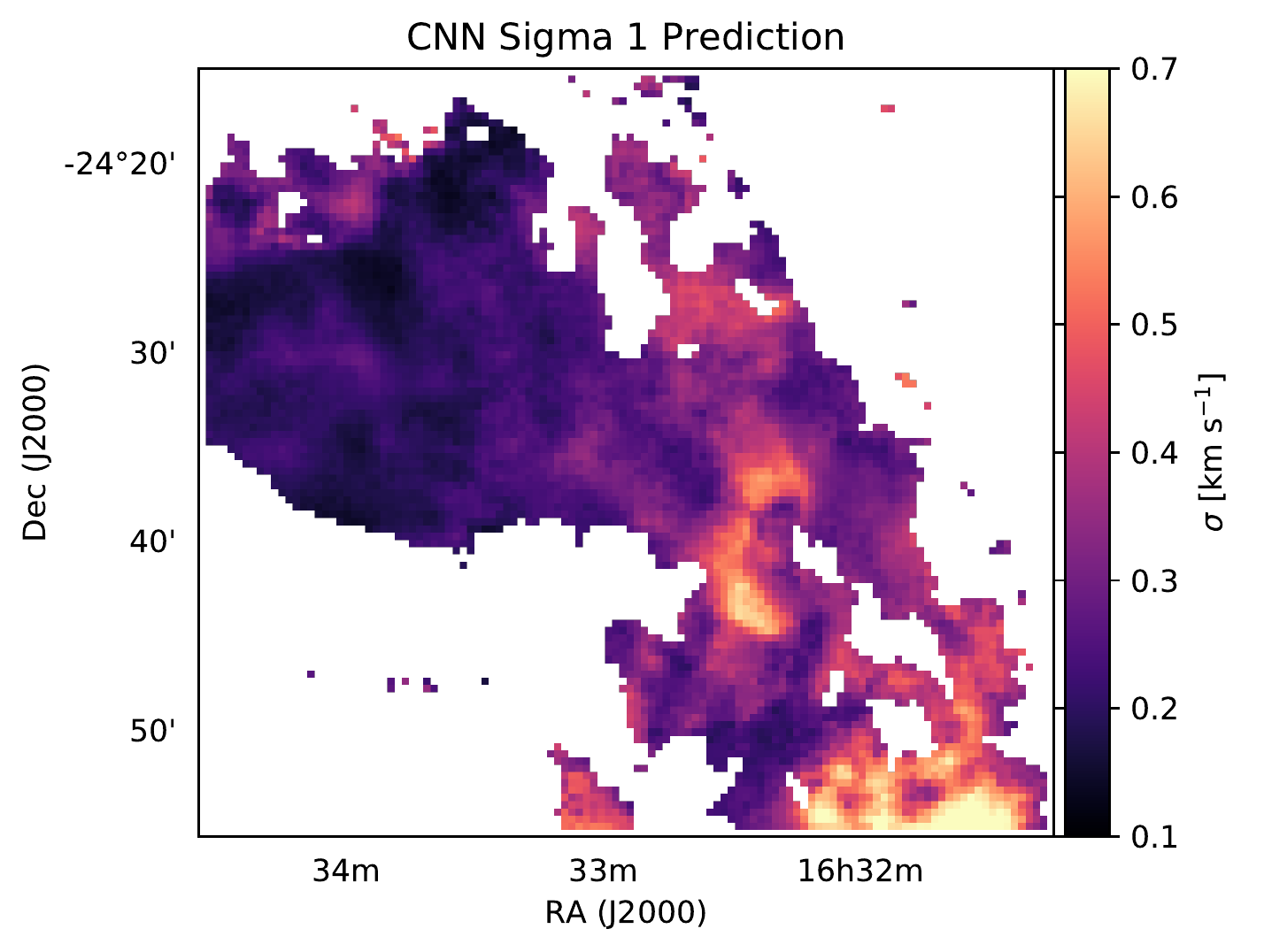}{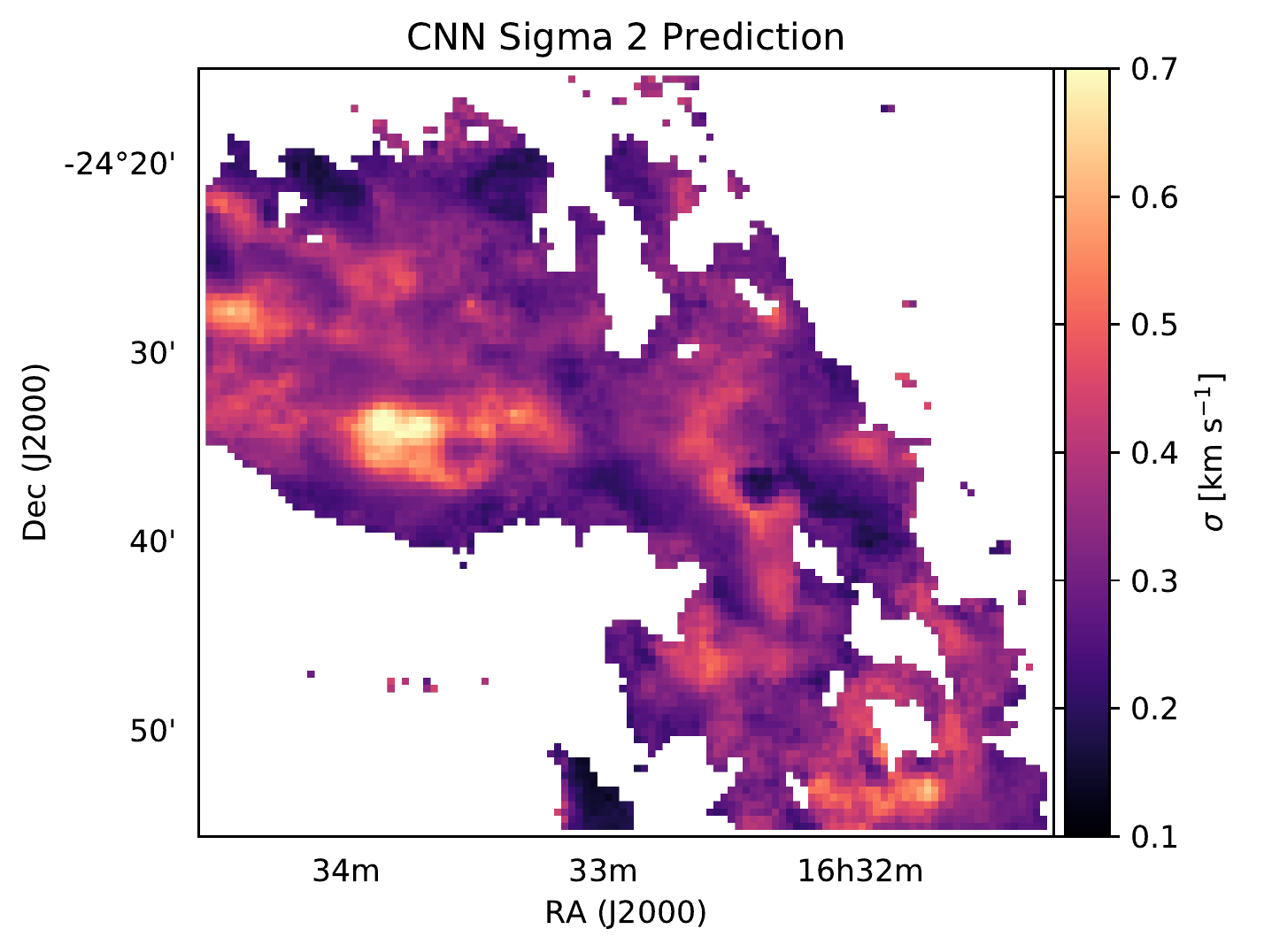}
\centering
\plottwo{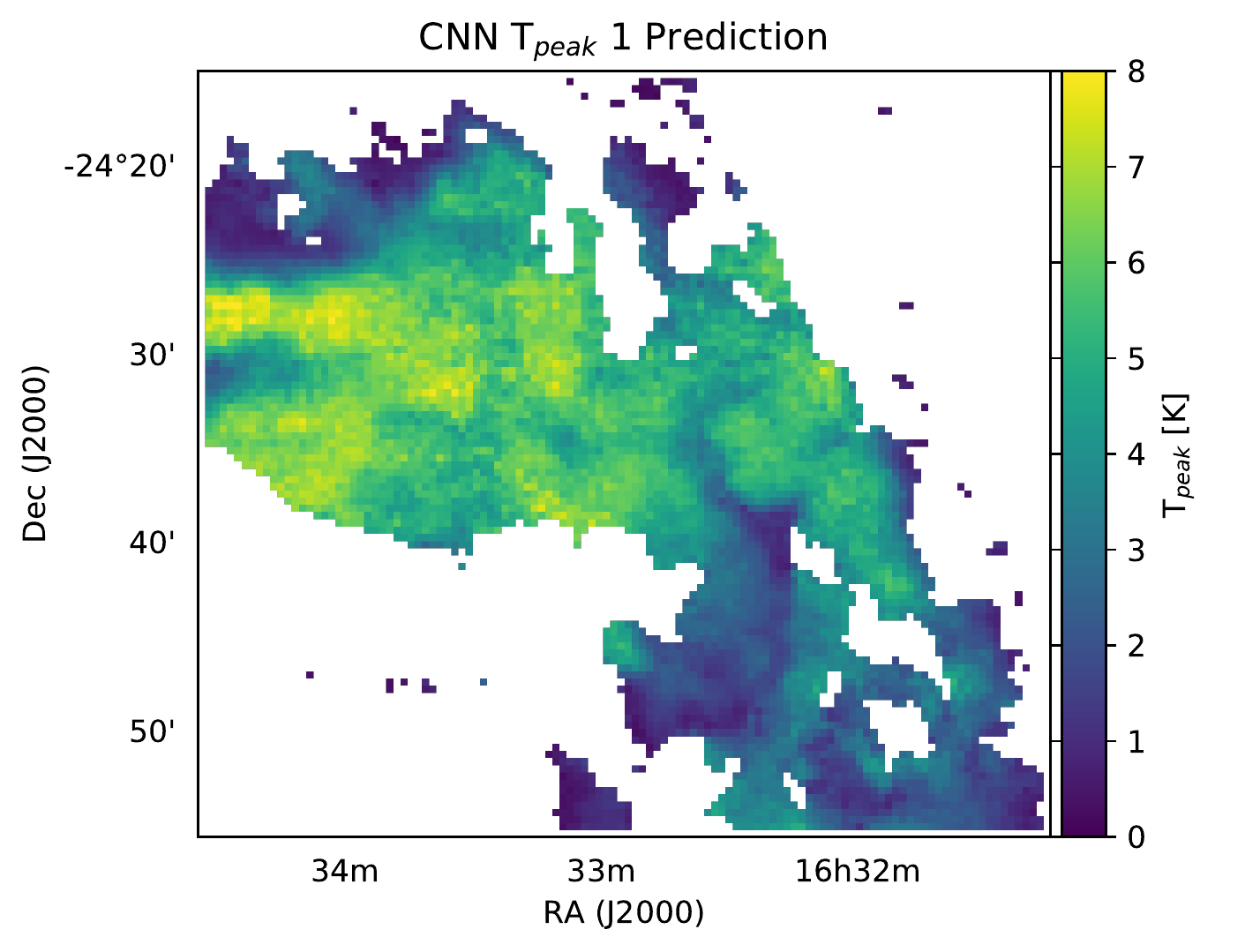}{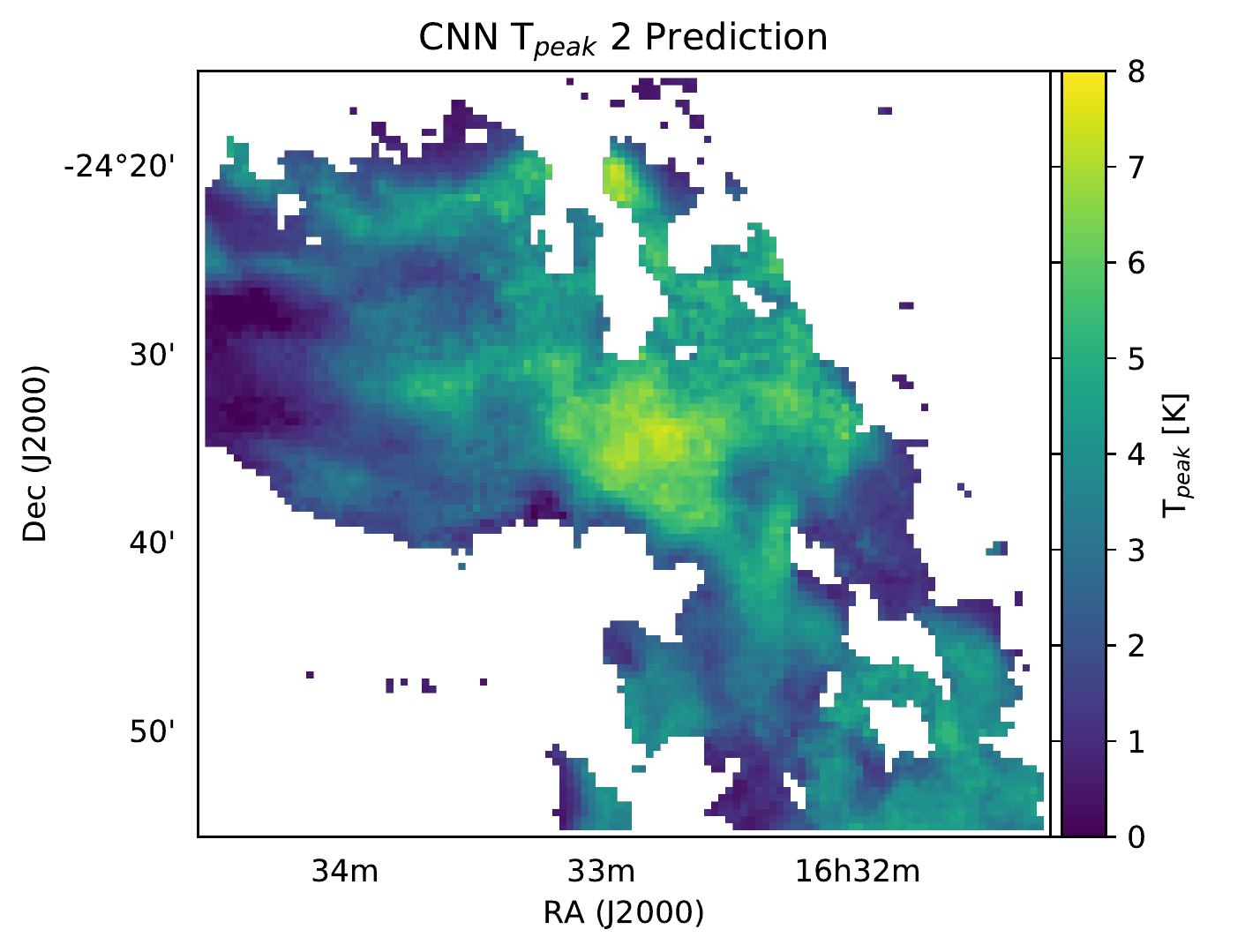}
\caption{Velocity centroid (top row), velocity dispersion (middle row), and peak brightness temperature (bottom row) predicted by CLOVER's regression CNN for all pixels predicted to be ``two-component'' class members by CLOVER's classification CNN for L1689.}
\label{multi_comps_sig}
\end{figure}

Figure \ref{multi_comps_sig} shows the result of running both CLOVER's classification and regression CNNs (i.e., the complete CLOVER method) on the full L1689 data cube.  In each panel, colored pixels represent those that were designated ``two-component'' class members by CLOVER's classification CNN.  The top row displays the predicted centroids for each pixel, the middle row shows the predicted dispersion, and the bottom row shows the predicted peak intensity.  The maps suggest a stronger gradient in centroid velocity for the lower-velocity component than the higher-velocity component, which is typically at $V_{LSR} > 4.0 $ km s$^{-1}$.  The lower-velocity component also tends to have smaller velocity dispersion, especially on the eastern side of the cloud. 

In addition, the full CLOVER classifications and parameter predictions take only $\sim154$ seconds on a single core of a 2.8 GHz Intel Core i7 CPU.  This is over an order of magnitude faster than the 3,209 seconds required for the $\chi^{2}$-minimization approach, which simultaneously provides classifications and parameter predictions, when run on the same CPU.

\clearpage

\clearpage 


\section{Classifying Spectra with Hyperfine Structure}
Hyperfine splitting is an additional mechanism that can cause an emission line to appear non-Gaussian.  The emission from the NH$_3$ (1,1) transition, for instance, is split across 18 different velocity components \cite[see, e.g.,][]{Ho_1983}.  For NH$_3$ (1,1), this splitting results in a central group of blended Gaussians with four satellite groups of blended Gaussians (two on each side). Such spectra would be problematic when making classifications with CLOVER as described in Section 4, since it would undoubtedly select the two-component class for every spectrum due to the multiple hyperfine groups.  

With ammonia being a popular tracer of modern large-scale molecular cloud mapping surveys \citep[e.g.,][]{Friesen_2017, Hogge_2018}, CLOVER would be much more useful to the star formation community if it were adaptable to transitions with hyperfine splitting.  Since the relative frequency separations of the NH$_3$ (1,1) hyperfine lines are well-known, however, we can train a new CNN to distinguish between the transition's intrinsic frequency separations and an actual second velocity-component source along the line of sight.  Here, we train such a CNN using synthetic NH$_3$ (1,1) spectra.

\subsection{Generating Synthetic NH$_3$ (1,1) Spectra}
3$\times$3 pixel cubes for 300,000 training samples (100,000 in each of the three training classes) were generated using the \texttt{cold$\_$ammonia} model generator within the \texttt{pyspeckit} Python package \citep{Ginsburg_2011}.  The generator creates NH$_3$ (1,1) emission models using the following input parameters that were randomly selected from the listed distributions:

\begin{itemize}
\item $T_K$ (kinetic gas temperature): uniformly distributed from $8-25$ K
\item $V_{off}$ (centroid velocity offset from spectrum center): $-2.5$ to 2.5 km s$^{-1}$, which is equivalent to channels $465 - 534$ of the 1000 channel spectrum.  Here, the spectral axis has been normalized so that each channel is separated by $\sim 0.07$ km s$^{-1}$.
\item log$N$ (logarithm of the NH$_3$ column density): uniformly distributed in log$_{10}$ space from $13-14.5$ log(cm$^{-2}$)
\item $\sigma_{nt}$ (non-thermal velocity dispersion): log-normally distributed in natural logarithm space with a 1-sigma range of $0.02-0.45$ km s$^{-1}$
\item $\sigma_{tot}$ (total velocity dispersion) = $\sqrt{\sigma_{nt}+0.08^2}$ km s$^{-1}$
\end{itemize} 

These distributions aim to mimic those seen in real NH$_3$ (1,1) observations by the Green Bank Ammonia Survey \citep{Friesen_2017} and KEYSTONE (Keown et al. 2019, submitted).  For the two-component class, two randomly chosen models were added together.  The velocity centroid of the second velocity component was drawn with respect to the first component (i.e., the range of possible centroids for individual components in the two-component samples is $-5$ km s$^{-1}$ to $5$ km s$^{-1}$). We also ensure that the centroids of the two components are separated by at least $1.0 \times \sigma_{max}$.

As described in Section 2.1, each model generated represented the central pixel in the 3$\times$3 pixel grid.  The outer pixels were filled by adding a perturbation to the central pixel model by drawing values from four normal distributions (one for each parameter) with mean of zero and variance of 0.2 K, 0.1 km s$^{-1}$, 0.1 km s$^{-1}$, and 0.01 cm$^{-2}$ for $T_K$, $V_{off}$, $\sigma_{nt}$, and log$N$, respectively. Random noise with an RMS of 0.1 K was also added to the cubes, a noise level that is typical for recent ammonia mapping surveys of nearby star-forming regions \citep{Friesen_2017, Keown_2017}.  For the noise class, only noise was added to an otherwise emission-free cube.

The final features for each sample are again the local (central pixel's spectrum) and global (averaged spectrum of all nine pixels) spectra, but in this case these have 1000 channels each to account for the hyperfine structure of the NH$_3$ (1,1) line that typically spreads over 500 channels.  We note also that the narrow range of possible centroid velocities for the NH$_3$ (1,1) synthetic spectra create a challenging training set since the majority of the samples have blended velocity components.  This choice was observationally motivated, however, since typical ammonia observations show few spectra with large centroid separations between each velocity component along the line of sight.

\subsection{Testing on Synthetic NH$_3$ (1,1) Spectra}
\begin{figure}[htb]
\plottwo{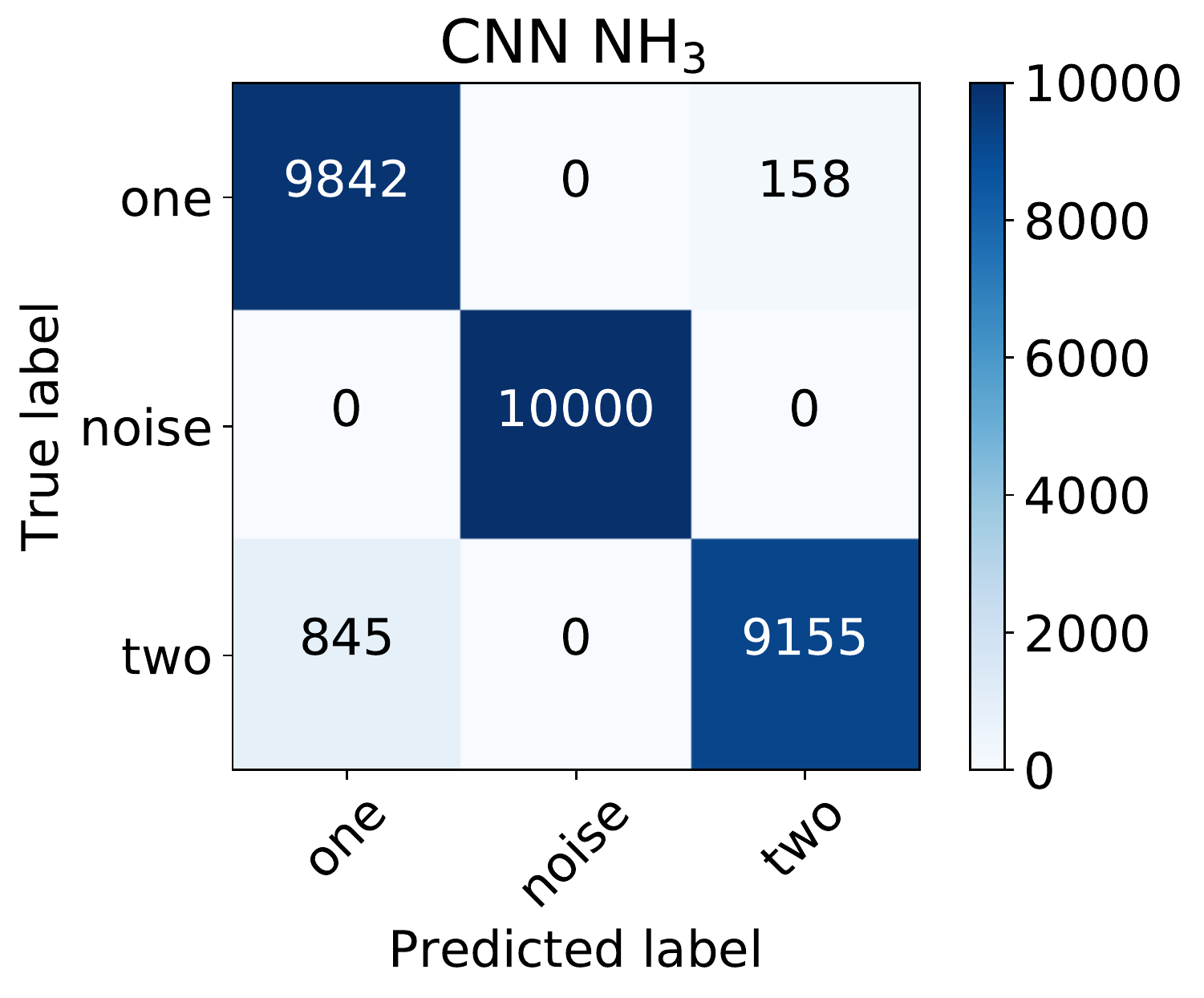}{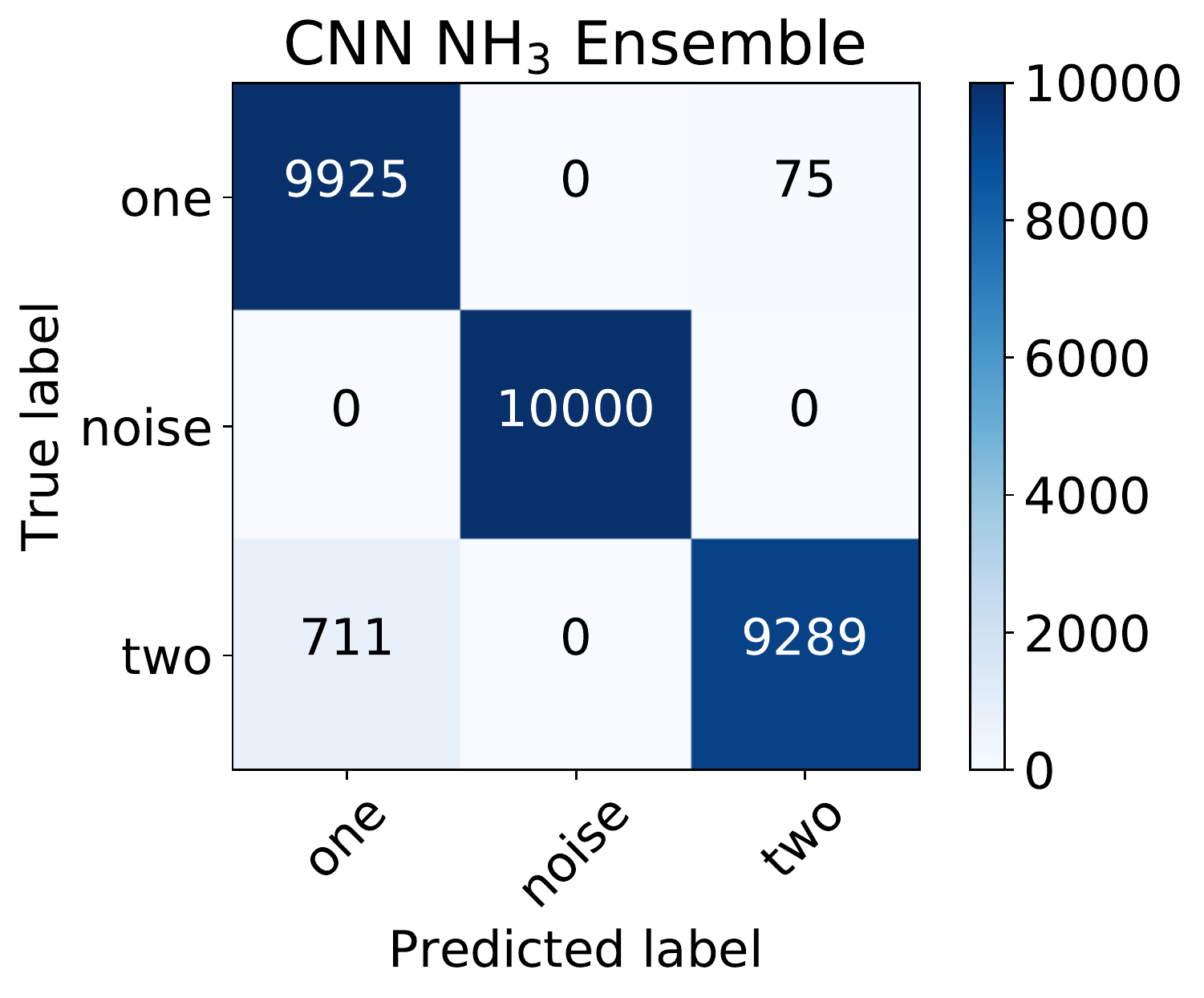}
\plottwo{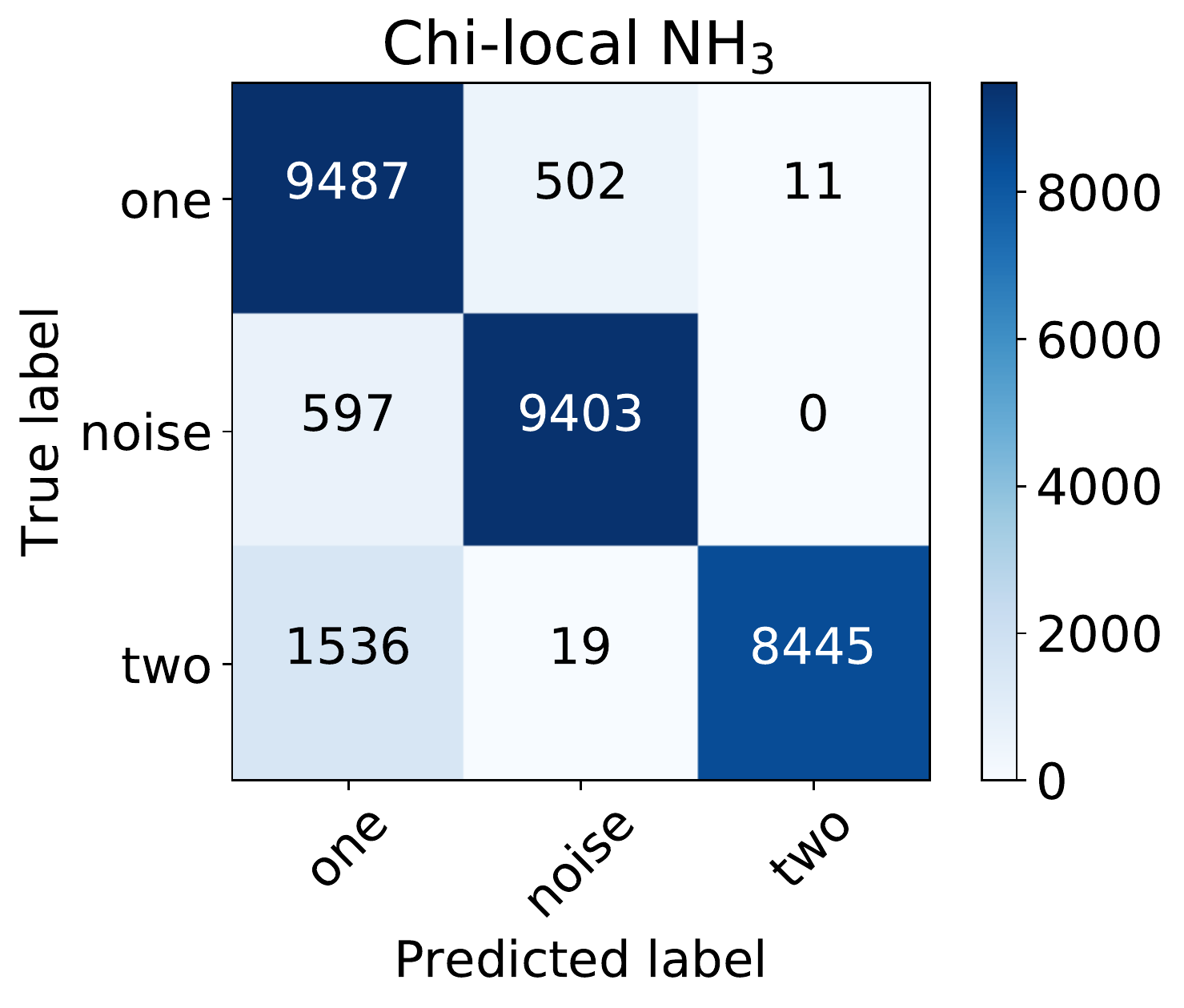}{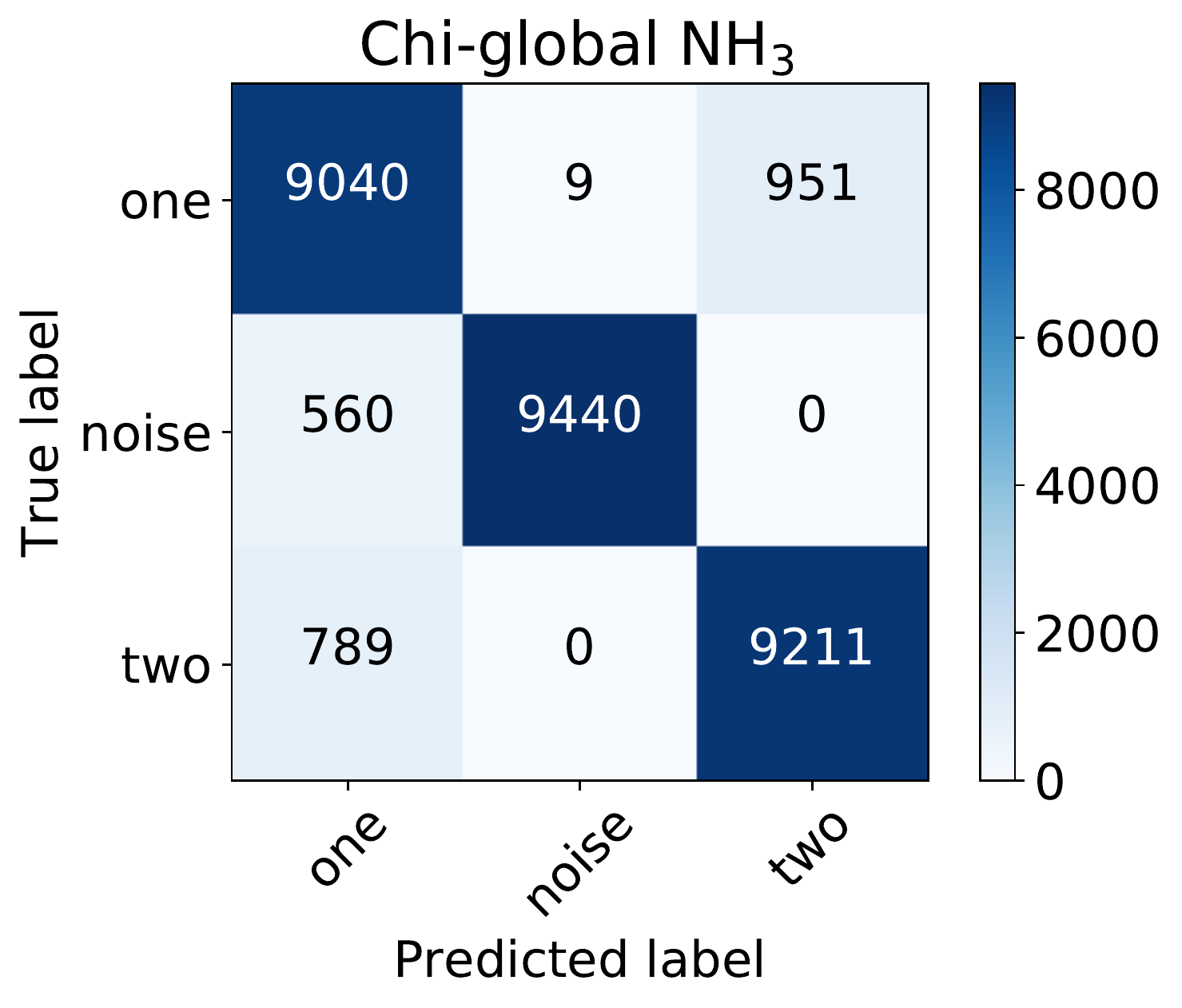}
\caption{Confusion matrices for a validation set of 30,000 synthetic NH$_3$ (1,1) spectra (10,000 in each class) classified by a single CNN (top left), an averaged ensemble of six CNNs (top right), traditional $\chi^2$-minimization on the ``local'' view spectrum (bottom left), and traditional $\chi^2$-minimization on the ``global'' view spectrum (bottom right).  The ``noise'' class for the $\chi^2$-minimization panels was selected based on a SNR threshold of 4.}
\label{NH3_cm}
\end{figure}

We adopt the same neural network architecture described in Section 3 to train the network.  An additional synthetic validation set of 90,000 samples (30,000 in each class) was also used to monitor model performance during training to implement early-stopping.  The trained CNN's performance on a separate synthetic test set of 30,000 additional samples (10,000 in each class) is shown in the top left panel of Figure \ref{NH3_cm}.  The CNN prediction accuracy is $\sim 98\%$, $100\%$, and $\sim 92\%$, for the one-component, noise-only, and two-component classes, respectively.  As shown in the right panel of Figure \ref{NH3_cm}, the prediction accuracy of ensemble averaging six independently trained CNNs improves to $\sim 99\%$, $100\%$, and $\sim 93\%$, for the one-component, noise-only, and two-component classes, respectively.  For this reason, all further comparisons and analysis of the NH$_3$ data will use the ensemble CNN.


To compare the ensemble CNN performance to traditional line fitting methods, we use the $\chi^2$-minimization model selection approach to classify each spectrum in the training set using the same technique described in Section 4.2.  The \texttt{cold$\_$ammonia} model generator within \texttt{pyspeckit} was used to generate one- a two-component models that were fit to the data using the $\chi^2$-minimization method.  The initial guesses for $T_K$, $\sigma_{nt}$, and log$N$ were set at 14 K, 0.3 km s$^{-1}$, and 13.5 cm$^{-2}$.  A grid of $V_{LSR}$ initial guesses were used, which included one guess centered on the peak intensity channel and increments of $\pm 0.4$, $\pm 1.3$, $\pm 2.2$, $\pm 3.1$, $\pm 4.0$, and $\pm 4.9$ km s$^{-1}$ offset from the peak intensity channel.   

\clearpage

The bottom panels in Figure \ref{NH3_cm} show that the CNN performance is better than the $\chi^2$-minimization model selection approach on the local (95$\%$, 94$\%$, and 85$\%$ for the one-component, noise-only, and two-component classes, respectively) and global (90$\%$, 94$\%$, and 92$\%$ for the one-component, noise-only, and two-component classes, respectively) spectra.  These results indicate that the $\chi^2$-global approach is actually susceptible to overfitting the hyperfine spectra, tending to classify incorrectly the one-component samples as two-component at a slightly higher rate than the $\chi^2$-local method.  In contrast, the ensemble CNN is more resilient to this overfitting while still incorporating the global spectrum as input. 

\subsection{Testing on Real NH$_3$ (1,1) Observations}
To demonstrate that CLOVER can accurately predict the class of real ammonia spectra, we utilize two NH$_3$ (1,1) cubes observed by the KFPA Examinations of Young STellar Object Natal Environments (KEYSTONE) survey (PI: James Di Francesco; Keown et al. 2019, submitted).  KEYSTONE used the 100m Green Bank Telescope to map ammonia emission across eleven of the nearest giant molecular clouds (0.9 kpc $< d <$ 3 kpc).  Here, we use the KEYSTONE observations of two clouds: 1) M17, which has a core of emission (M17SW) with obvious multiple velocity components, and 2) MonR2, which is composed mainly of single velocity component spectra.  To match the size of the spectra used to train CLOVER's CNNs, the ammonia cubes are clipped to 1000 channels along the spectral axis.  

\begin{figure}[htb]
\epsscale{0.85}
\plotone{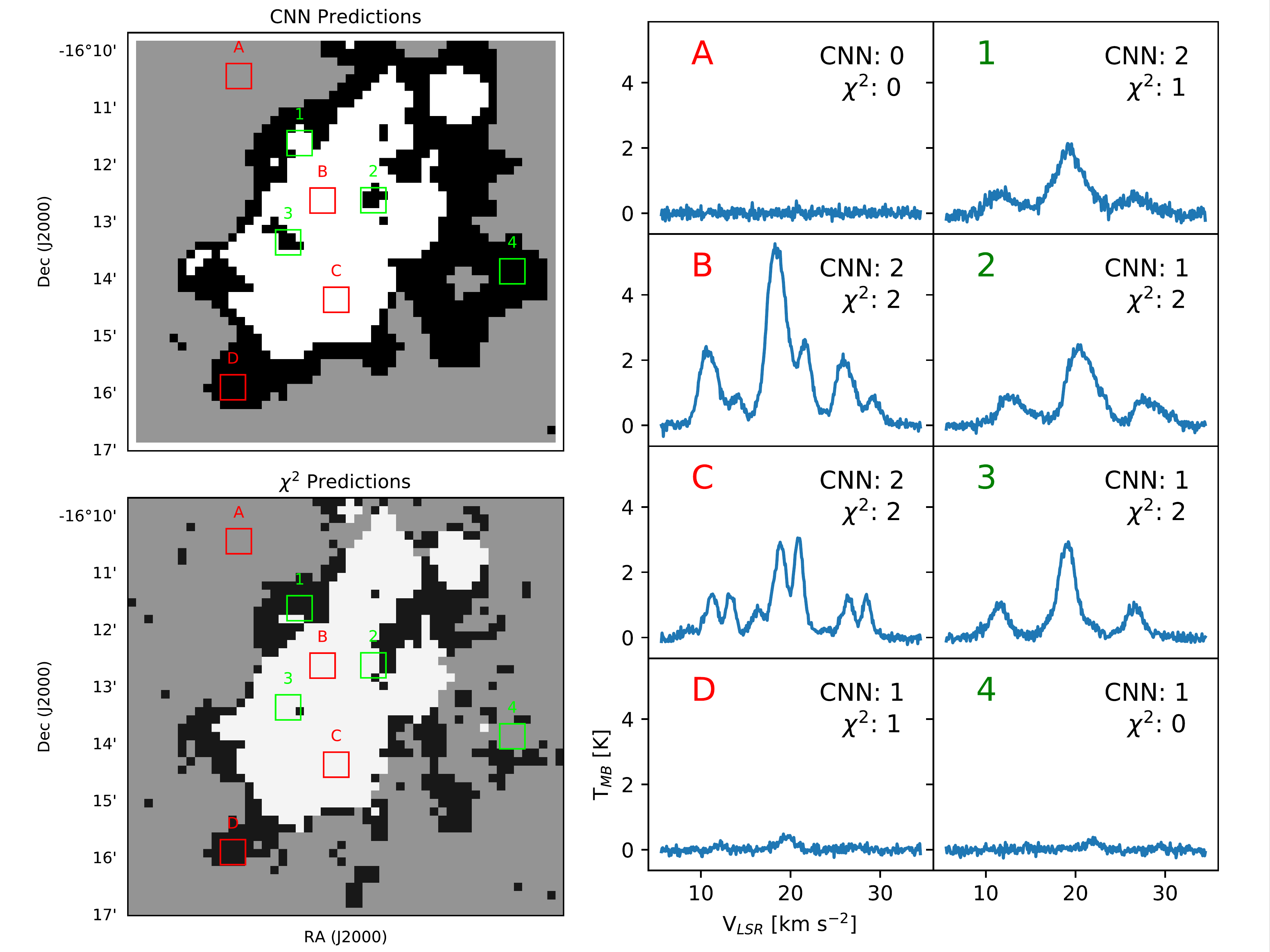}
\caption{Left panels: segmentation of a NH$_3$ (1,1) spectral cube observation of M17 into three classes: single velocity component spectrum (black), multiple velocity component spectrum (white), and noise (grey) using CLOVER's CNN ensemble (top) and traditional $\chi^{2}$-minimization model fitting (bottom).  Right panels: The ``global'' view spectra extracted from the observed spectral cube at the positions of the 3$\times$3 pixel windows overlaid onto the left panels.  Red letters denote positions where CLOVER and the $\chi^{2}$ technique agree in their class predictions, while the green numbers show positions where they disagree.  The spectra in all panels have been clipped around the central three hyperfine groups. The text in the upper right corner of each panel shows the class predicted by CLOVER and the $\chi^{2}$ technique for that spectrum, where 2=two-component, 1=one-component, and 0=noise.}
\label{M17_NH3}
\end{figure}

\begin{figure}[htb]
\epsscale{0.85}
\plotone{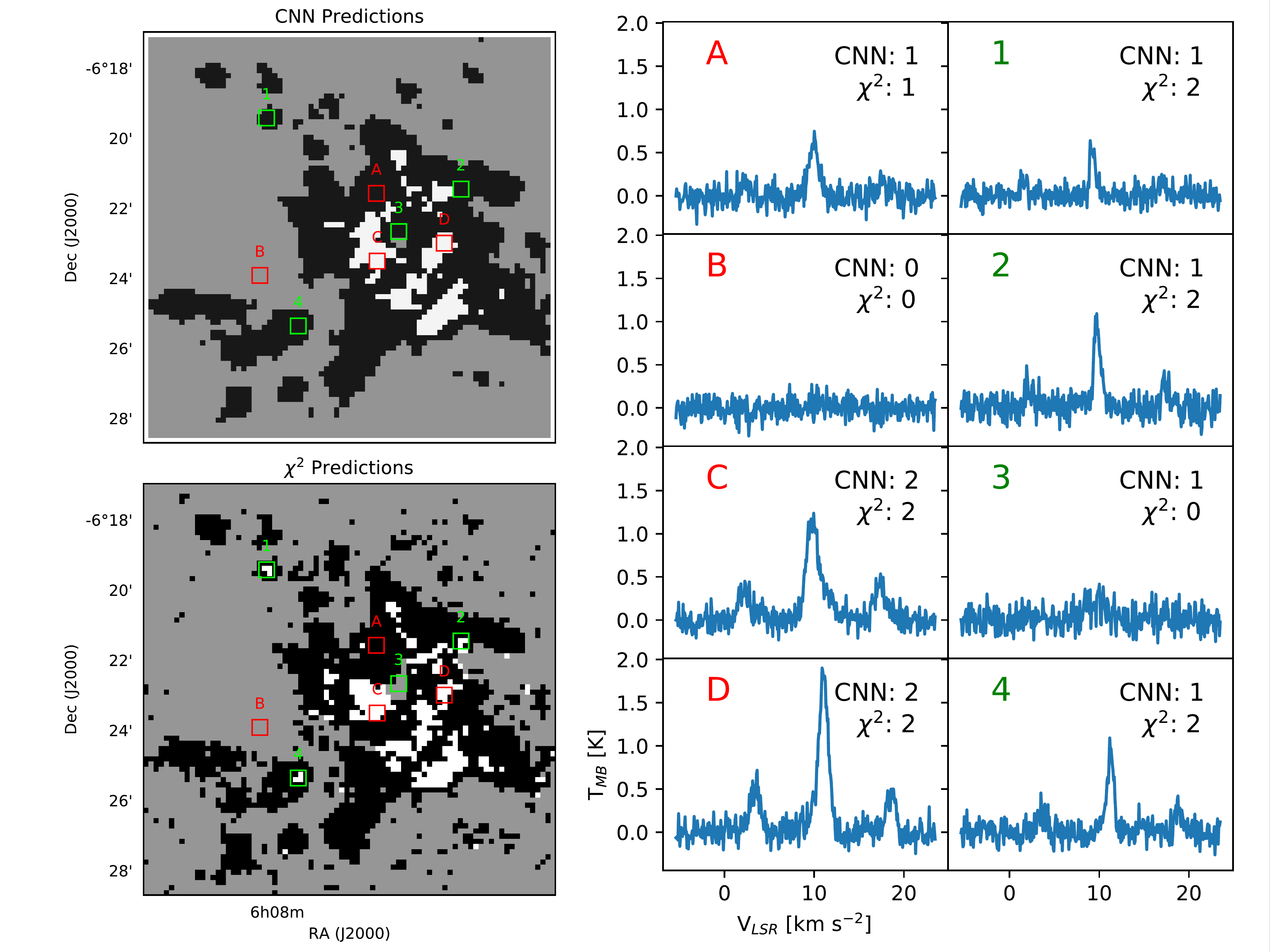}
\caption{Same as Figure \ref{M17_NH3} for MonR2.}
\label{MonR2_NH3}
\end{figure}

Following the method described in Section 4.3, predictions are made for each pixel using both CLOVER's ensemble CNN and the $\chi^2$-minimization technique.  Figures \ref{M17_NH3} and \ref{MonR2_NH3} show the resulting segmentation maps for M17 and MonR2, respectively.  Similar to the results of CLOVER's non-hyperfine classifications, we see clear cases where CLOVER's hyperfine classifications are more robust than the $\chi^2$-minimization technique across all three classes.  In particular, CLOVER appears to provide better noise classifications and be more resilient to overfitting the spectra than the $\chi^2$-minimization technique.  

There is also evidence that CLOVER is able to identify spectra with more than two-components (e.g., three or more velocity components).  For instance, labeled spectrum C in Figure \ref{M17_NH3} shows a location in M17 that clearly has three velocity components.  Even without including three-component spectra in the training set, CLOVER is able to correctly identify that the spectrum has more than one velocity component. 

To test robustly how CLOVER will classify three-component spectra that it receives as input, we perform a three-component classification test similar to the test described in Section 4.4.  An additional test set of 3,000 synthetic three-component NH$_3$ (1,1) samples were created by injecting three synthetic spectra into the test cubes by creating models at random from the distributions listed in Section 6.1.  For these 3000 synthetic three-component samples, CLOVER classifies 2945 ($\sim 98\%$) as ``two-component'' and 55 ($\sim 2\%$) as ``one-component.''  This result suggests that CLOVER's two-component class can be thought of as ``multi-component'' (i.e., emission with more than one velocity component), which is similar to the result found for CLOVER's non-hyperfine classification discussed in Section 4.4.



Furthermore, CLOVER is again remarkably faster at making classifications than the $\chi^2$-minimization technique.  CLOVER's predictions for M17 and MonR2 take 82 seconds and 170 seconds, respectively, on a single CPU core.  In comparison, the full $\chi^2$-minimization technique requires 3918 seconds for M17 and 8435 seconds for MonR2 with the computations run in parallel on eight CPU cores.  This implies CLOVER's classifications provide several orders of magnitude in speed improvements over traditional methods.

\section{Predicting NH$_3$ (1,1) Kinematics}
Hyperfine splitting also poses a challenge for predicting the kinematics of spectra when using CLOVER.  The predictions from CLOVER's regression CNN discussed in Section 5, for example, become unreliable for transitions with hyperfine splitting since the emission is implicitly split across multiple lines with distinct centroid velocities.  To overcome this issue, we train an additional regression CNN to predict the velocity centroids, velocity dispersions, and peak intensities for NH$_3$ (1,1) spectra with two velocity components.  A training set of 300,000 synthetic two-component NH$_3$ (1,1) spectra was generated as described in Section 6.1.  The labels for the training set were a six-number array including the values of $V_{off}$, $\sigma_{tot}$, and the peak intensity ($T_{peak}$) for both of the velocity components.  

\begin{figure}[htb]
\epsscale{1.0}
\plotone{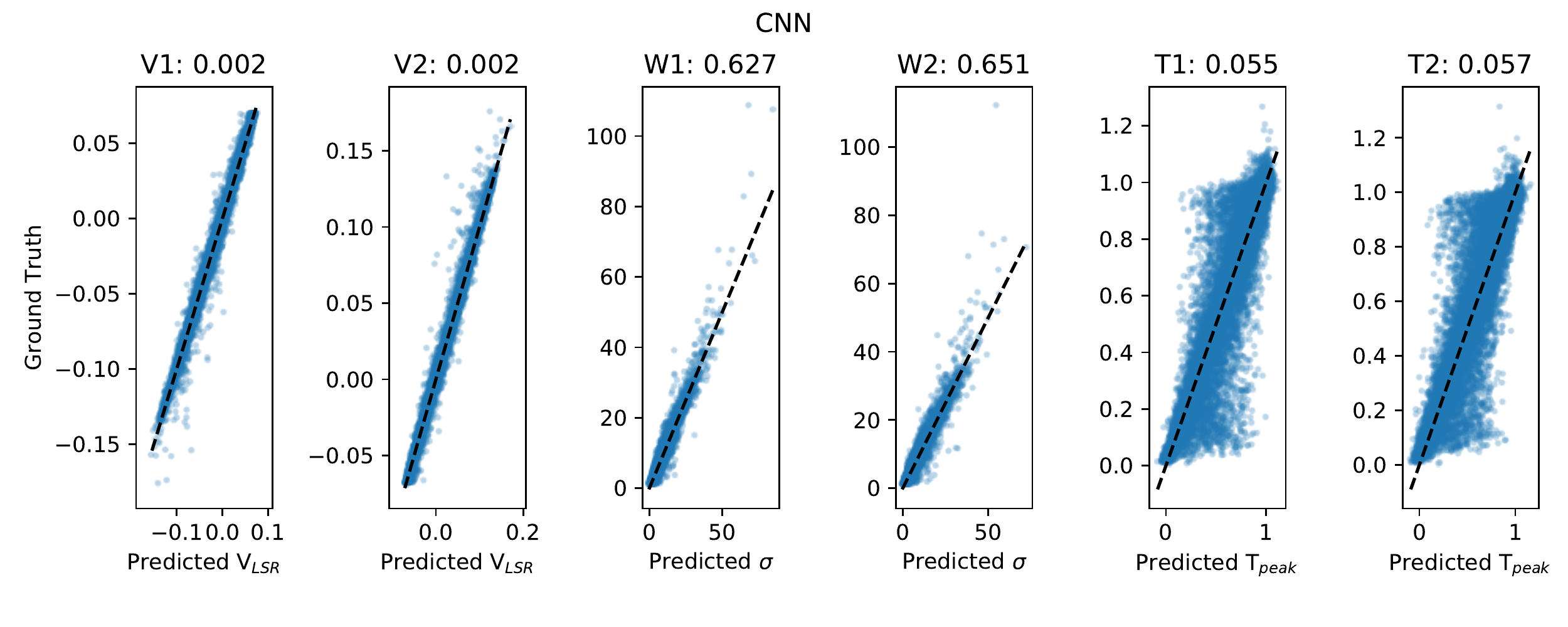}
\caption{Velocity centroid (two left panels), dispersion (two middle panels), and peak intensity (two right panels) predictions by CLOVER's trained NH$_3$ (1,1) regression CNN versus the ``ground-truth'' for the low-velocity component (V1, W1, T1) and high-velocity component (V2, W2, T2) for the 30,000 two-component spectra in the synthetic test set.  The dashed lines show a one-to-one correspondence. In all panels, the centroid velocities are normalized between $-1$ km s$^{-1}$ and 1 km s$^{-1}$.  The velocity dispersion units are the number of channels in the spectrum.  The subtitle above each panel shows the mean absolute error for that parameter.}
\label{NH3_pred_params}
\end{figure}

The performance of the trained network on a validation set of 30,000 additional synthetic spectra is shown in Figure \ref{NH3_pred_params}.  The mean absolute errors for the validation set are $\sim0.002$ for centroid velocity, $\sim0.6$ for velocity dispersion, and $\sim0.06$ for peak intensity.  Since these MAEs have been calculated after normalizing the velocity centroids, dispersions, and peak intensities in the same way as those for the non-hyperfine regression CNN, the MAEs between the two models can be directly compared.  Although the MAEs for velocity dispersion are smaller for the non-hyperfine regression CNN ($\sim0.4$), its centroid velocity and peak intensity MAEs are larger ($\sim0.01$ and $\sim0.064$).  These differences are to be expected since the hyperfine training set was generated using slightly different parameter distributions than the non-hyperfine training set.  

The horizontal flaring at large and small $T_{peak}$ values seen in Figure \ref{NH3_pred_params} also indicates that the hyperfine peak intensity predictions have a slight degeneracy at large and small values.  This effect is also likely related to the way in which the hyperfine training set was generated.  For example, the non-hyperfine training set generator ensured that the velocity components for two-component samples were separated by at least $1.5 \times \sigma_{max}$ (see Section 2.1).  For the hyperfine training set, we instead chose the minimum centroid separation to be $1.0 \times \sigma_{max}$ to probe closer velocity component separations.  This alteration leads to a slightly higher fraction of the hyperfine samples being indistinguishable from single velocity component spectra.  A degeneracy in the peak intensity predictions for those samples is created because it becomes unclear which of the blended components is brighter. 

\begin{figure}[htb]
\epsscale{1.1}
\plottwo{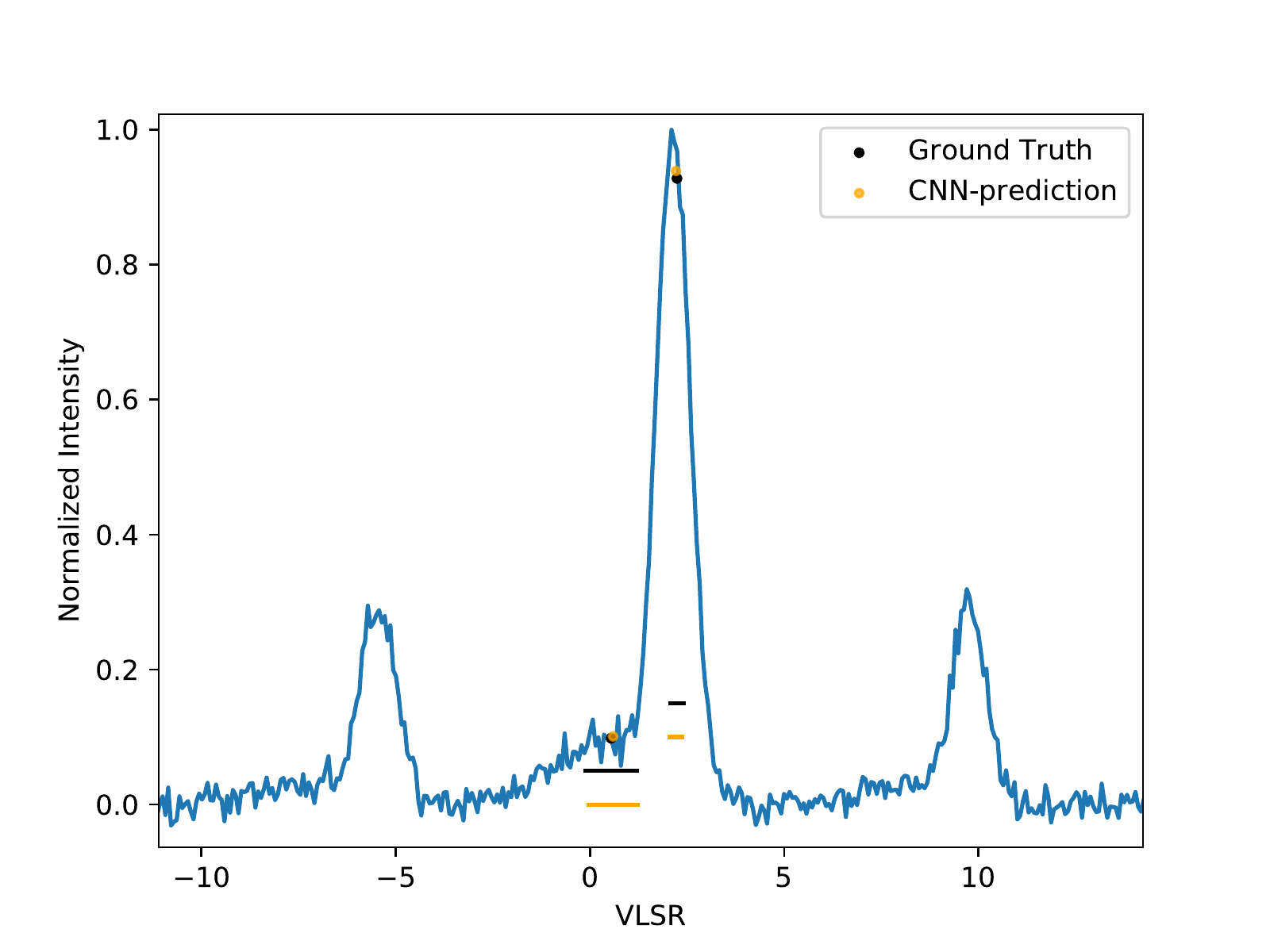}{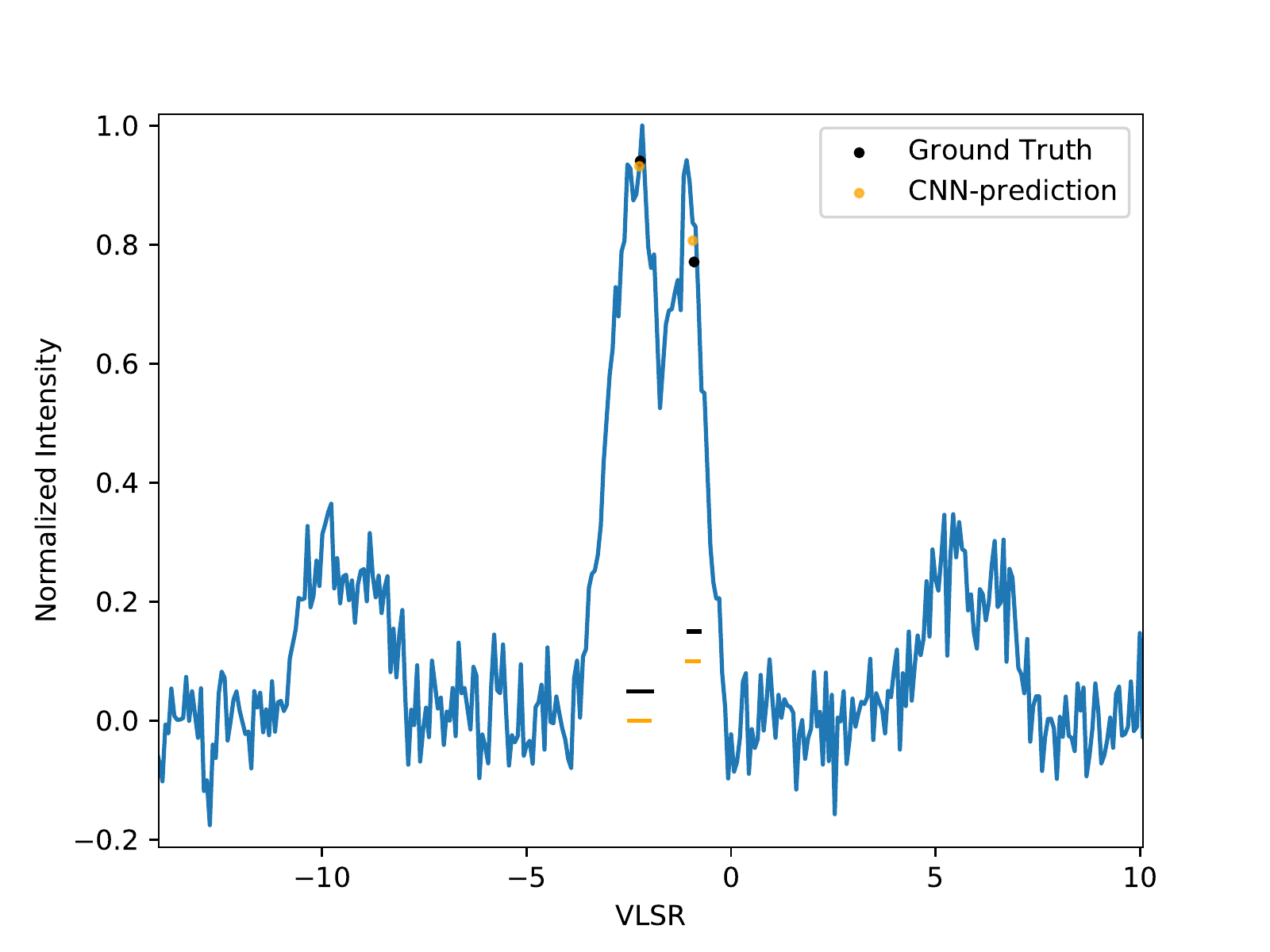}
\plottwo{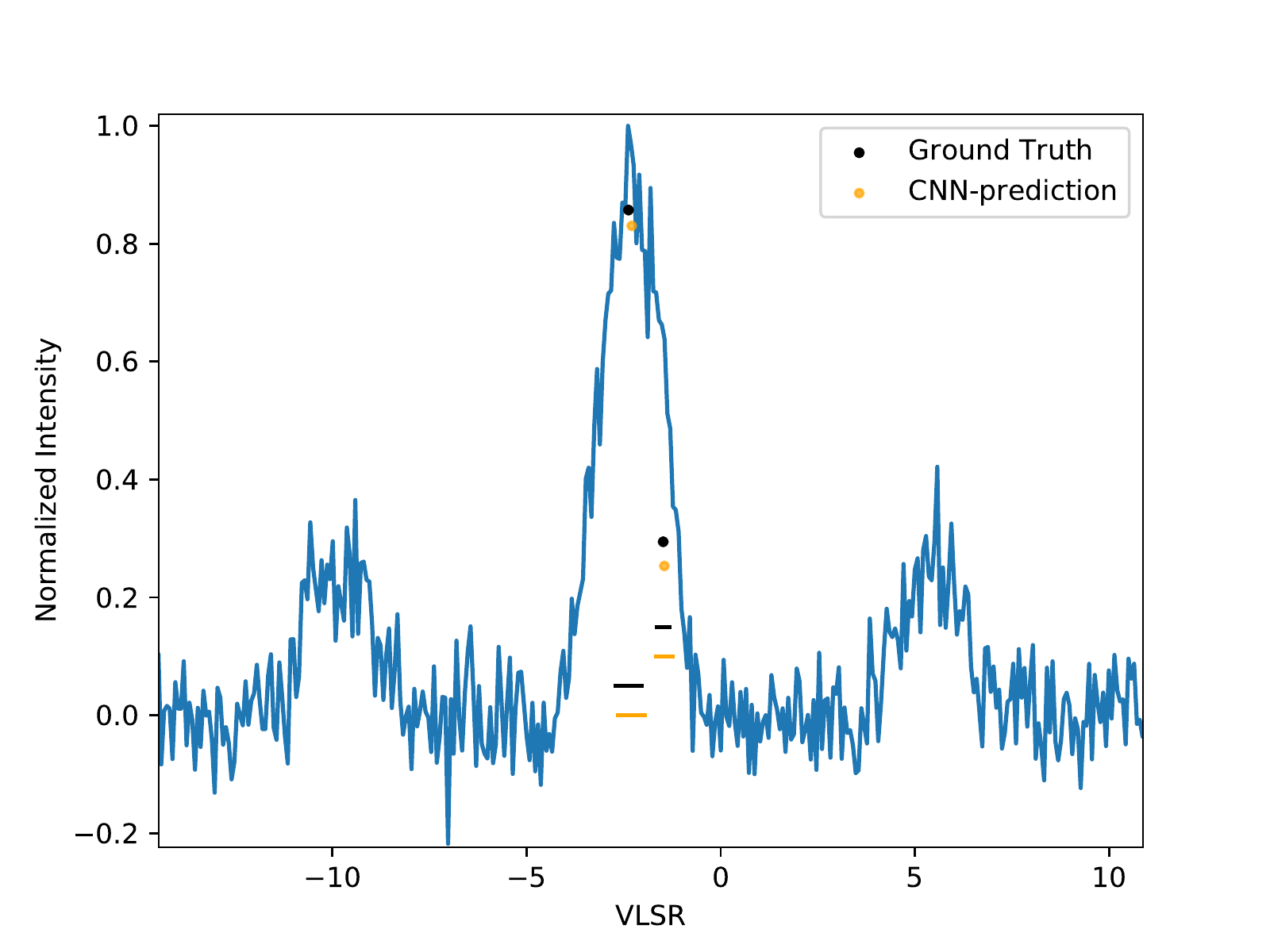}{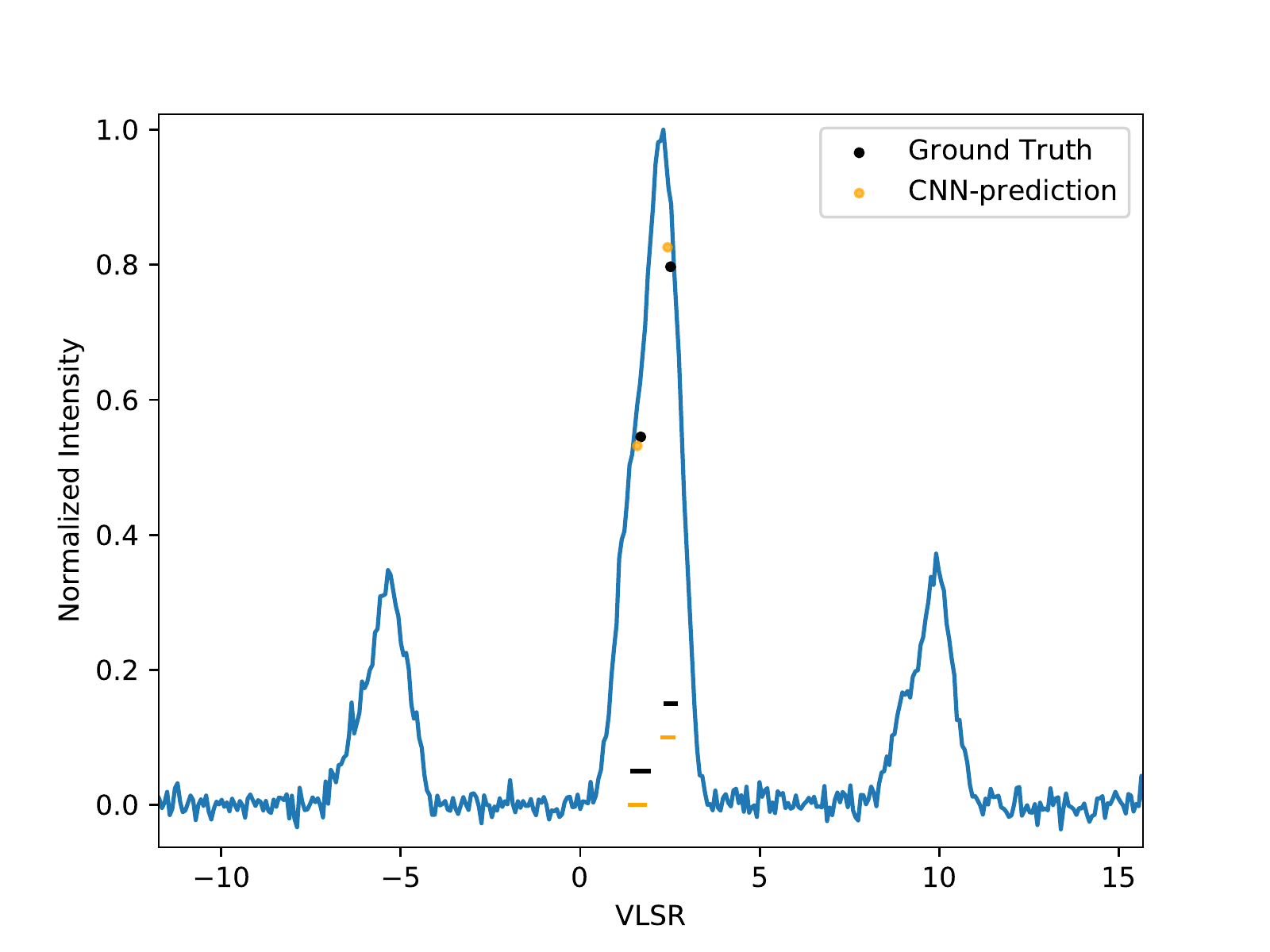}
\centering
\plottwo{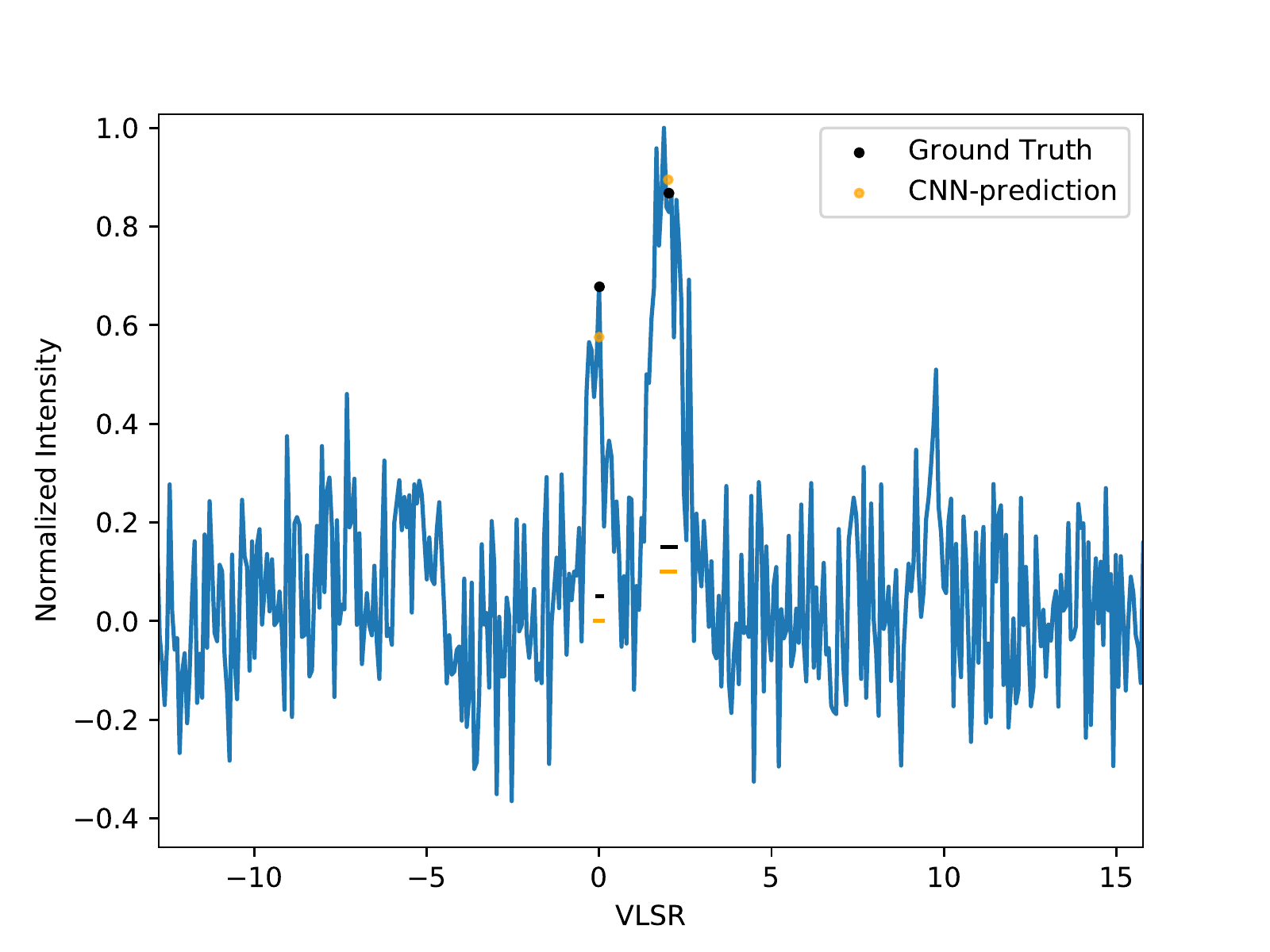}{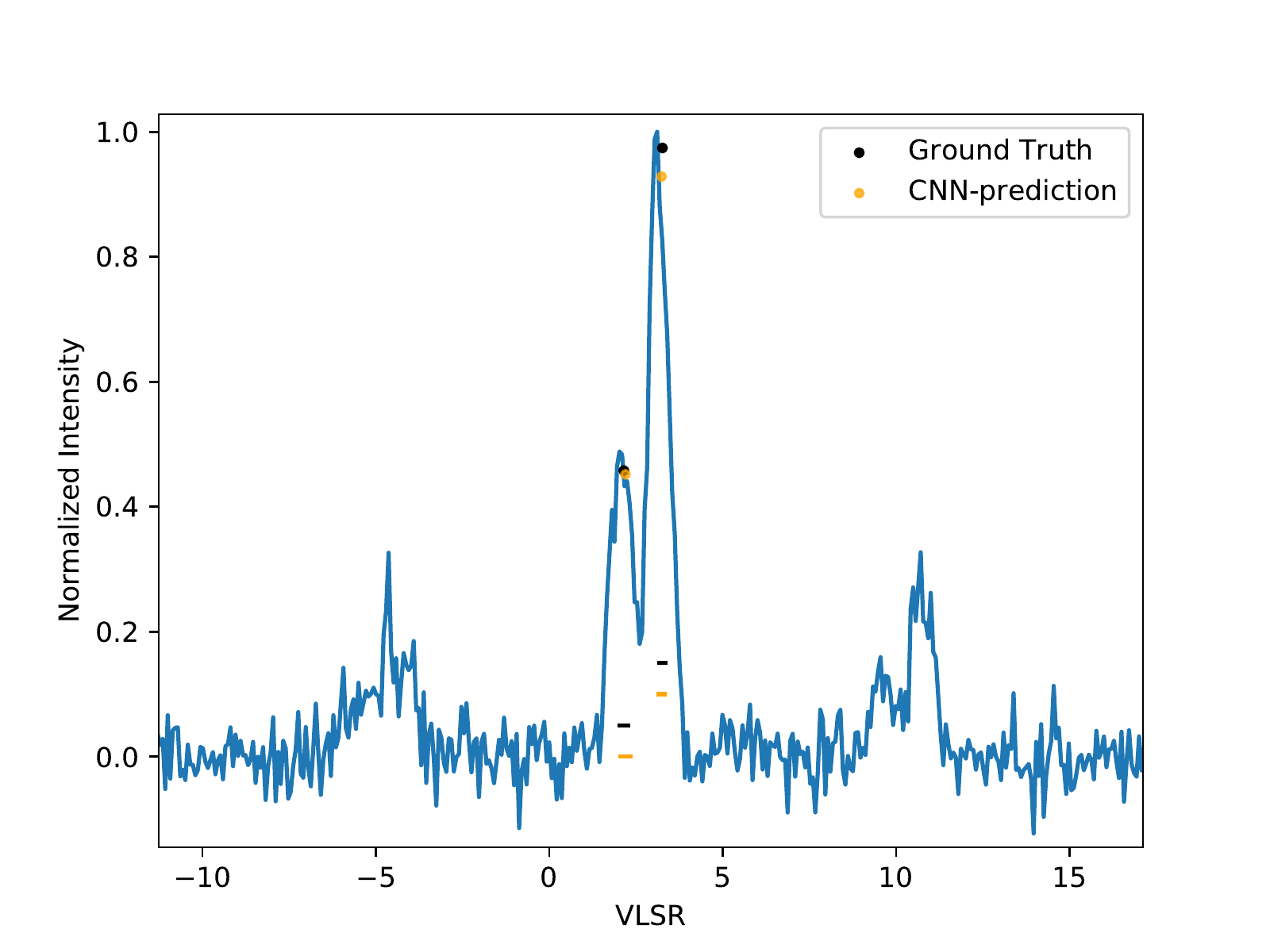}
\caption{Example predictions by CLOVER on previously unseen spectra from the hyperfine synthetic test set.  The horizontal bars show the positions of each velocity component's centroid and dispersion for the ``ground-truth'' (black) and CLOVER predictions (orange) overlaid onto the ``local'' spectrum.  The black and orange dots show the peak intensity for each velocity component for the ground-truth and CLOVER predictions, respectively.  In all panels, the central three hyperfine groups are shown.}
\label{NH3_preds}
\end{figure}

Figure \ref{NH3_preds} displays CLOVER's centroid, dispersion, and peak intensity predictions overlaid onto the local spectra for six unseen samples included in the synthetic test set.  In most cases, CLOVER's predictions are well-matched to the ground-truth values used to create the samples.  Even for blended components (middle panels in Figure \ref{NH3_preds}) and those with shallow wings (top left panel in Figure \ref{NH3_preds}), CLOVER can accurately recover the underlying kinematics.

These tests prove that CNNs can be trained to not only classify spectra with hyperfine structure and multiple velocity components, but also predict with high accuracy the kinematics of the emitting gas.  Moreover, this method can easily be adjusted to incorporate other molecular tracers of interest that exhibit hyperfine splitting (e.g., HCN, N$_2$H$^+$, etc.).  Although the current implementation of CLOVER considers only the one- versus two-component classes of emission, the method could also be generalized to emission with three- or more velocity components. 


\clearpage

\section{Improving Virial Analyses with CLOVER}
CLOVER's two-component spectral classifications and kinematics predictions can be used to improve existing analyses that neglect the presence of multiple velocity components.  For instance, many virial stability analyses of star-forming structures rely on velocity dispersions measured from models that assume a single velocity component along the line of sight \cite[e.g.,][]{Keown_2017, Kirk_2017, Chen_2018, Kerr_2019}.  When multiple velocity components are present along the line of sight, however, models assuming a single velocity component typically fit the observed spectrum with a much wider velocity dispersion than would be obtained by using a model with multiple velocity components.  Since the virial parameter is proportional to $\sigma^2$, the wider  velocity dispersions measured from one-component fits have a significant impact on the stability measurement of a given structure.  

\begin{figure}[htb]
\epsscale{1.2}
\plottwo{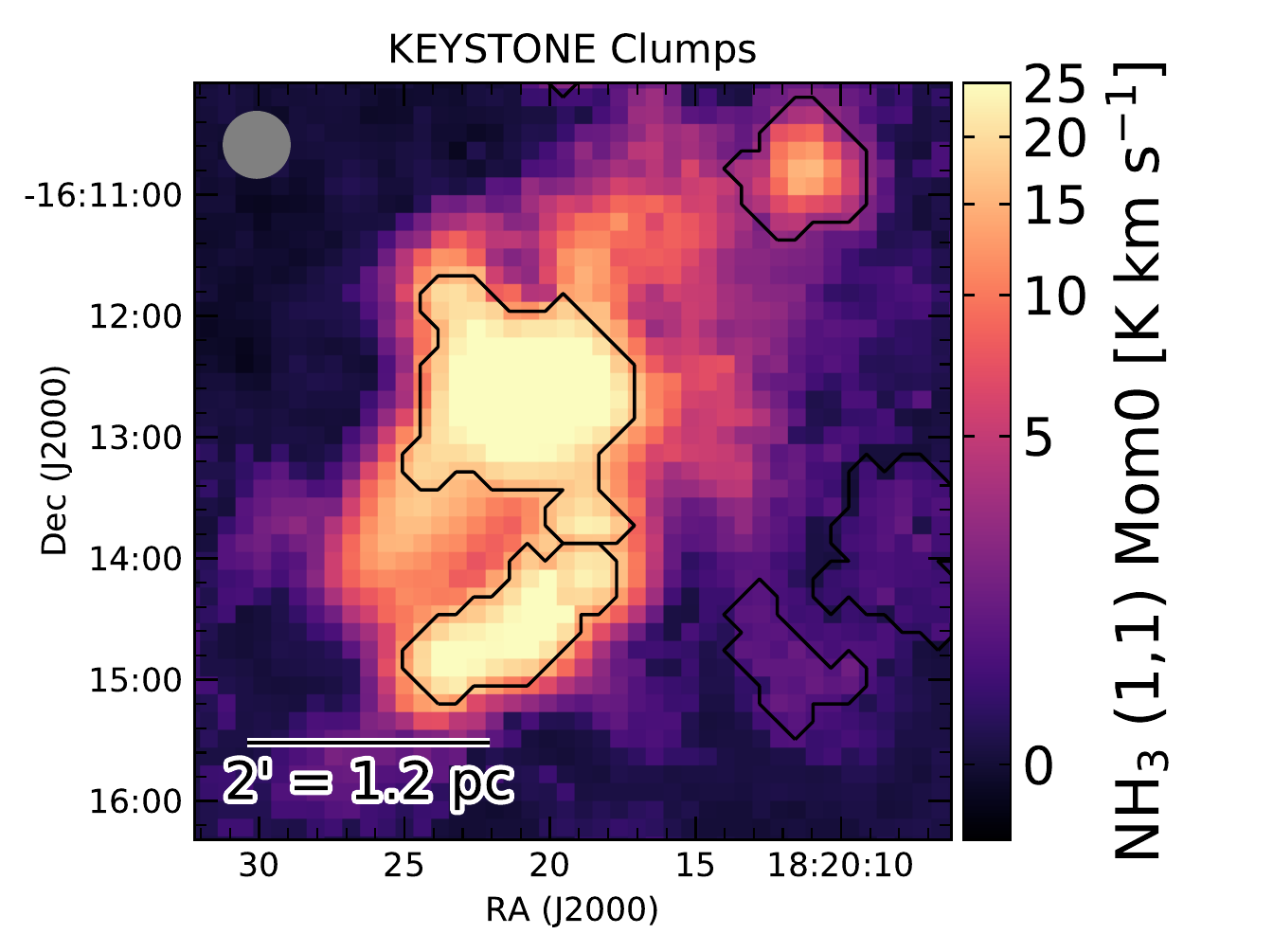}{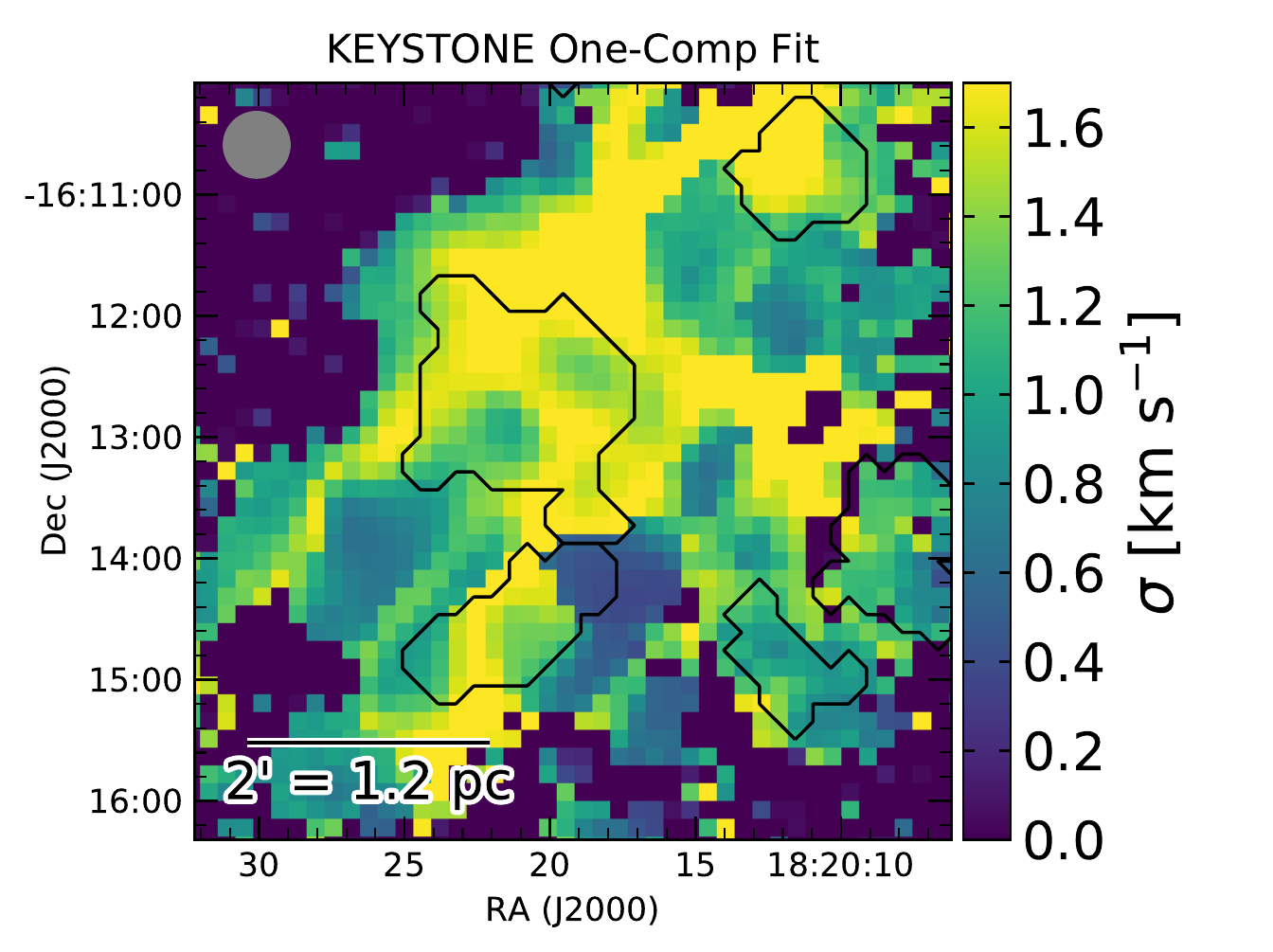}
\plottwo{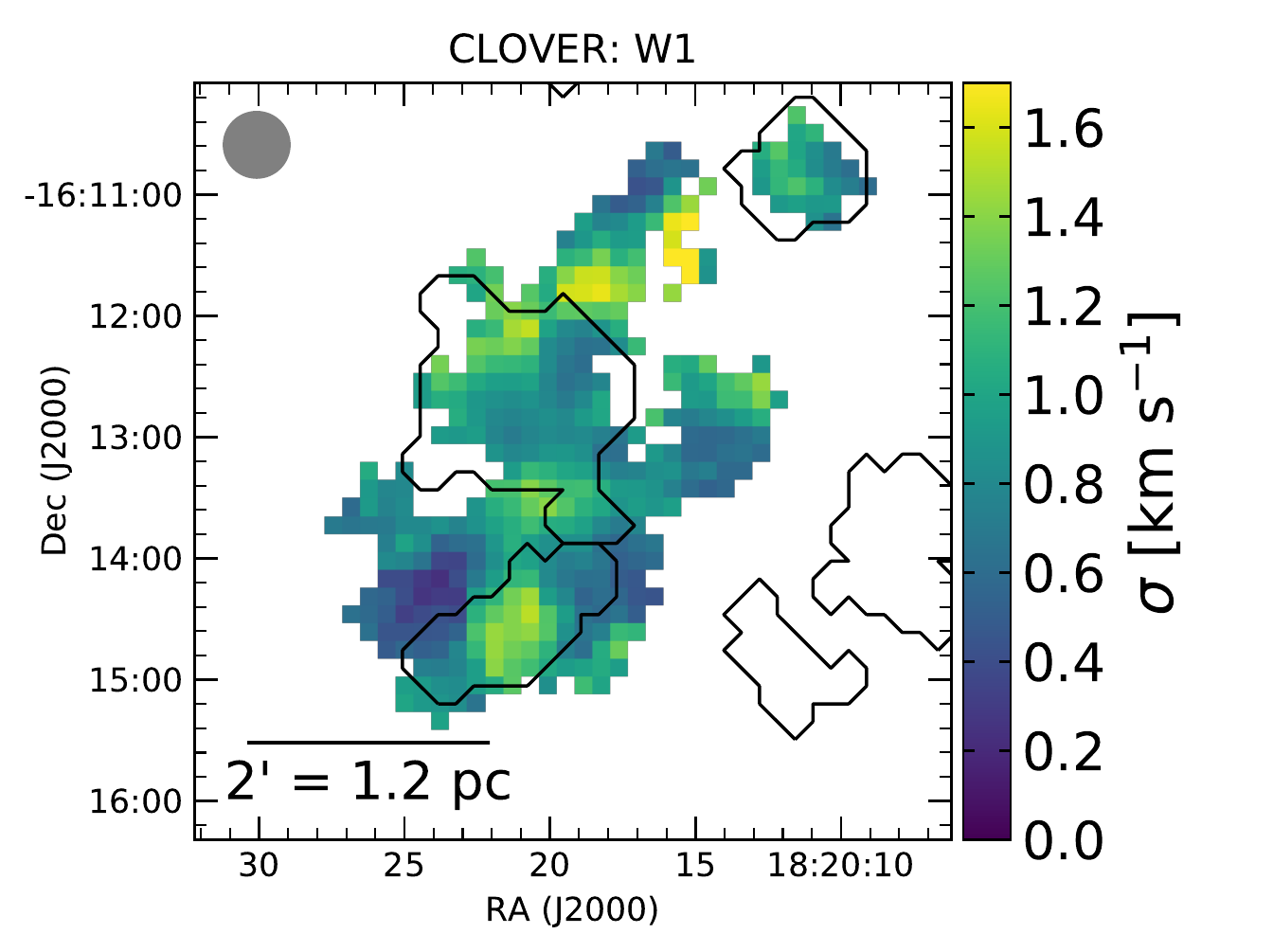}{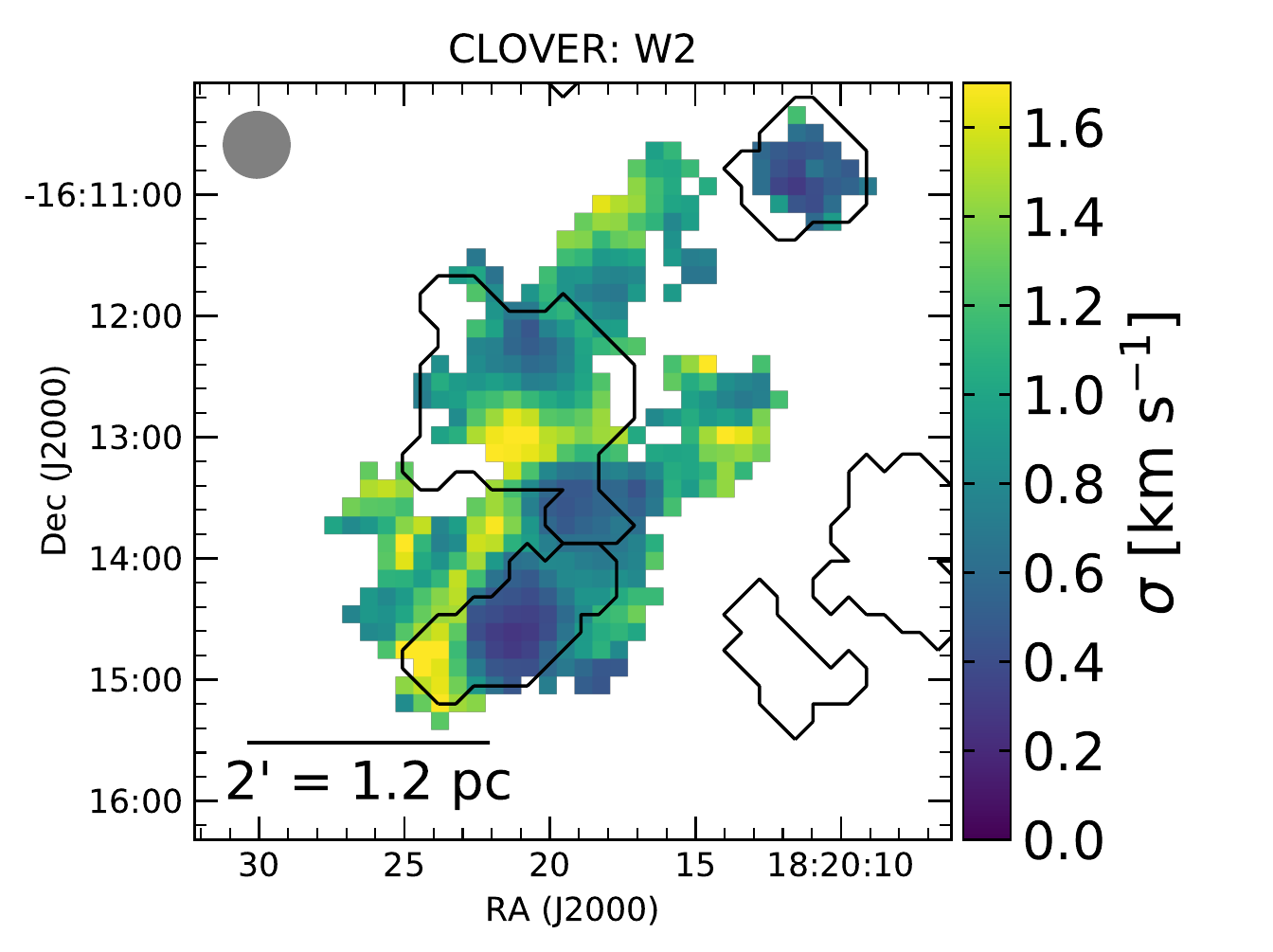}
\caption{Top left: NH$_3$ (1,1) integrated intensity map of the M17SW region observed by KEYSTONE with dendrogram-identified clumps overlaid as black contours (Keown et al. 2019, submitted).  Top right: KEYSTONE velocity dispersion measurements from modeling the NH$_3$ (1,1) emission with one velocity component.  Bottom row: Velocity dispersion measured by CLOVER for pixels classified as ``two-component.''}
\label{M17_leaves}
\end{figure}

To demonstrate CLOVER's ability to improve virial analyses, we use two-component velocity dispersions measured by CLOVER to update the virial analysis of M17SW by Keown et al. (2019, submitted).  The Keown et al. analysis used the KEYSTONE NH$_3$ (1,1) integrated intensity maps of M17 to identify dense gas clumps, which are shown as black contours in Figure \ref{M17_leaves}.  Virial parameters were calculated by Keown et al. for each structure using a velocity dispersion map (top right panel of Figure \ref{M17_leaves}) measured from an ammonia model assuming a single velocity component along the line of sight.  The velocity dispersion maps predicted by CLOVER for pixels identified as two-component in M17SW are shown in the bottom row of Figure \ref{M17_leaves}. Figure \ref{M17_leaves} clearly shows that the one-component fit produces larger velocity dispersions than the two velocity components identified by CLOVER.

Three of the clumps identified by Keown et al. fall on pixels identified as two-component by CLOVER.  Here, we re-calculate the virial parameters for these three structures using the same mass, average temperature, and radius measured by Keown et al., but replace the average velocity dispersion with values measured from the CLOVER velocity dispersion maps.  Although this approach neglects mass and/or size differences in the multiple structures along the line of sight, it serves as a test to see how much the two-component kinematics might affect their calculated virial parameters. 

\begin{figure}[htb]
\epsscale{0.7}
\plotone{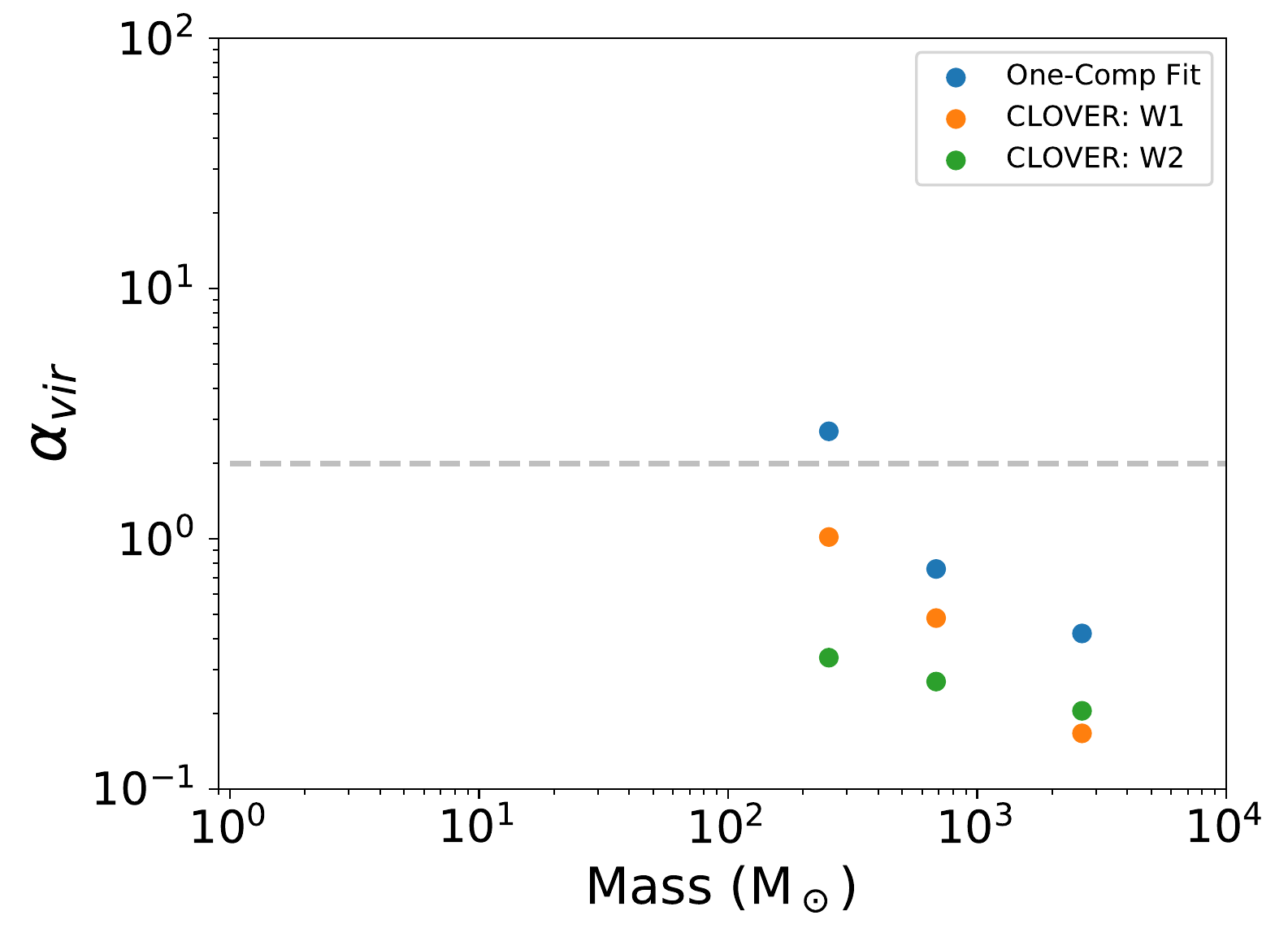}
\caption{Virial parameter versus mass for the three dendrogram-identified clumps in Figure \ref{M17_leaves} falling on pixels classified as ``two-component'' by CLOVER.  Blue shows the virial parameters derived by Keown et al. (2019, submitted) using the KEYSTONE velocity dispersions from a one-component fit to the NH$_3$ (1,1).  Orange and green show the virial parameters derived when using the CLOVER velocity dispersion predictions for each identified velocity component (W1 and W2).  Sources falling below the grey dashed line are gravitationally bound when neglecting magnetic fields and external pressure.}
\label{alpha}
\end{figure}

Following the method described in Keown et al., each structure's average velocity dispersion is calculated as the average of all pixels falling within its 2D mask shown in Figure \ref{M17_leaves}.  The average is weighted by the integrated intensity map such that $\sigma_{v, avg} = w_1\sigma_1 + w_2\sigma_2 \cdots w_n\sigma_n$, where $w_n$ and $\sigma_n$ are the fraction of the source's integrated intensity and value of the velocity dispersion, respectively, for pixel $n$.  Since CLOVER predicts two velocity dispersions for every pixel (one for each velocity component), we calculate two virial parameters for each structure based on the weighted average velocity dispersion measured in each map.  Figure \ref{alpha} compares these new virial parameters with the original values presented in Keown et al. (2019, submitted).

As expected, the virial parameters using the CLOVER velocity dispersions are lower than the Keown et al. measurements.  Specifically, the CLOVER-measured virial parameters are a factor of 1.5 - 8 times lower depending on the structure and which velocity component map is used.  The lowest mass structure also moves from the upper ``gravitationally unbound'' half of the plot to the lower ``gravitationally bound'' half when using the CLOVER measurements.  Although only three structures are analyzed here, this example shows the usefulness of CLOVER for virial analyses that include structures with multiple velocity components along the line of sight.

\section{Summary}
We present a new method for identifying emission line spectra with two velocity components by training an ensemble of convolutional neural networks (CNNs) using synthetic spectral cubes.  The networks predict the class of $3\times3$ pixel windows, utilizing the spatial information of pixels adjacent to the central pixel to make a prediction.  The trained network ensemble has classification accuracies of $99.92 \pm 0.02 \%$, $100\%$, and $96.72 \pm 0.18\%$ for one-component, noise-only, and two-component synthetic spectral windows.  This performance is a significant improvement over traditional line fitting approaches that do not consider the spatial information in adjacent pixels.  The ensemble CNN's high classification performance was also demonstrated on real spectral cubes, which revealed that the ensemble CNN is able to segment accurately real observations into each of the three training set classes.  Moreover, the speed with which the ensemble CNN makes its classifications was measured to be over an order of magnitude faster than a traditional line fitting approach. 

A regression CNN is also trained to extract kinematics directly from the spectra identified as two-component class members by the ensemble CNN classifications.  We show that the regression CNN has high prediction accuracy for two-component spectra that exhibit large centroid velocity separations and those that are blended.  The combination of the ensemble and regression CNNs provides a quick way to measure accurately kinematics from two-component spectra.  Named Convnet Line-fitting Of Velocities in Emission-line Regions (CLOVER), this combination unlocks a new method to analyze large spectral cubes of emission lines from star-forming molecular clouds. 

After testing CLOVER on observations of four different molecular emission lines from five distinct star-forming regions observed by three separate observatories, it is clear that its predictions can be generalized to many data sets.  In particular, we show that the method can be applied to transitions with hyperfine splitting.  The versatility and speed of CLOVER's predictions make it an attractive option for signal versus noise segmentation and line fitting for large-scale spectral mapping surveys.  The higher accuracy kinematics measurements provided by CLOVER also make it a useful tool for improving virial stability analyses of star-forming structures.  CLOVER is publicly available as a Python package called \texttt{astroclover} at \url{https://github.com/jakeown/astroclover/}.

\section*{Acknowledgments}
We would like to thank Stella Offner for helpful recommendations regarding CLOVER's network architecture and training set.  We also thank the anonymous referee for their thoughtful comments that have undoubtedly improved our manuscript. JK, JDF, ER, and MCC acknowledge the financial support of a Discovery Grant from NSERC of Canada.  The Green Bank Observatory is a facility of the National Science Foundation operated under cooperative agreement by Associated Universities, Inc.  The James Clerk Maxwell Telescope is operated by the East Asian Observatory on behalf of The National Astronomical Observatory of Japan; Academia Sinica Institute of Astronomy and Astrophysics; the Korea Astronomy and Space Science Institute; Center for Astronomical Mega-Science (as well as the National Key R$\&$D Program of China with No. 2017YFA0402700). Additional funding support is provided by the Science and Technology Facilities Council of the United Kingdom and participating universities in the United Kingdom and Canada. This research made use of Astropy (\url{http://www.astropy.org}), a community-developed core Python package for Astronomy \citep{Astropy_2013}, pyspeckit (\url{https://pyspeckit.readthedocs.io/en/latest/}), a Python spectroscopic analysis and reduction toolkit \citep{Ginsburg_2011}, and Keras, a Python API for neural networks \citep{chollet2015keras}. 

{\it Facility:} \facility{GBT}, \facility{JCMT}, \facility{FCRAO}


\section*{Appendix}
\begin{appendix}
\section{Installing and Using CLOVER}
CLOVER is publicly available for use as a Python package called \texttt{astroclover}.  Here, we provide a brief description of the installation and usage instructions for the package.  

Users must first ensure that they have Python 3 and all required packages installed.  This can easily be done by installing the \texttt{Anaconda} Python package manager at \url{https://www.anaconda.com/distribution/}, which has many of \texttt{astroclover's} package dependencies pre-installed.  \texttt{Anaconda} version 4.6.11 or later is recommended for \texttt{astroclover}, but other \texttt{Anaconda} versions have not been tested.  

Once \texttt{Anaconda} is installed, users can run the following commands in a Linux or Mac terminal to setup a new environment and install the remaining packages required for \texttt{astroclover}:

\begin{itemize}
\item conda create $-$n clover$\_$env python$=\!3.6$ anaconda
\item conda activate clover$\_$env
\item pip install tensorflow$==\!1.8.0$ keras$==\!2.2.0$ spectral$\_$cube
\end{itemize}  The first two commands will setup an \texttt{Anaconda} virtual environment named \texttt{clover$\_$env}, which must be entered when running \texttt{astroclover} by running the \texttt{conda activate clover$\_$env} command.  The last command installs \texttt{tensorflow} version 1.8.0, \texttt{keras} version 2.2.0, and \texttt{spectral$\_$cube}, which are the CLOVER package dependencies not included by default in \texttt{Anaconda}.  Although Python version 3.6 is recommended, \texttt{astroclover} has also been tested on Python 2.7 for users that wish to use Python version 2.

After successfully setting up the \texttt{Anaconda} environment, users can clone or download \texttt{astroclover} at \url{https://github.com/jakeown/astroclover}.  This will create a new directory called \textit{astroclover} at the download location.  Users must enter this directory and run the following command from their \texttt{Anaconda} environment:

\begin{itemize}
\item python download$\_$models.py
\end{itemize} which will download the trained convolutional neural networks that CLOVER uses from a remote directory into the user's local \textit{astroclover} directory.  The 14 files are $\sim12$ GB in total.

Once the neural network files have been downloaded, users must ensure the spectral cube they input into CLOVER is formatted properly.  CLOVER's predictions require FITS data cubes with position-position-spectral axes.  A spectral axis of 500 channels is required for Gaussian emission lines and 1000 channels for NH$_3$ (1,1).  If the cube a user inputs into CLOVER is smaller than those sizes, CLOVER will add random noise channels to each end of the spectral axis up to the required size.  If the input cube's spectral axis is larger than the required input size, CLOVER will clip channels from each end of the spectral axis until the required size is obtained.

It is also recommended that the centroids of the emission lines in an input cube be located within the central $\sim275$ channels for Gaussian emission lines and the central $\sim140$ channels for NH$_3$ (1,1).  These bounds are set by the range of possible centroids used to train CLOVER.  If a cube has large centroid velocity gradients that cause some of the emission lines to fall outside these bounds, it is recommended that users split their cube into sub-cubes so that all emission is within the aforementioned channel bounds.

To run CLOVER on a prepared data cube, simply use the \texttt{predict(f=your$\_$cube$\_$name.fits)} function in the \texttt{clover.py} script. If the cube is NH$_3$ (1,1), add \texttt{nh3=True} in the call to \texttt{predict()} (e.g., \texttt{predict(f$=$your$\_$nh3$\_$cube.fits, nh3$=$True)}).  For example, if the user is predicting on a NH$_3$ (1,1) cube using an iPython session within the \textit{astroclover} directory, they would use the following commands:

\begin{itemize}
\item import clover
\item clover.predict(f$=$your$\_$nh3$\_$cube.fits, nh3$=$True)
\end{itemize}

The classification step uses an ensemble of six independently trained CNNs to make the final class prediction.  These six predictions can be done in parallel by specifying the number of desired parallel processes.  For example, to run all six predictions at once, use \texttt{predict(f$=$your$\_$nh3$\_$cube.fits, nproc=6)}.  

CLOVER will output its classification map and parameter predictions as individual FITS files.  In total, up to eight files are generated:
\begin{enumerate}
\item input$\_$name $+ `\_$clover.fits' - cube after the spectral axis has been corrected (not generated if input cube already has proper spectral length)
\item input$\_$name $+ `\_$class.fits' - predicted class of each pixel (2=two-component, 1=noise, 0=one-component)
\item input$\_$name $+ `\_$vlsr1.fits' - predicted centroid velocity of component with lowest centroid
\item input$\_$name $+ `\_$vlsr2.fits' - predicted centroid velocity of component with highest centroid
\item input$\_$name $+ `\_$sig1.fits' - predicted velocity dispersion of component with lowest centroid
\item input$\_$name $+ `\_$sig2.fits' - predicted velocity dispersion of component with highest centroid
\item input$\_$name $+ `\_$tpeak1.fits' - predicted peak intensity of component with lowest centroid
\item input$\_$name $+ `\_$tpeak2.fits' - predicted peak intensity of component with highest centroid
\end{enumerate} where input$\_$name is the name of the FITS file input into CLOVER.  

Please refer to \url{https://github.com/jakeown/astroclover} for the most up-to-date install and usage instructions since new features may be developed in the future.  
\end{appendix}

\bibliographystyle{apj}
\bibliography{Cepheus_GAS_stability}

\begin{thebibliography}{}
\expandafter\ifx\csname natexlab\endcsname\relax\def\natexlab#1{#1}\fi

\bibitem[{{Astropy Collaboration} {et~al.}(2013){Astropy Collaboration},
  {Robitaille}, {Tollerud}, {Greenfield}, {Droettboom}, {Bray}, {Aldcroft},
  {Davis}, {Ginsburg}, {Price-Whelan}, {Kerzendorf}, {Conley}, {Crighton},
  {Barbary}, {Muna}, {Ferguson}, {Grollier}, {Parikh}, {Nair}, {Unther},
  {Deil}, {Woillez}, {Conseil}, {Kramer}, {Turner}, {Singer}, {Fox}, {Weaver},
  {Zabalza}, {Edwards}, {Azalee Bostroem}, {Burke}, {Casey}, {Crawford},
  {Dencheva}, {Ely}, {Jenness}, {Labrie}, {Lim}, {Pierfederici}, {Pontzen},
  {Ptak}, {Refsdal}, {Servillat}, \& {Streicher}}]{Astropy_2013}
{Astropy Collaboration}, {Robitaille}, T.~P., {Tollerud}, E.~J., {et~al.} 2013,
  \aap, 558, A33

\bibitem[{{Chen} {et~al.}(2019{\natexlab{a}}){Chen}, {Pineda}, {Goodman},
  {Burkert}, {Offner}, {Friesen}, {Myers}, {Alves}, {Arce}, {Caselli},
  {Chac{\'o}n-Tanarro}, {Chen}, {Di Francesco}, {Ginsburg}, {Keown}, {Kirk},
  {Martin}, {Matzner}, {Punanova}, {Redaelli}, {Rosolowsky}, {Scibelli}, {Seo},
  {Shirley}, {Singh}, \& {The GAS Collaboration}}]{Chen_2018}
{Chen}, H.~H.-H., {Pineda}, J.~E., {Goodman}, A.~A., {et~al.}
  2019{\natexlab{a}}, \apj, 877, 93

\bibitem[{{Chen} {et~al.}(2019{\natexlab{b}}){Chen}, {Zhang}, {Wright},
  {Busquet}, {Lin}, {Liu}, {Olguin}, {Sanhueza}, {Nakamura}, \&
  {Palau}}]{Chen_2019}
{Chen}, H.-R.~V., {Zhang}, Q., {Wright}, M.~C.~H., {et~al.} 2019{\natexlab{b}},
  \apj, 875, 24

\bibitem[{{Chen} {et~al.}(in prep){Chen}, {Di Francesco}, {Rosolowsky},
  {Pineda}, \& {Friesen}}]{Chen_prep}
{Chen}, M., {Di Francesco}, J., {Rosolowsky}, E., {Pineda}, J., \& {Friesen},
  R. in prep, \apj

\bibitem[{Chollet {et~al.}(2015)}]{chollet2015keras}
Chollet, F., {et~al.} 2015, Keras, \url{https://keras.io}

\bibitem[{{Clarke} {et~al.}(2018){Clarke}, {Whitworth}, {Spowage},
  {Duarte-Cabral}, {Suri}, {Jaffa}, {Walch}, \& {Clark}}]{Clarke_2018}
{Clarke}, S.~D., {Whitworth}, A.~P., {Spowage}, R.~L., {et~al.} 2018, \mnras,
  479, 1722

\bibitem[{Devaraj {et~al.}(2014)Devaraj, Church, Cleary, Frayer, Gawande,
  Goldsmith, Gundersen, Harris, Kangaslahti, Readhead, Reeves, Samoska, Sieth,
  \& Larkoski}]{Sieth_2014}
Devaraj, K., Church, S., Cleary, K., {et~al.} 2014, 1--1

\bibitem[{{Fabbro} {et~al.}(2018){Fabbro}, {Venn}, {O'Briain}, {Bialek},
  {Kielty}, {Jahandar}, \& {Monty}}]{Fabbro_2018}
{Fabbro}, S., {Venn}, K.~A., {O'Briain}, T., {et~al.} 2018, \mnras, 475, 2978

\bibitem[{{Friesen} {et~al.}(2013){Friesen}, {Medeiros}, {Schnee}, {Bourke},
  {Francesco}, {Gutermuth}, \& {Myers}}]{Friesen_2013}
{Friesen}, R.~K., {Medeiros}, L., {Schnee}, S., {et~al.} 2013, \mnras, 436,
  1513

\bibitem[{{Friesen} {et~al.}(2017){Friesen}, {Pineda}, {co-PIs}, {Rosolowsky},
  {Alves}, {Chac{\'o}n-Tanarro}, {How-Huan Chen}, {Chun-Yuan Chen}, {Di
  Francesco}, {Keown}, {Kirk}, {Punanova}, {Seo}, {Shirley}, {Ginsburg},
  {Hall}, {Offner}, {Singh}, {Arce}, {Caselli}, {Goodman}, {Martin}, {Matzner},
  {Myers}, {Redaelli}, \& {The GAS Collaboration}}]{Friesen_2017}
{Friesen}, R.~K., {Pineda}, J.~E., {co-PIs}, {et~al.} 2017, \apj, 843, 63

\bibitem[{{Ginsburg} \& {Mirocha}(2011)}]{Ginsburg_2011}
{Ginsburg}, A., \& {Mirocha}, J. 2011, {PySpecKit: Python Spectroscopic
  Toolkit}, Astrophysics Source Code Library, ascl:1109.001

\bibitem[{{Gottschalk} {et~al.}(2012){Gottschalk}, {Kothes}, {Matthews},
  {Landecker}, \& {Dent}}]{Gottschalk_2012}
{Gottschalk}, M., {Kothes}, R., {Matthews}, H.~E., {Landecker}, T.~L., \&
  {Dent}, W.~R.~F. 2012, \aap, 541, A79

\bibitem[{{Hampton} {et~al.}(2017){Hampton}, {Medling}, {Groves}, {Kewley},
  {Dopita}, {Davies}, {Ho}, {Kaasinen}, {Leslie}, {Sharp}, {Sweet}, {Thomas},
  {Allen}, {Bland-Hawthorn}, {Brough}, {Bryant}, {Croom}, {Goodwin}, {Green},
  {Konstantantopoulos}, {Lawrence}, {L{\'o}pez-S{\'a}nchez}, {Lorente},
  {McElroy}, {Owers}, {Richards}, \& {Shastri}}]{Hampton_2017}
{Hampton}, E.~J., {Medling}, A.~M., {Groves}, B., {et~al.} 2017, \mnras, 470,
  3395

\bibitem[{{Henshaw} {et~al.}(2013)}]{Henshaw_2013}
{Henshaw}, J.~D., {et~al.} 2013, \mnras, 428, 3425

\bibitem[{{Henshaw} {et~al.}(2016){Henshaw}, {Longmore}, {Kruijssen}, {Davies},
  {Bally}, {Barnes}, {Battersby}, {Burton}, {Cunningham}, {Dale}, {Ginsburg},
  {Immer}, {Jones}, {Kendrew}, {Mills}, {Molinari}, {Moore}, {Ott}, {Pillai},
  {Rathborne}, {Schilke}, {Schmiedeke}, {Testi}, {Walker}, {Walsh}, \&
  {Zhang}}]{Henshaw_2016}
{Henshaw}, J.~D., {Longmore}, S.~N., {Kruijssen}, J.~M.~D., {et~al.} 2016,
  \mnras, 457, 2675

\bibitem[{{Ho} {et~al.}(2016){Ho}, {Medling}, {Groves}, {Rich}, {Rupke},
  {Hampton}, {Kewley}, {Bland-Hawthorn}, {Croom}, {Richards}, {Schaefer},
  {Sharp}, \& {Sweet}}]{Ho_2016}
{Ho}, I.~T., {Medling}, A.~M., {Groves}, B., {et~al.} 2016, \apss, 361, 280

\bibitem[{{Ho} \& {Townes}(1983)}]{Ho_1983}
{Ho}, P.~T.~P., \& {Townes}, C.~H. 1983, \araa, 21, 239

\bibitem[{Hochreiter {et~al.}(2001)Hochreiter, Bengio, \&
  Frasconi}]{Hochreiter_2001}
Hochreiter, S., Bengio, Y., \& Frasconi, P. 2001, in Field Guide to Dynamical
  Recurrent Networks, ed. J.~Kolen \& S.~Kremer (IEEE Press)

\bibitem[{{Hogge} {et~al.}(2018){Hogge}, {Jackson}, {Stephens}, {Whitaker},
  {Foster}, {Camarata}, {Anish Roshi}, {Di Francesco}, {Longmore}, {Loughnane},
  {Moore}, {Rathborne}, {Sanhueza}, \& {Walsh}}]{Hogge_2018}
{Hogge}, T., {Jackson}, J., {Stephens}, I., {et~al.} 2018, \apjs, 237, 27

\bibitem[{Kauffmann {et~al.}(2013)Kauffmann, Pillai, \&
  Goldsmith}]{Kauffmann_2013}
Kauffmann, J., Pillai, T., \& Goldsmith, P.~F. 2013, ApJ, 779, 185

\bibitem[{{Keown} {et~al.}(2016){Keown}, {Schnee}, {Bourke}, {Di Francesco},
  {Friesen}, {Caselli}, {Myers}, {Williger}, \& {Tafalla}}]{Keown_2016}
{Keown}, J., {Schnee}, S., {Bourke}, T.~L., {et~al.} 2016, \apj, 833, 97

\bibitem[{{Keown} {et~al.}(2017){Keown}, {Di Francesco}, {Kirk}, {Friesen},
  {Pineda}, {Rosolowsky}, {Ginsburg}, {Offner}, {Caselli}, {Alves},
  {Chac{\'o}n-Tanarro}, {Punanova}, {Redaelli}, {Seo}, {Matzner}, {Chun-Yuan
  Chen}, {Goodman}, {Chen}, {Shirley}, {Singh}, {Arce}, {Martin}, \&
  {Myers}}]{Keown_2017}
{Keown}, J., {Di Francesco}, J., {Kirk}, H., {et~al.} 2017, \apj, 850, 3

\bibitem[{{Kerr} {et~al.}(2019){Kerr}, {Kirk}, {Di Francesco}, {Keown}, {Chen},
  {Rosolowsky}, {Offner}, {Friesen}, {Pineda}, {Shirley}, {Redaelli},
  {Caselli}, {Punanova}, {Seo}, {Alves}, {Chac{\'o}n-Tanarro}, \&
  {Chen}}]{Kerr_2019}
{Kerr}, R., {Kirk}, H., {Di Francesco}, J., {et~al.} 2019, \apj, 874, 147

\bibitem[{{Kingma} \& {Ba}(2014)}]{Kingma_2014}
{Kingma}, D.~P., \& {Ba}, J. 2014, arXiv e-prints, arXiv:1412.6980

\bibitem[{{Kirk} {et~al.}(2013){Kirk}, {Myers}, {Bourke}, {Gutermuth},
  {Hedden}, \& {Wilson}}]{Kirk_2013}
{Kirk}, H., {Myers}, P.~C., {Bourke}, T.~L., {et~al.} 2013, \apj, 766, 115

\bibitem[{{Kirk} {et~al.}(2017){Kirk}, {Friesen}, {Pineda}, {Rosolowsky},
  {Offner}, {Matzner}, {Myers}, {Di Francesco}, {Caselli}, {Alves},
  {Chac{\'o}n-Tanarro}, {Chen}, {Chun-Yuan Chen}, {Keown}, {Punanova}, {Seo},
  {Shirley}, {Ginsburg}, {Hall}, {Singh}, {Arce}, {Goodman}, {Martin}, \&
  {Redaelli}}]{Kirk_2017}
{Kirk}, H., {Friesen}, R.~K., {Pineda}, J.~E., {et~al.} 2017, \apj, 846, 144

\bibitem[{{Lee} {et~al.}(1999){Lee}, {Myers}, \& {Tafalla}}]{Lee_1999}
{Lee}, C.~W., {Myers}, P.~C., \& {Tafalla}, M. 1999, \apj, 526, 788

\bibitem[{{Lindner} {et~al.}(2015){Lindner}, {Vera-Ciro}, {Murray},
  {Stanimirovi{\'c}}, {Babler}, {Heiles}, {Hennebelle}, {Goss}, \&
  {Dickey}}]{Lindner_2015}
{Lindner}, R.~R., {Vera-Ciro}, C., {Murray}, C.~E., {et~al.} 2015, \aj, 149,
  138

\bibitem[{{Lombardi} {et~al.}(2008){Lombardi}, {Lada}, \&
  {Alves}}]{Lombardi_2008}
{Lombardi}, M., {Lada}, C.~J., \& {Alves}, J. 2008, \aap, 480, 785

\bibitem[{{Lovas}(2004)}]{Lovas_2004}
{Lovas}, F.~J. 2004, Journal of Physical and Chemical Reference Data, 33, 177

\bibitem[{{Morgan} {et~al.}(2008){Morgan}, {White}, {Lockman}, {Bryerton},
  {Saini}, {Norrod}, {Simon}, {Srikanth}, {Anderson}, \&
  {Pisano}}]{Morgan_2008}
{Morgan}, M., {White}, S., {Lockman}, J., {et~al.} 2008, in Union Radio
  Scientifique Internationale XXIX General Assembly, J02p1

\bibitem[{{Pattle} {et~al.}(2015){Pattle}, {Ward-Thompson}, {Kirk}, {White},
  {Drabek-Maunder}, {Buckle}, {Beaulieu}, {Berry}, {Broekhoven-Fiene},
  {Currie}, {Fich}, {Hatchell}, {Kirk}, {Jenness}, {Johnstone}, {Mottram},
  {Nutter}, {Pineda}, {Quinn}, {Salji}, {Tisi}, {Walker-Smith}, {Francesco},
  {Hogerheijde}, {Andr{\'e}}, {Bastien}, {Bresnahan}, {Butner}, {Chen},
  {Chrysostomou}, {Coude}, {Davis}, {Duarte-Cabral}, {Fiege}, {Friberg},
  {Friesen}, {Fuller}, {Graves}, {Greaves}, {Gregson}, {Griffin}, {Holland},
  {Joncas}, {Knee}, {K{\"o}nyves}, {Mairs}, {Marsh}, {Matthews},
  {Moriarty-Schieven}, {Rawlings}, {Richer}, {Robertson}, {Rosolowsky},
  {Rumble}, {Sadavoy}, {Spinoglio}, {Thomas}, {Tothill}, {Viti}, {Wouterloot},
  {Yates}, \& {Zhu}}]{Pattle_2015}
{Pattle}, K., {Ward-Thompson}, D., {Kirk}, J.~M., {et~al.} 2015, \mnras, 450,
  1094

\bibitem[{{Pattle} {et~al.}(2017){Pattle}, {Ward-Thompson}, {Kirk}, {Di
  Francesco}, {Kirk}, {Mottram}, {Keown}, {Buckle}, {Beaulieu}, {Berry},
  {Broekhoven-Fiene}, {Currie}, {Fich}, {Hatchell}, {Jenness}, {Johnstone},
  {Nutter}, {Pineda}, {Quinn}, {Salji}, {Tisi}, {Walker-Smith}, {Hogerheijde},
  {Bastien}, {Bresnahan}, {Butner}, {Chen}, {Chrysostomou}, {Coud{\'e}},
  {Davis}, {Drabek-Maunder}, {Duarte-Cabral}, {Fiege}, {Friberg}, {Friesen},
  {Fuller}, {Graves}, {Greaves}, {Gregson}, {Holland}, {Joncas}, {Knee},
  {Mairs}, {Marsh}, {Matthews}, {Moriarty-Schieven}, {Mowat}, {Rawlings},
  {Richer}, {Robertson}, {Rosolowsky}, {Rumble}, {Sadavoy}, {Thomas},
  {Tothill}, {Viti}, {White}, {Wouterloot}, {Yates}, \& {Zhu}}]{Pattle_2017}
---. 2017, \mnras, 464, 4255

\bibitem[{{Peretto} {et~al.}(2014){Peretto}, {Fuller}, {Andr{\'e}},
  {Arzoumanian}, {Rivilla}, {Bardeau}, {Duarte Puertas}, {Guzman Fernandez},
  {Lenfestey}, {Li}, {Olguin}, {R{\"o}ck}, {de Villiers}, \&
  {Williams}}]{Peretto_2014}
{Peretto}, N., {Fuller}, G.~A., {Andr{\'e}}, P., {et~al.} 2014, \aap, 561, A83

\bibitem[{{Pineda} {et~al.}(2010){Pineda}, {Goodman}, {Arce}, {Caselli},
  {Foster}, {Myers}, \& {Rosolowsky}}]{Pineda_2010}
{Pineda}, J.~E., {Goodman}, A.~A., {Arce}, H.~G., {et~al.} 2010, \apjl, 712,
  L116

\bibitem[{{Ridge} {et~al.}(2006){Ridge}, {Di Francesco}, {Kirk}, {Li},
  {Goodman}, {Alves}, {Arce}, {Borkin}, {Caselli}, {Foster}, {Heyer},
  {Johnstone}, {Kosslyn}, {Lombardi}, {Pineda}, {Schnee}, \&
  {Tafalla}}]{Ridge_2006}
{Ridge}, N.~A., {Di Francesco}, J., {Kirk}, H., {et~al.} 2006, \aj, 131, 2921

\bibitem[{{Riener} {et~al.}(2019){Riener}, {Kainulainen}, {Henshaw}, {Orkisz},
  {Murray}, \& {Beuther}}]{Riener_2019}
{Riener}, M., {Kainulainen}, J., {Henshaw}, J.~D., {et~al.} 2019, \aap, 628,
  A78

\bibitem[{{Schlafly} {et~al.}(2014){Schlafly}, {Green}, {Finkbeiner}, {Rix},
  {Bell}, {Burgett}, {Chambers}, {Draper}, {Hodapp}, {Kaiser}, {Magnier},
  {Martin}, {Metcalfe}, {Price}, \& {Tonry}}]{Schlafly_2014}
{Schlafly}, E.~F., {Green}, G., {Finkbeiner}, D.~P., {et~al.} 2014, \apj, 786,
  29

\bibitem[{{Schnee} {et~al.}(2013){Schnee}, {Brunetti}, {Di Francesco},
  {Caselli}, {Friesen}, {Johnstone}, \& {Pon}}]{Schnee_2013}
{Schnee}, S., {Brunetti}, N., {Di Francesco}, J., {et~al.} 2013, \apj, 777, 121

\bibitem[{{Schneider} {et~al.}(2006){Schneider}, {Bontemps}, {Simon}, {Jakob},
  {Motte}, {Miller}, {Kramer}, \& {Stutzki}}]{Schneider_2006}
{Schneider}, N., {Bontemps}, S., {Simon}, R., {et~al.} 2006, \aap, 458, 855

\bibitem[{{Schneider} {et~al.}(2010){Schneider}, {Motte}, {Bontemps},
  {Hennemann}, {di Francesco}, {Andr{\'e}}, {Zavagno}, {Csengeri},
  {Men'shchikov}, {Abergel}, {Baluteau}, {Bernard}, {Cox}, {Didelon}, {di
  Giorgio}, {Gastaud}, {Griffin}, {Hargrave}, {Hill}, {Huang}, {Kirk},
  {K{\"o}nyves}, {Leeks}, {Li}, {Marston}, {Martin}, {Minier}, {Molinari},
  {Olofsson}, {Panuzzo}, {Persi}, {Pezzuto}, {Roussel}, {Russeil}, {Sadavoy},
  {Saraceno}, {Sauvage}, {Sibthorpe}, {Spinoglio}, {Testi}, {Teyssier},
  {Vavrek}, {Ward-Thompson}, {White}, {Wilson}, \&
  {Woodcraft}}]{Schneider_2010}
{Schneider}, N., {Motte}, F., {Bontemps}, S., {et~al.} 2010, \aap, 518, L83

\bibitem[{Schwarz(1978)}]{Schwarz_1978}
Schwarz, G. 1978, The Annals of Statistics, 6, 461

\bibitem[{{Seo} {et~al.}(2015){Seo}, {Shirley}, {Goldsmith}, {Ward-Thompson},
  {Kirk}, {Schmalzl}, {Lee}, {Friesen}, {Langston}, {Masters}, \&
  {Garwood}}]{Seo_2015}
{Seo}, Y.~M., {Shirley}, Y.~L., {Goldsmith}, P., {et~al.} 2015, \apj, 805, 185

\bibitem[{{Shallue} \& {Vanderburg}(2018)}]{Shallue_2018}
{Shallue}, C.~J., \& {Vanderburg}, A. 2018, \aj, 155, 94

\bibitem[{{Sohn} {et~al.}(2007){Sohn}, {Lee}, {Park}, {Lee}, {Myers}, \&
  {Lee}}]{Sohn_2007}
{Sohn}, J., {Lee}, C.~W., {Park}, Y.-S., {et~al.} 2007, \apj, 664, 928

\bibitem[{{Sokolov} {et~al.}(2017){Sokolov}, {Wang}, {Pineda}, {Caselli},
  {Henshaw}, {Tan}, {Fontani}, {Jim{\'e}nez-Serra}, \& {Lim}}]{Sokolov_2017}
{Sokolov}, V., {Wang}, K., {Pineda}, J.~E., {et~al.} 2017, \aap, 606, A133

\bibitem[{{Teimoorinia} \& {Keown}(2018)}]{Teimoorinia_2018}
{Teimoorinia}, H., \& {Keown}, J. 2018, \mnras, 478, 3177

\bibitem[{{Van Oort} {et~al.}(2019){Van Oort}, {Xu}, {Offner}, \&
  {Gutermuth}}]{Van_2019}
{Van Oort}, C.~M., {Xu}, D., {Offner}, S. S.~R., \& {Gutermuth}, R.~A. 2019,
  \apj, 880, 83

\bibitem[{{Xu} {et~al.}(2011){Xu}, {Moscadelli}, {Reid}, {Menten}, {Zhang},
  {Zheng}, \& {Brunthaler}}]{Xu_2011}
{Xu}, Y., {Moscadelli}, L., {Reid}, M.~J., {et~al.} 2011, \apj, 733

\bibitem[{{Zhang} {et~al.}(2018){Zhang}, {Gajjar}, {Foster}, {Siemion},
  {Cordes}, {Law}, \& {Wang}}]{Zhang_2018}
{Zhang}, Y.~G., {Gajjar}, V., {Foster}, G., {et~al.} 2018, \apj, 866, 149

\end{thebibliography}

\end{document}